\documentclass[preprint2,trackchanges]{aastex62}
\usepackage{xspace}
\usepackage{url}
\usepackage{amsmath}

\newcommand{\z}[1]{$z \sim {#1}$}
\newcommand{\FSPS}{{\sc FSPS}\xspace}
\newcommand{\pFSPS}{{\tt \textbf{python-fsps}}\xspace}
\newcommand{\CloudyFSPS}{{\tt \textbf{CloudyFSPS}}\xspace}

\newcommand{\Cloudy}{\textsc{Cloudy}\xspace}

\newcommand{\hii}{H\,{\sc ii}\xspace}
\newcommand{\nii}{[N\,{\sc ii}]\xspace}
\newcommand{\niii}{[N\,{\sc iii}]\xspace}
\newcommand{\sii}{[S\,{\sc ii}]\xspace}

\newcommand{\oiii}{[O\,{\sc iii}]\xspace}
\newcommand{\oii}{[O\,{\sc ii}]\xspace}

\newcommand{\neiii}{[Ne\,{\sc iii}]\xspace}
\newcommand{\heii}{[He\,{\sc ii}]\xspace}

\newcommand{\civ}{C\,{\sc iv}\xspace}
\newcommand{\SiuII}{[Si\,{\sc ii}]\xspace}
\newcommand{\SiuIII}{[Si\,{\sc iii}]\xspace}
\newcommand{\alII}{[Al\,{\sc ii}]\xspace}
\newcommand{\mgii}{[Mg\,{\sc ii}]\xspace}
\newcommand{\ciii}{C\,{\sc iii}]\xspace}
\newcommand{\cii}{[C\,{\sc ii}]\xspace}
\newcommand\vs{\ensuremath{\mathrm{vs.}}\xspace}
\newcommand{\sigmaL}{\ensuremath{\sigma_{\mathrm{line}}}\xspace}
\newcommand{\sigmaR}{\ensuremath{\sigma_{\mathrm{ratio}}}\xspace}
\newcommand{\sigmaRU}{\ensuremath{\sigma_{\mathrm{ratio,}\,\mathcal{U}}}\xspace}
\newcommand{\sigmaRZ}{\ensuremath{\sigma_{\mathrm{ratio,}\,Z}}\xspace}
\newcommand{\sigmaU}{\ensuremath{\sigma_{\mathcal{U}}}\xspace}
\newcommand{\sigmaZ}{\ensuremath{\sigma_{Z}}\xspace}
\newcommand{\femline}{\ensuremath{f_{\mathrm{line}}}\xspace}

\newcommand\Msun{\ensuremath{\mathrm{M_{\sun}}}\xspace}
\newcommand{\ha}{\ensuremath{\mathrm{H\alpha}}\xspace}
\newcommand{\hb}{\ensuremath{\mathrm{H\beta}}\xspace}
\newcommand{\Myr}{$\,$Myr\xspace}
\newcommand{\logten}{\ensuremath{\log_{10}}}
\newcommand{\logz}{\ensuremath{\logten \mathrm{Z}/\mathrm{Z}_{\sun}}\xspace}
\newcommand{\logZeq}[1]{\ensuremath{\logten \mathrm{Z}/\mathrm{Z}_{\sun} = #1}}
\newcommand{\ang}{\ensuremath{\mbox{\AA}}\xspace}

\newcommand{\U}{\ensuremath{\mathcal{U}_{0}}\xspace}
\newcommand{\logU}{\ensuremath{\logten \mathcal{U}_0}}
\newcommand{\logUeq}[1]{\ensuremath{\logten \mathcal{U}_0 = #1}}
\newcommand{\kms}{\ensuremath{\;\mathrm{km}\;\mathrm{s}^{-1}}\xspace}
\newcommand{\fcont}{\ensuremath{\mathcal{F}_{\mathrm{cont}}(\lambda)}\xspace}
\newcommand{\fline}{\ensuremath{\mathcal{F}_{\mathrm{line}}(\lambda)}\xspace}

\begin{document}
\title{Stellar and nebular diagnostics in the UV for star-forming galaxies}
\correspondingauthor{Nell Byler}
\email{ebyler@uw.edu}
\author[0000-0002-7392-3637]{Nell Byler}
\affil{Department of Astronomy, University of Washington, Box 351580, Seattle, WA 98195, USA}
\affil{Research School of Astronomy and Astrophysics, The Australian National University, Cotter Road, ACT 2611, Australia}
\affil{ARC Centre of Excellence for All Sky Astrophysics in 3 Dimensions (ASTRO 3D), Australia}
\author[0000-0002-1264-2006]{Julianne J. Dalcanton}
\affil{Department of Astronomy, University of Washington, Box 351580, Seattle, WA 98195, USA}
\author[0000-0002-1590-8551]{Charlie Conroy}
\affil{Department of Astronomy, Harvard University, Cambridge, MA, USA}
\author[0000-0002-9280-7594]{Benjamin D. Johnson}
\affil{Department of Astronomy, Harvard University, Cambridge, MA, USA}
\author[0000-0003-2184-1581]{Emily M. Levesque}
\affil{Department of Astronomy, University of Washington, Box 351580, Seattle, WA 98195, USA}
\author[0000-0002-4153-053X]{Danielle A. Berg}
\affil{Center for Gravitation, Cosmology and Astrophysics, Department of Physics, University of Wisconsin Milwaukee, 1900 East Kenwood Boulevard, Milwaukee, WI 53211, USA}

\begin{abstract}

There is a long history of using optical emission and absorption lines to constrain the metallicity and ionization parameters of gas in galaxies. However, comparable diagnostics are less well-developed for the UV. Here, we assess the diagnostic potential of both absorption and emission features in the UV and evaluate the diagnostics against observations of local and high redshift galaxies. We use the FSPS nebular emission model of \citet{Byler+2017}, extended to include emission predictions in the UV, to evaluate the metallicity sensitivity of established UV stellar absorption indices, and to identify those that include a significant contribution from nebular emission. We present model UV emission line fluxes as a function of metallicity and ionization parameter, assuming both instantaneous bursts and constant SFRs. We identify combinations of strong emission lines that constrain metallicity and ionization parameter, including [C{\sc iii}]$\lambda1907$, C{\sc iii}]$\lambda1909$, O{\sc iii}]$\lambda1661,1666$, Si{\sc iii}]$\lambda1883,1892$, C{\sc iv}$\lambda$1548,1551, N{\sc ii}]$\lambda1750,1752$, and Mg{\sc ii}$\lambda2796$, and develop UV versions of the canonical ``BPT'' diagram. We quantify the relative contribution from stellar wind emission and nebular line emission to diagnostic line ratios that include the C{\sc iv}$\lambda$1548,1551 lines, and also develop an observationally motivated relationship for N and C enrichment that improves the performance of photoionization models. We summarize the best diagnostic choices and the associated redshift range for low-, mid-, and high-resolution rest-UV spectroscopy in preparation for the launch of JWST.

\end{abstract}
\section{Introduction} \label{sec:intro}

In the optical spectra of star-forming galaxies, emission lines from ionized gas can be a factor of ten times brighter than the stellar continuum. These emission lines can also reveal key physical properties of the gas and its ionizing source. For gas that is ionized by stars, various combinations of emission lines are frequently used to estimate the gas-phase metallicity of a galaxy. Popular combinations of emission lines include the ``R23'' \citep{Pagel+1979} and ``N2''\citep{Pettini+2004} diagnostics, which use ratios of strong emission lines from Oxygen, Nitrogen, and Hydrogen that have been calibrated by theoretical models \citep[e.g.,][]{McGaugh+1991, Kewley+2002}.

With the historical accessibility of optical observations, it is a natural development that the most well-used nebular diagnostics are based in the optical portion of the spectrum. Unfortunately, however, by \z{3}, the emission lines required for traditional optical diagnostics have been redshifted out of the wavelength window accessible to ground-based near-IR observatories. Luckily, strong nebular emission lines exist across the spectrum, from the UV through the IR, which can potentially be used as equally informative gas diagnostics. Diagnostics based on UV emission lines are of particular interest, since they are applicable to galaxies at high redshift.

The UV also offers a number of informative absorption features as well. A galaxy's spectrum is typically dominated by light from young, massive stars in addition to the emission from the ionized gas surrounding them. While there are fewer emission lines in the UV than in the optical, we gain additional diagnostic features from the numerous stellar UV absorption lines. Many of these line are produced in the atmospheres of young massive stars that should share the same metallicity as the gas producing the emission lines. Because these same O- and B-type stars are responsible for producing the ionizing radiation incident to the gas cloud, it is possible to jointly analyze absorption and emission features when inferring the metallicity and ionization state of a galaxy. However, doing so requires that one self-consistently links the temporal and metallicity-dependent changes in the stellar absorption features to the nebular emission lines.

In this paper, we take advantage of the work of \citet{Byler+2017}, who paired the population synthesis models from the Flexible Stellar Population Synthesis code \citep[\FSPS; ][]{Conroy+2009, Conroy+2010} with photoionization models from \Cloudy to self-consistently model the flux from star-forming galaxies. We build on the UV absorption indices identified in \citet{Leitherer+2011}, focusing on the wind-driven and photospheric absorption lines, and the UV emission features identified in \citet{Erb+2010}. We present the most promising combinations of nebular emission line ratios and stellar absorption features in the UV that can be used to estimate stellar and nebular metallicities for galaxies at high and low redshift.

We describe the stellar and nebular model in \S\ref{sec:model}. In \S\ref{sec:modspec} we identify combinations of emission and absorption features in the UV that track the gas-phase metallicity. In \S\ref{sec:obs} we compare the most promising diagnostics to galaxies locally and at redshift \z{3} and in \S\ref{sec:feas} discuss the utility of the various diagnostics in the context of current and future observatories.

\section{Description of the stellar and nebular model} \label{sec:model}

\subsection{The stellar model}\label{sec:model:stellar}

For stellar population synthesis, we use the Flexible Stellar Population Synthesis package \citep[\FSPS;][]{Conroy+2009, Conroy+2010} via the Python interface, \pFSPS \citep{pythonFSPSdfm}. We adopt a Kroupa initial mass function \citep[IMF;][]{Kroupa+2001} with an upper and lower mass limit of 120\Msun and 0.08\Msun, respectively. We discuss our choice for evolutionary tracks and spectral library in detail below; for all other SPS parameters, we use the default parameters in \FSPS\footnote{GitHub commit hash \texttt{4e1b3f5}}.

\subsubsection{Evolutionary Tracks}
We use the MESA Isochrones \& Stellar Tracks \citep[MIST;][]{Dotter+2016, Choi+2016}, single-star stellar evolutionary models which include the effect of stellar rotation. The evolutionary tracks are computed using the publicly available stellar evolution package Modules for Experiments in Stellar Astrophysics \citep[MESA v7503;][]{Paxton+2011,Paxton+2013, Paxton+2015} and the isochrones are generated using Aaron Dotter's publicly available \texttt{iso} package on github\footnote{https://github.com/dotbot2000/iso}. A complete description of the models can be found in \citet{Choi+2016}, however, we describe some relevant model properties below.

The MIST models cover ages from $10^5$ to $10^{10.3}$ years, initial masses from $0.1$ to $300\,$\Msun, and metallicities in the range of $-2.0 \leq$ $[\mathrm{Z}/\mathrm{H}]$ $\leq 0.5$. Abundances are solar-scaled, assuming the \citet{Asplund+2009} protosolar birth cloud bulk metallicity, for a reference solar value of $\mathrm{Z}_{\odot} = 0.0142$.

The MIST models include the effect of rotation, which is particularly important for massive star evolution. We briefly describe the implementation of rotation in the MIST models below. The reader is referred to Section 3 of \citet{Choi+2016} for a comprehensive description. As explained in \citet{Choi+2017}, the effects of rotation appear in the MESA stellar evolution calculations in three main ways. First, rotation decreases the gravitational acceleration via the centrifugal force, which in turn affects the stellar structure. Second, rotation can promote extra mixing in the interior, boosting the transport of chemicals and angular momentum. MESA adopts the common approach of treating the chemical and angular momentum transport in a diffusion approximation. Finally, rotation enhances mass loss. MESA adopts the formulation from \citet{Langer+1998}, where the mass loss rate $\dot{M}$ is multiplied by a factor that increases dramatically as the surface angular velocity $\Omega$ approaches critical, or break-up, angular velocity.

The MIST models assume an initial rotation rate of  $\nu_{\mathrm{ZAMS}}/\nu_{\mathrm{crit}} = 0.4$, meaning that the the surface velocity is set to 40\% of the critical, or break-up, velocity. This rotation rate is also adopted in the Geneva models \citep{Ekstrom+2012}\footnote{Note that $\nu_{\mathrm{crit}}$ and $\Omega_{\mathrm{crit}}$ are defined differently in MESA and in the Geneva models. In the Geneva models, the equatorial radius is 1.5 times larger than the polar radius when $\Omega = \Omega_{\mathrm{crit}}$ but this distinction is not made in MESA. See Section 2.1 in \citet{Georgy+2013} for more details.}, and is motivated by observations of young B stars \citep{Huang+2010} and theoretical work on rotation rates in massive stars \citep{Rosen+2012}. Rotation is not included for stars with $M_{i} \leq 1.2$\Msun to match the slow rotation rate observed in the Sun and other low mass stars. The rotation rate is gradually increased from 0 to $\nu_{\mathrm{ZAMS}}/\nu_{\mathrm{crit}} = 0.4$ over the stellar mass range $1.2-1.8$\Msun. This gradual ramp-up produces velocities of order $10-25$\kms in this mass range, in agreement with observed rotation rates of MS and evolved stars in the mass range 1.2-1.5\Msun \citep{Wolff+1997, Canto+2011}, see \citet{Choi+2016} for details.

\citet{Choi+2016} compared the rotating MIST and Geneva evolutionary tracks and found that at fixed stellar mass, the Geneva models are hotter and more luminous at the end of their main sequence lifetimes, implying that rotational mixing is more efficient in the Geneva models. In general, the Geneva rotating models have enhanced main sequence lifetimes compared to the default MIST model; however, main sequence lifetimes in Geneva and MIST agree to within 10-15\% at solar metallicity. For higher mass stars the MIST models have MS lifetimes that fall between those of the non-rotating and rotating Geneva models.

In general, rotating stellar populations have higher bolometric luminosities and sustain harder ionizing radiation fields over a longer period of time compared to non-rotating populations \citep[e.g.,][]{Levesque+2012, Choi+2017}. While a full analysis of UV properties as a function of rotation rate is beyond the scope of this work, we note that the rotating MIST models adopted here produce ionizing luminosities that are consistent with recent observations of young nuclear star clusters. We also refer the reader to \citet{Choi+2017} for an in-depth investigation on the ionizing properties of massive rotating populations for a range of initial rotation rates ($\nu_{\mathrm{ZAMS}}/\nu_{\mathrm{crit}}$ from 0.0 to 0.6) and composite populations with a distribution of rotation rates.

For hot luminous stars, line-driven winds can cause significant mass loss. For stars with $T_{\mathrm{eff}}>1.1\times10^4\,$K and surface hydrogen mass fraction $X_{\mathrm{surf}}>0.4$, the mass loss prescription from \citet{Vink+2000, Vink+2001} are adopted. If the star loses a considerable amount of its outer hydrogen layer ($X_{\mathrm{surf}}<0.4$) and becomes a Wolf-Rayet (WR) star, the \citet{Nugis+2000} wind prescription is used instead. For cooler stars with effective temperatures below $10^4\,$K, the wind prescription from \citet{deJager+1988} is used. As noted in a recent review, these prescriptions fail to account for clumpiness and inhomogeneities in outflows, and may overestimate mass loss rates by a factor of 2 to 3 \citep{Smith+2014}.

The MIST models used in this work do not account for binary stars. This is an additional source of uncertainty, since O and B-type stars are observed to have a high binary fraction, between 0.6 and 1.0 \citep{Sana+2012, Kobulnicky+2014}, and the presence of a binary companion can dramatically alter the evolutionary path of a massive star \citep[e.g.,][]{deMink+2013, Smith+2014}. The Binary Population and Spectral Synthesis models \citep[BPASS;][]{Eldridge+2009, Eldridge+2017} include binary interactions and have found that binary mass transfer can substantially increase the ionizing photon output from a stellar population and prolong ionizing photon production for $\sim10$\,\Myr over single-star models \citep{Ma+2016, Xiao+2018}.

\subsubsection{Spectral Library}

We combine the MIST tracks with a new, high resolution theoretical spectral library (C3K; Conroy, Kurucz, Cargile, Castelli, \emph{in prep}) based on Kurucz stellar atmosphere and spectral synthesis routines \citep[ATLAS12 and SYNTHE,][]{Kurucz+2005}. The spectra use the latest set of atomic and molecular line lists and include both lab and predicted lines. The grid was computed assuming the \citet{Asplund+2009} solar abundance scale and a constant microturbulent velocity of 2\kms.

We require high resolution model spectra, since stellar photospheric lines have equivalent widths ${\sim}1$-$3$\ang. Typical stellar wind lines have equivalent widths (EWs) of ${\sim}5$-$15$\ang. We use a version of the C3K spectral library that has $R=1000$ from $\lambda=100$-$1500$\ang and $R=10,000$ from $\lambda=1500$\ang - $1\mu$m.

For very hot stars and stars in rapidly evolving evolutionary phases, we use alternative spectral libraries. We highlight the libraries used for very hot stars, since these will be responsible for providing the radiation necessary to ionize hydrogen. For main sequence stars with temperatures above 25,000$\,$K (O- and B-type stars) we use spectra from M. Ng, G. Taylor \& J.J. Eldridge (priv. comm),  as described in \citet{Eldridge+2017}, computed using WM-Basic \citep{Pauldrach+2001}. For WR stars, we use the spectral library from \citet{Smith+2002} using CMFGEN \citep{Hillier+2001}, which includes both WN and WC subtypes. Stars classified as WR stars with surface C/O ratio ${>}1$ are labeled as WC, while WR stars with surface C/O ratio ${\leq}1$ are labeled as WN.

\subsection{The nebular model}\label{sec:model:neb}

We use the \Cloudy nebular model implemented within \FSPS, \CloudyFSPS \citep{cloudyFSPSv1}, to generate spectra that include nebular line and continuum emission. Calculations were performed with version 13.03 of \Cloudy, last described by \citet{Ferland+2013}. The model is described in detail in \citet{Byler+2017}, and summarized briefly here. The nebular model is a grid in (1) simple stellar population (SSP) age, (2) SSP and gas-phase metallicity, and (3) ionization parameter, \U, a dimensionless quantity that gives the ratio of ionizing photons to the total hydrogen density.

The model uses \FSPS to generate spectra from coeval clusters of stars, each with a single age and metallicity (SSPs). Using the photoionization code \Cloudy, the SSP is used as the ionization source for the gas cloud and the gas-phase metallicity is scaled to the metallicity of the SSP. For each SSP of age $t$ and metallicity $Z$, photoionization models are run at different ionization parameters, \U, from \logUeq{-4} to \logUeq{-1} in steps of 0.5. The resultant line and continuum emission is normalized by the number of ionizing photons calculated from the input ionizing spectrum. The normalized line and continuum emission are recorded in separate look-up tables.

For a given SSP $(t, Z)$ and specified \U{}, \FSPS returns the associated line and continuum emission associated with that grid-point from the look-up table. This approach maintains the model self-consistency, such that the nebular emission is added to the same spectrum that was used to ionize the gas cloud. \FSPS removes the ionizing photons from the SED to enforce energy balance; in this work we assume an ionizing photon escape fraction of zero.

The set of 128 emission lines used in \citet{Byler+2017} included 31, 51, and 45 lines in the UV, optical, and IR, respectively. To fully explore the emission properties in the UV, we have created an ``extended'' line list that includes 381 emission lines in total, with 142 emission lines in the UV ($\lambda \leq 3000$\ang), 122 emission lines in the optical ($3000$\ang $< \lambda \leq $ 10,000\ang), and 117 in the IR ($\lambda \geq$ 10,000\ang). While this list is not exhaustive, most of these additional emission lines are relatively weak and there are thus no quantitative differences in the synthesized colors for populations using the 128 line list and populations using the 381 line list. The full list of emission lines is given in Appendix~\ref{appdx:lines}, Table~\ref{tab:emLines}.

\subsubsection{Gas Phase Abundances}

We assume that the gas phase metallicity scales with the metallicity of the stellar population, given that the metallicity of the most massive stars should be identical to the metallicity of the gas cloud from which the stars formed.

For most elements we use the solar abundances from \citet{Grevesse+2010}, based on the results from \citet{Asplund+2009}, and adopt the depletion factors specified by \citet{Dopita+2013}.

We treat nitrogen and carbon differently, however, due to the likelihood that both these elements scale non-linearly with metallicity \citep[e.g.,][]{VilaCostas+1993, Henry+2000, Berg+2016}. Nitrogen has known secondary nucleosynthetic production at high metallicity, wherein nitrogen is dredged up during the bottleneck step of the CNO cycle, which is directly dependent on metallicity \citep{Cowley+1995}. In the case of carbon, additional carbon is returned to the ISM through metallicity-dependent processes such as stellar winds \citep{Garnett+1999}. This is not a nucleosynthetic process and thus, carbon said to have a ``pseudo-secondary'' production process.

To set the relationship between N/H and O/H we use the following equation:
\begin{equation}\label{eq:nitrogen}
\begin{aligned}
    \log_{10}&(\mathrm{N}/\mathrm{O}) = \\
    & -1.5 + \log\left( 1 + e^{\frac{12 + \log_{10}(\mathrm{O}/\mathrm{H})-8.3}{0.1}}\right),
\end{aligned}
\end{equation}

and for C/H and O/H we assume:
\begin{equation}\label{eq:carbon}
\begin{aligned}
    \log_{10}&(\mathrm{C}/\mathrm{O}) = \\
    & -0.8 + 0.14\cdot\left(12 + \log_{10}(\mathrm{O}/\mathrm{H})-8.0\right) \\
    & + \log\left( 1 + e^{\frac{12 + \log_{10}(\mathrm{O}/\mathrm{H})-8.0}{0.2}}\right),
\end{aligned}
\end{equation}

These choices are largely modeled after the empirical fits used by \citet{Dopita+2013}. However, we have found that the N/H and C/H relationship used in \citet{Dopita+2013} is too steep at the lowest metallicities. We discuss this issue at length in Appendix~\ref{appdx:CO}, given that the predicted emission lines are sensitive to the abundance prescription applied, especially for important coolants like C, N, and O.

The abundance for each element and applied depletion factors at solar metallicity are given in Table~\ref{tab:solarAbunds}.

\begin{deluxetable}{lcc}[b!]
\tabletypesize{\footnotesize}
\tablecolumns{3}
\tablecaption{Elemental abundances and adopted depletion factors $D$ for each element in the nebular model at solar metallicity, which has $Z=0.0142$ ($\log_{10}$(O/H) = -3.31 or $12+\log_{10}$(O/H) = 8.69).}
\tablehead{
\colhead{Element} &
\colhead{$\log_{10}(\mathrm{E}/\mathrm{H})$} &
\colhead{$\log_{10}(D)$}
}
\startdata
H   & 0	& 0 \\
He  & -1.01 & 0 \\
C   & -3.57 & -0.30 \\
N   & -4.60 & -0.05 \\
O   & -3.31 & -0.07 \\
Ne  & -4.07 & 0 \\
Na  & -5.75 & -1.00 \\
Mg  & -4.40 & -1.08 \\
Al  & -5.55 & -1.39 \\
Si  & -4.49 & -0.81 \\
S   & -4.86 & 0 \\
Cl  & -6.63 & -1.00 \\
Ar  & -5.60 & 0 \\
Ca  & -5.66 & -2.52 \\
Fe  & -4.50 & -1.31 \\
Ni  & -5.78 & -2.00 \\
\enddata
\tablecomments{Solar abundances are from \citet{Grevesse+2010} and depletion factors are from \citet{Dopita+2013}.}
\label{tab:solarAbunds}
\end{deluxetable}

\section{Model Spectra} \label{sec:modspec}

\subsection{Emission and absorption features in the UV} \label{sec:mod:spec}

The UV portion of a galaxy spectrum is rich in absorption features, including stellar absorption features and absorption from the interstellar gas and circumgalactic medium \citep{Leitherer+2011}. While there has been work highlighting the utility of UV interstellar absorption features \citep[e.g.,][]{Rix+2004} as metallicity diagnostics, in this work we focus primarily on stellar features produced by winds and photospheric absorption, since these will link the metallicity-dependent changes in the stellar population with the metallicity-dependent changes in the gas. For short-lived massive stars, the metallicity of the stars should be nearly identical to that of the surrounding natal gas cloud.

\begin{figure*}
  \begin{center}
    \includegraphics[width=\linewidth]{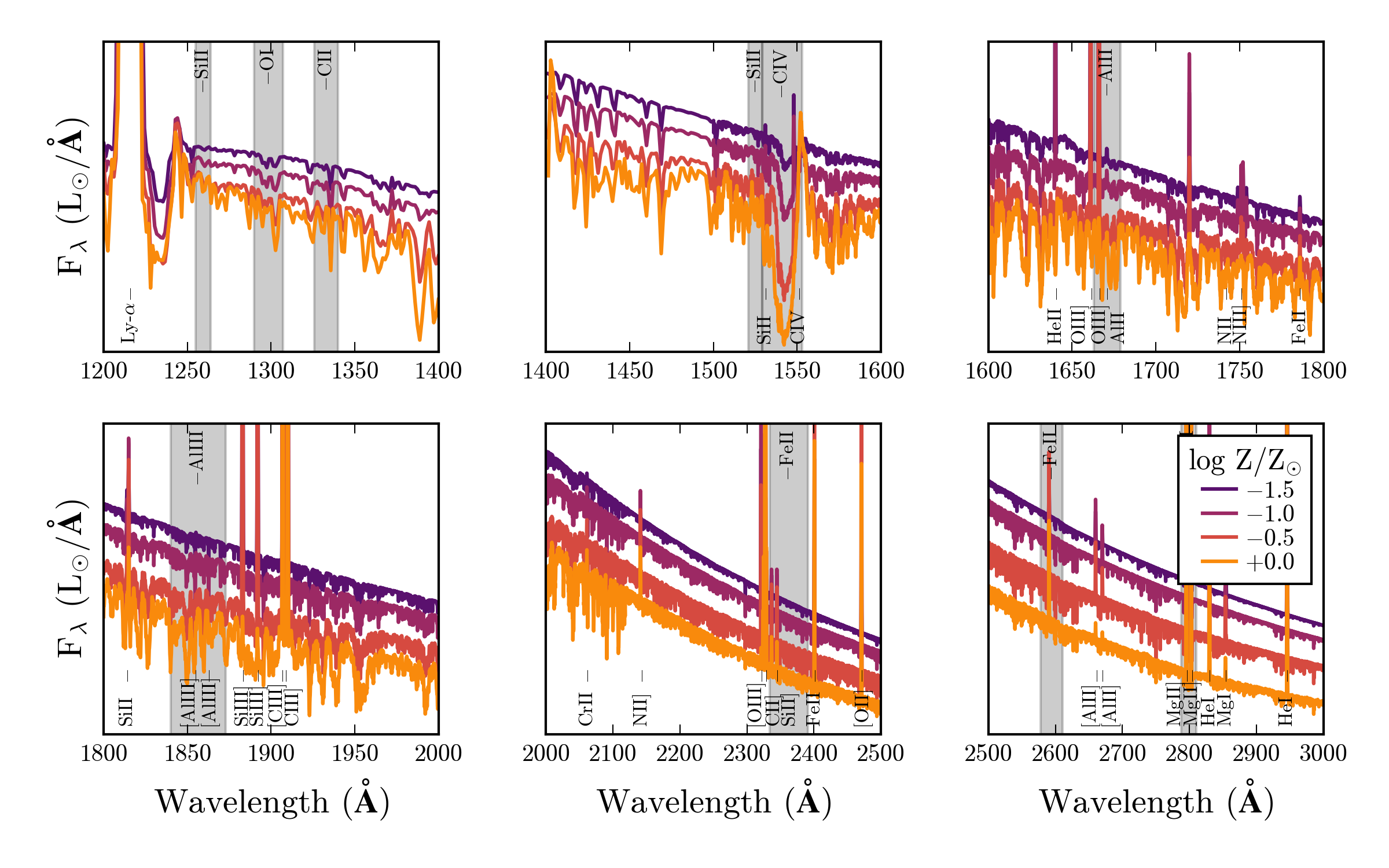}
    \caption{The spectrum for a stellar population with constant SFR at 10\Myr with \logUeq{-2.5} at \logZeq{-1.5},$-1.0$,$-0.5$, and $0.0$. The shaded regions highlight the L11 absorption indices, labeled at the top of the axes. The labels at the bottom of the axes identify strong emission lines included in the \FSPS nebular model. The shape of the nebular continuum and the strength of the emission lines vary with age and metallicity as the thermodynamic properties of the gas cloud change.}
    \label{fig:FullSpec}
  \end{center}
\end{figure*}

To highlight various emission and absorption features, in Fig.~\ref{fig:FullSpec} we show example spectra for a population with constant SF over 10\Myr at \logZeq{-1.5},$-1.0$,$-0.5$, and $0.0$ over the wavelength range 1275-3100\ang. In general, the depth and number of photospheric absorption lines increase with increasing metallicity. Emission from line-driven winds is stronger in high metallicity models as well, like the  C\textsc{iv} wind feature near 1550\ang.

\begin{figure*}
  \begin{center}
    \includegraphics[width=\linewidth]{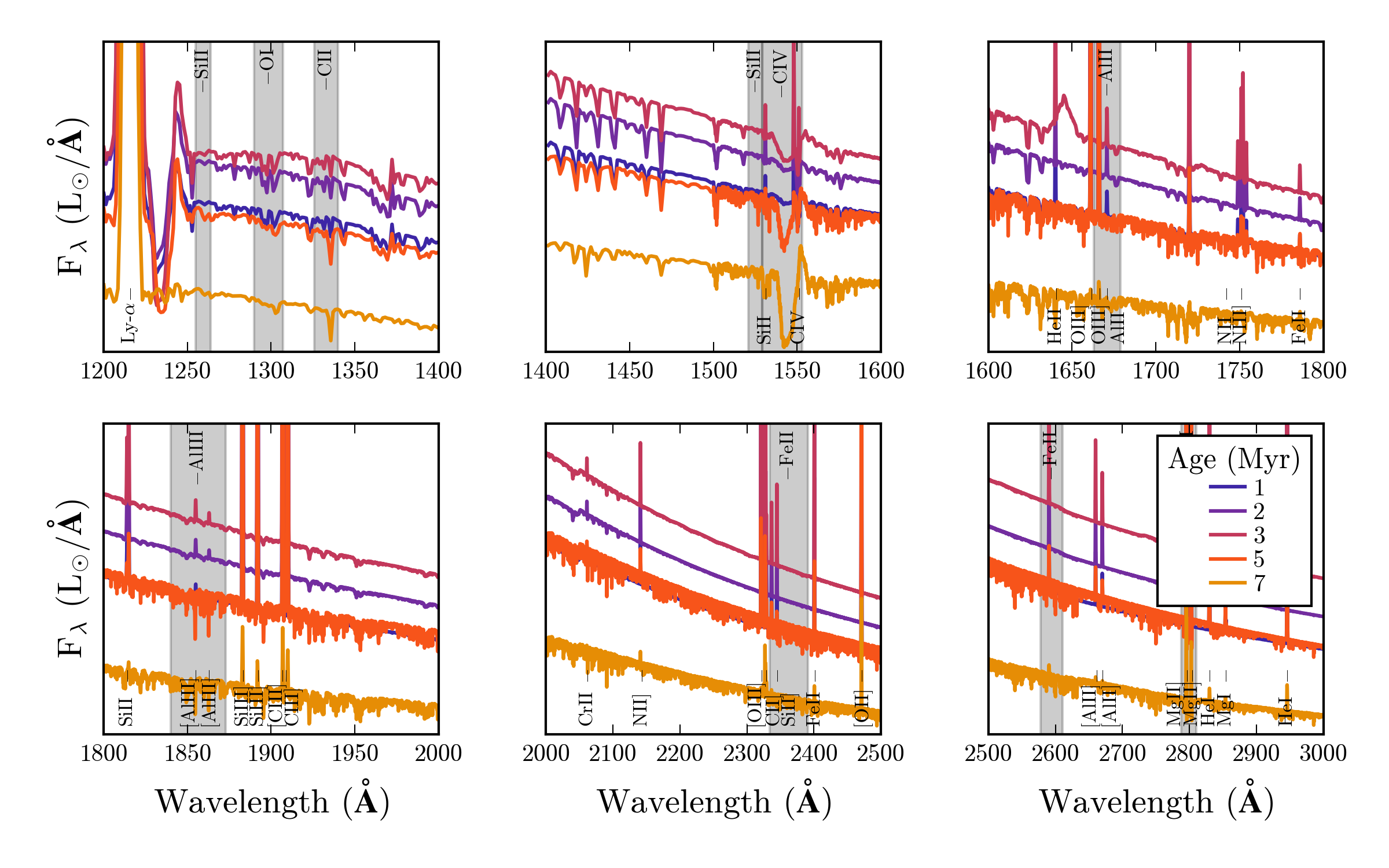}
    \caption{The spectrum for \logZeq{-0.5} instantaneous burst populations with \logUeq{-2.5} at 1, 2, 3, 5 and 7\Myr. The shaded regions highlight the L11 absorption indices, labeled at the top of the axes. The labels at the bottom of the axes identify strong emission lines included in the \FSPS nebular model. The shape of the nebular continuum and the strength of the emission lines vary with age and metallicity as the thermodynamic properties of the gas cloud change.}
    \label{fig:FullSpecAge}
  \end{center}
\end{figure*}

We show a model with \logZeq{-0.5} at several different ages in Fig.~\ref{fig:FullSpecAge} to demonstrate the evolution of the UV spectrum with time. Nebular line and continuum emission is strongest between 1 and 3\Myr, with significant line emission sustained to 7\Myr. The time dependence of some stellar wind features is apparent, with the strength of the \civ $\lambda1551$ wind emission increasing until 7\Myr. Broad \heii$\lambda\,1640$ emission can be seen briefly at 3\Myr due to the presence of WR stars.

In Figs.~\ref{fig:FullSpec} \& \ref{fig:FullSpecAge} we have highlighted the set of 12 UV line indices defined by \citet{Leitherer+2011} in grey. Most of these indices are dominated by lines from interstellar gas, which are not included in this model. Exceptions include the {\tt SiIV\_1400} and {\tt CIV\_1550} indices, which are typically dominated by stellar-wind lines, and {\tt AlIII\_1670}, which is a blend of wind and interstellar lines. The depth of these features changes with metallicity, since the efficiency of hot star winds is metallicity-dependent.

We also highlight important emission lines, noted by their species name below the continuum. The strength of the various emission lines will depend on the gas phase metallicity, the ionization parameter, and the hardness of the ionizing spectra, which will change with the age and metallicity of the stellar population responsible for ionizing the surrounding gas.

The nebular model as implemented within \FSPS self-consistently predicts the line and continuum emission for a stellar population with an arbitrary star formation history and chemical evolution history. However, in the construction of diagnostics, it is often useful to highlight the limiting cases: the instantaneous burst and continuous star formation. For all diagnostics presented in this work, we thus present a model with an instantaneous burst of star formation 1\Myr ago and a model with constant star formation over 10\Myr. We justify this choice at the end of this section in \ref{sec:mod:age}.

\subsection{Measuring emission and absorption line features}\label{sec:mod:meas}

One of the challenges associated with measuring the strength of absorption lines is defining the continuum. In the UV this choice is especially difficult, since metals in the stellar atmosphere absorb heavily, masking the true continuum level. Without knowledge of the true continuum level, equivalent width measurements will only place lower limits on the absolute line strength.

The high density of metal lines in the UV presents another difficulty for absorption line equivalent width measurements, as absorption line indices often cover multiple blended absorption features. While blended features may still track the overall metallicity, many prominent absorption lines have complicated dependencies on stellar spectral type or individual abundance patterns \citep{Maraston+2009}.

To measure absorption feature equivalent widths, we adopt the set of 12 UV line indices defined by \citet{Leitherer+2011}, referred to as L11 in the remainder of this work. For each index, L11 defined a central bandpass and two flanking continuum bandpasses, given in Table~\ref{tab:EWdefs}.

Following L11, the continuum flux in a region is defined as the median value of the flux within the bandpass, and is assigned to the midpoint wavelength. The continuum, \fcont, is then taken to be the straight line connecting the midpoints of the continuum bands on either side of a line. We then calculate the equivalent width with:

\begin{equation}
\mathrm{EW\;(\ang)} = \int^{\lambda_f}_{\lambda_i} \frac{\mathcal{F}_{\lambda,\,\mathrm{cont}}(\lambda) - \mathcal{F}_{\lambda,\,\mathrm{line}}(\lambda)}{\mathcal{F}_{\lambda,\,\mathrm{cont}}(\lambda)}\;d\lambda ,
\end{equation}
where \fline is the flux in the central bandpass spanning $\lambda_i < \lambda < \lambda_f$.

Much of the work analyzing the UV absorption line behavior of massive stars and stellar populations only considered the effects of stellar absorption and interstellar absorption \citep[e.g.,][]{Rix+2004, Maraston+2009, Leitherer+2011,Zetterlund+2015}. This assumption is reasonable for comparisons against observations of globular clusters and some young massive star clusters. However, the UV spectrum of star forming galaxies contains significant contributions from nebular lines and continuum emission from ionized gas.

\begin{figure*}
  \begin{center}
    \includegraphics[width=0.9\linewidth]{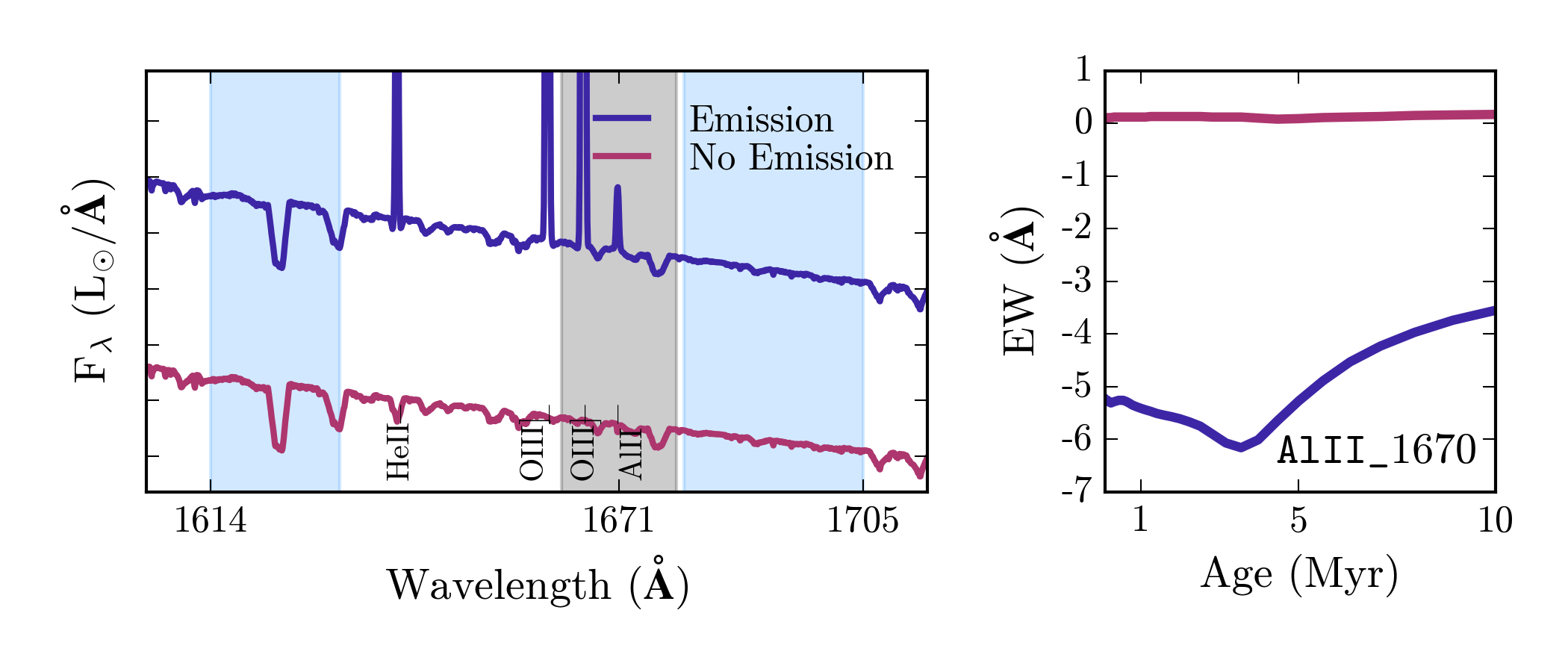}
    \caption{Including nebular line and continuum emission changes equivalent widths measured in standard UV absorption indices. \emph{Left:} The UV spectrum surrounding the Al{\sc \,iii} index for models with and without nebular emission. \emph{Right:} The {\tt AlIII\_1670} index equivalent width as a function of time for single burst models with and without nebular emission at solar metallicity. Nebular continuum emission increases the overall continuum level. Nebular line emission occurs throughout the spectrum, and in the case of {\tt AlIII\_1670}, there are emission lines included in the feature bandpass.}
    \label{fig:EWnebular}
  \end{center}
\end{figure*}

The nebular continuum emits significantly in the UV blueward of the Balmer break\footnote{The nebular continuum can account for $\sim5-10\%$ of the total flux between 900 and 1800\ang, and as much as $\sim20\%$ of the total flux between 1800 and 4000\ang\citep{Byler+2017}.}, which acts to increase the continuum level for the entire spectrum, reducing the equivalent width compared to a pure stellar prediction. In addition, prominent emission lines from the gas surrounding star-forming regions fall within many of the defined line and continuum bandpasses (as seen in Fig.~\ref{fig:EWnebular}), which also affects the calculated equivalent widths.

We show the effect of adding in nebular line and continuum emission to a solar-metallicity single-burst stellar population in Fig.~\ref{fig:EWnebular}, where we plot the UV spectrum near the absorption line and continuum bandpasses for the L11 {\tt AlIII\_1670} absorption index. When nebular emission is included, the spectrum becomes brighter and contains line emission throughout the plotted wavelength range, including in the continuum and feature bandpasses. Nebular emission changes the expected EWs by 10-50\% in most cases, but by as much as an order of magnitude for cases like \texttt{MgII\_2800} where the central bandpass covers three emission lines. The effect of including nebular emission is most important for young populations; for bursts older than 10\Myr, the difference in the absorption index EW when compared to the stellar only EW is less than 5\%.

In what follows, we include the effects of both emission and absorption when computing absorption line equivalent widths.For nebular emission line ratios, we use the full predicted flux from each line and do not reduce the emission line flux due to underlying stellar absorption, since this is already accounted for in most observational comparisons.

\begin{deluxetable*}{lcccl}
\tabletypesize{\footnotesize}
\tablecolumns{6}
\tablecaption{UV Line Indices from \citet{Leitherer+2011}}
\tablehead{
\colhead{Index Name} &
\colhead{Index Bandpass} &
\colhead{Blue Continuum} &
\colhead{Red Continuum} &
\colhead{Features Included}
}
\startdata
{\tt SiII\_1260} & $1255-1264$ & $1268-1286$ & $1268-1286$ & Si{\sc \,ii}$\lambda\,$1260; S{\sc \,ii}$\,\lambda\,$1259\\
{\tt OI\_SiII\_1303} & $1290-1307$ & $1268-1286$ & $1308-1324$ & O{\sc \,i}$\,\lambda\,$1302; Si{\sc \,ii}$\,\lambda\,$1304; Si{\sc \,iii}$\,\lambda\,$1295\\
{\tt CII\_1335} & $1326-1340$ & $1308-1324$ & $1348-1378$ & C{\sc \,ii}$\,\lambda\,$1334; C{\sc \,ii*}$\,\lambda\,$1335\\
{\tt SiIV\_1400} & $1379-1405$ & $1348-1378$ & $1433-1460$ & Si{\sc \,iv}$\,\lambda\,1393,1402$\\
{\tt SiII\_1526} & $1521-1529$ & $1460-1495$ & $1572-1599$ & Si{\sc \,ii}$\,\lambda\,$1526\\
{\tt CIV\_1550}\tablenotemark{e} & $1529-1553$ & $1460-1495$ & $1583-1599$ & C{\sc \,iv}$\,\lambda\,$1548$^e$,1550$^e$; Si{\sc \,ii*}$\,\lambda\,$1533\\
{\tt FeII\_1608} & $1600-1613$ & $1583-1599$ & $1614-1632$ & Fe{\sc \,ii}$\,\lambda\,$1608\\
{\tt AlII\_1670}\tablenotemark{e} & $1663-1679$ & $1614-1632$ & $1680-1705$ & Al{\sc \,ii}$\,\lambda\,1670^e$; ([O{\sc \,iii}]$\,\lambda\,1666$, Al{\sc \,ii}$\,\lambda\,1670$)\\
{\tt AlIII\_1860}\tablenotemark{e} & $1840-1873$ & $1815-1839$ & $1932-1948$ & Al{\sc \,iii}$\,\lambda\,$1854,1862\\
{\tt FeII\_2370}\tablenotemark{e} & $2334-2391$ & $2267-2290$ & $2395-2450$ & Fe{\sc \,ii}$\,\lambda\,$2344,2374,2382\\
{\tt FeII\_2600}\tablenotemark{e} & $2578-2611$ & $2525-2572$ & $2613-2674$ & Mn{\sc \,ii}$\,\lambda\,$2576,2594,2606; Fe{\sc \,ii}$\,\lambda\,$2586,2600\\
{\tt MgII\_2800}\tablenotemark{e} & $2788-2810$ & $2720-2785$ & $2812-2842$ & Mg{\sc \,ii}$\,\lambda\,$2796$^e$,2803\\
\enddata
\tablenotetext{e}{One or more emission lines falls within a defined bandpass for this index}
\tablecomments{The last column includes only stellar absorption features. If present, nebular emission lines are indicated in parentheses. Lines that can appear in emission or absorption are flagged with $e$.}
\label{tab:EWdefs}
\end{deluxetable*}


\subsection{Age sensitivity}\label{sec:mod:age}

The same O- and B-type stars that produce the absorption lines in Figs.~\ref{fig:FullSpec} \& \ref{fig:FullSpecAge} are also responsible for providing the ionizing radiation incident to the surrounding natal gas cloud, producing nebular line and continuum emission. These stars are short-lived, thus most of the UV spectral features are age-dependent. For nebular emission models, it is common to provide model grids at the two limiting cases of star formation: models where star formation occurs in one instantaneous burst and models where the star formation is constant with time \citep[e.g.,][]{Dopita+2000, Kewley+2001, Dopita+2006, Levesque+2012, Dopita+2013, Gutkin+2016}.

\begin{figure*}
  \begin{center}
    \includegraphics[width=\linewidth]{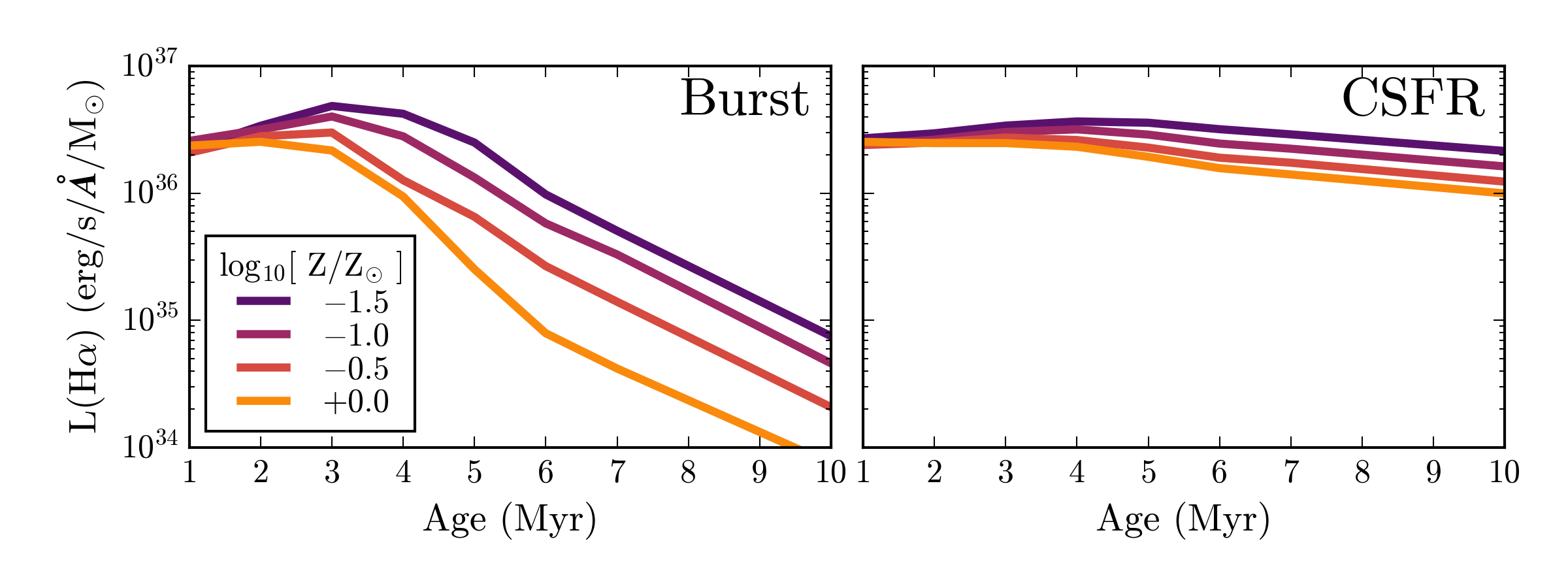}
    \caption{The time evolution of \ha emission as a function of metallicity for models assuming an instantaneous burst (\emph{left}) and a constant SFR (1\Msun/year; \emph{right}) with \logUeq{-2.5}. \ha fluxes are normalized to 1\Msun. Nebular emission is strongest at young ages, with the bulk of the emission from populations 3-5\Myr and younger. Models with a constant SFR reach a ``steady-state'' after several Myr, where the rate of stars forming and the rate of stars dying is in equilibrium.}
    \label{fig:EmTimeEvol}
  \end{center}
\end{figure*}

Instantaneous burst models are only useful for young stellar populations and very recent starbursts. In the left panel of Fig.~\ref{fig:EmTimeEvol} we show the time evolution of \ha for instantaneous burst models at several metallicities. As expected, the emission is strongest at young ages, and eventually disappears entirely as the massive star population required to ionize the surrounding gas evolves off of the main sequence (MS). The exact timescales associated with strong nebular emission will depend on the underlying stellar evolutionary models. For models that do not include stellar rotation or multiplicity, significant nebular emission persists for at most 3-4\Myr. Models that include stellar rotation can extend the lifetime of nebular emission up to 5-7\Myr \citep[e.g., MIST, used in this work;][]{Byler+2017, Choi+2017}. For evolutionary tracks that include the effects of stellar multiplicity, the lifetime of nebular emission can extend to greater than 10\Myr\citep[e.g., BPASS;][]{Eldridge+2012, Stanway+2014, Xiao+2018}.

For models with continuous star formation, the rate at which stars die and are born eventually equilibrates, and the ionizing spectrum reaches a ``steady-state'' and the luminosity traces the SFR. In \citet{Byler+2017}, we showed that this ``steady-state'' was reached around ${\sim}4$\Myr for non-rotating stellar models. The right panel of Fig.~\ref{fig:EmTimeEvol} shows the time evolution of \ha emission for the CSFR models (1\Msun/year). For the rotating MIST models used in this work, the steady state is reached between 7-10 Myr, which is consistent with the findings of \citet{Jaskot+2016}, who found that the rotating Geneva and BPASS models reached equilibrium much later than single-star models, as late as 20\Myr.

In the remainder of this work, we use the 1\Myr instantaneous burst and the 10\Myr CSFR models to represent the young starburst and ``steady-state'' limits. We note that diagnostics including absorption or stellar features will be much more sensitive to the exact SFH over 1-300\Myr timescales than diagnostics using emission line ratios. However, beyond the instantaneous burst and the “steady-state” limits, model parameter space increases massively. While a potentially interesting area to explore, it is outside the scope of this work.

\subsection{Absorption line trends with model age and metallicity} \label{sec:mod:abs}

\begin{figure*}
  \begin{center}
    \includegraphics[width=0.9\linewidth]{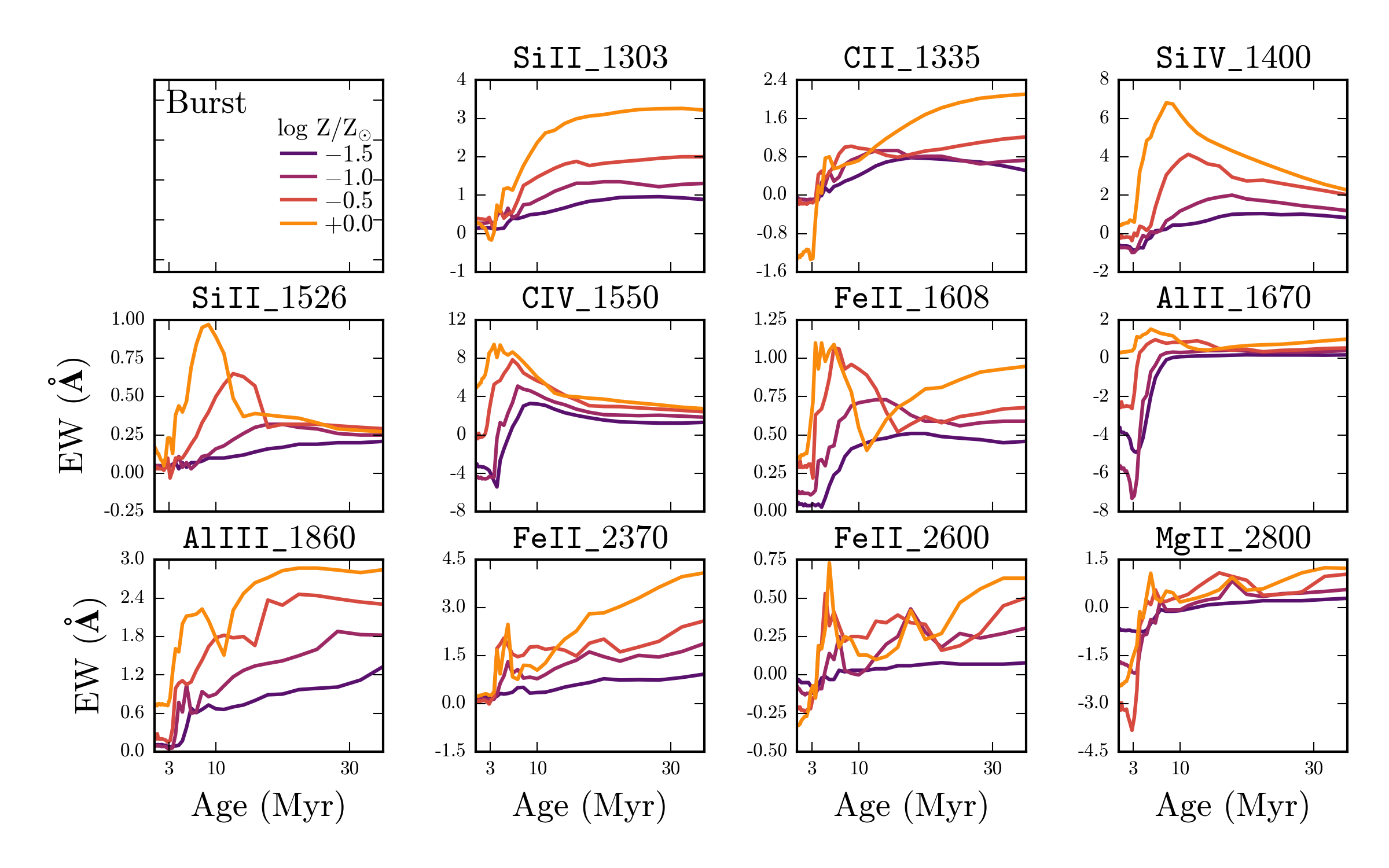}\\
    \includegraphics[width=0.9\linewidth]{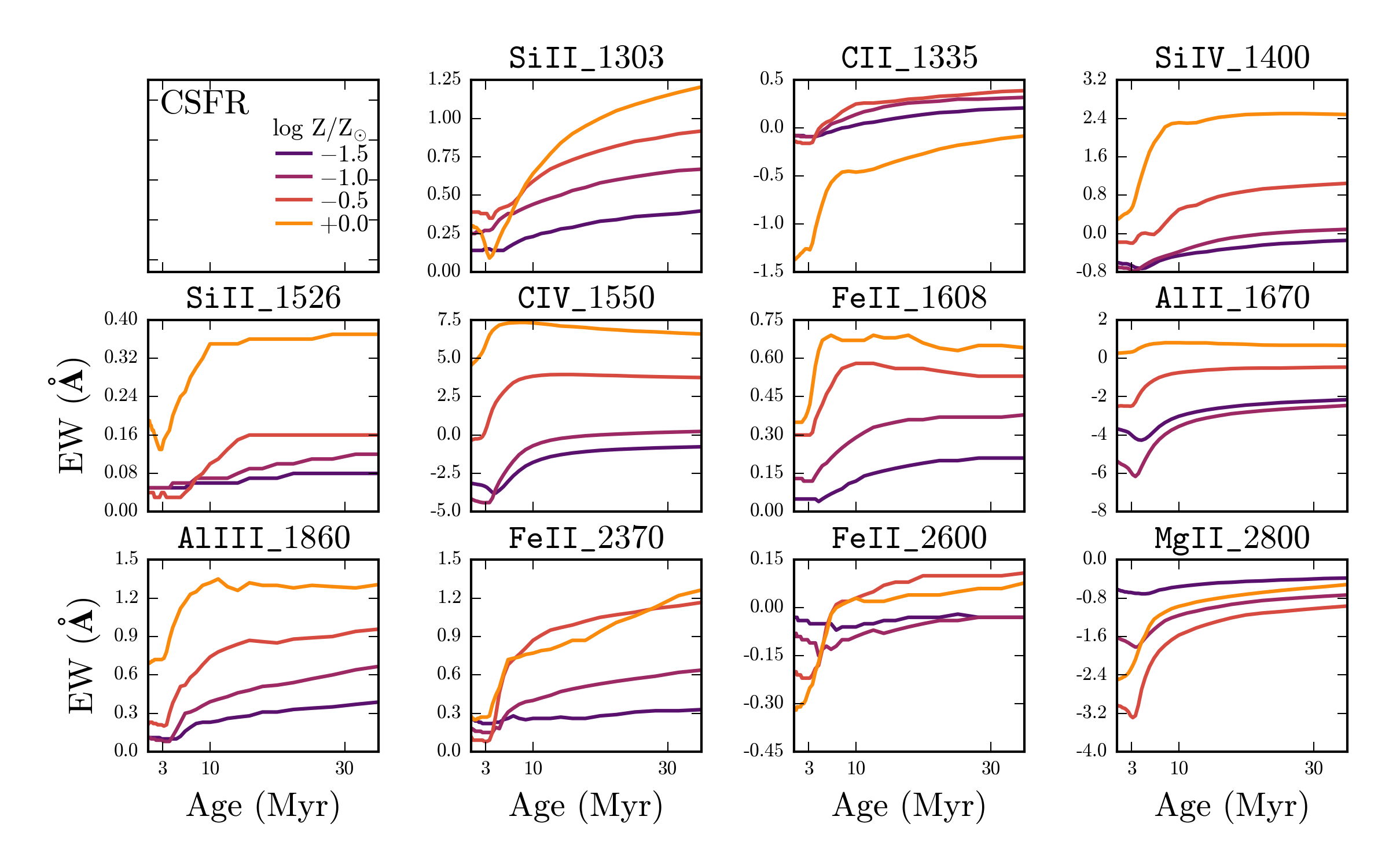}
    \caption{The time evolution of absorption index equivalent widths at \logZeq{-1.5}, $-1.0$, $-0.5$, and $+0.0$, for single-aged populations (top) and a constant SFR (bottom). The most promising absorption features show a large dispersion in equivalent width with metallicity. The C{\sc \,iv}$\,\lambda$1551\ang feature (middle row, second column) correlates strongly with metallicity at young ages.}
    \label{fig:EWTimeEvol}
  \end{center}
\end{figure*}

In Fig.~\ref{fig:EWTimeEvol} we show the L11 absorption indices for our model spectra as a function of age for a single burst (top) and a constant SFR (bottom). The time and metallicity evolution of these absorption indices is frequently complex, due to their dependence on a combination of stellar photospheres, nebular emission, and stellar winds.

The indices that probe stellar wind features (\texttt{SiIV\_1400}, \texttt{CIV\_1550}) are strongest at young ages and gradually decrease with time, as the massive stars evolve off the main sequence. In particular, the C{\sc \,iv} index at 1550\ang shows equivalent widths as high as 12\ang in the first few million years, making it both easy to detect and highly age-sensitive.

Fig.~\ref{fig:EWTimeEvol} also shows how each absorption index changes with metallicity. In general, the high metallicity models have larger equivalent widths, as line blanketing in the upper atmosphere of O- and B-type stars reduces the amount of emergent flux at high metallicities. This effect is clearest in the \texttt{Fe\_2370} index, which shows dense photospheric absorption in the highest metallicity models, distinguishing low-metallicity stellar populations from the high metallicity stellar populations.

For young, massive stars, radiatively-driven winds produce significant mass loss on the main sequence \citep{Kudritzki+2000}. These winds are metallicity dependent; \citet{Vink+2001} estimate that mass loss rates scale as $\dot{M}\sim\mathrm{Z}^{0.69,0.64}$ for O- and B-type stars. Thus, absorption indices that include stellar winds like \texttt{CIV\_1550} and \texttt{SiIV\_1400} will be strongly dependent on metallicity. For populations 10\Myr and younger, these indices have equivalent widths that vary by factors of 2-6 between different metallicity models.

The \texttt{AlIII\_1670} index is one of the indices that inadvertently includes emission lines in the feature bandpass ([O{\sc \,iii}]$\,\lambda\,1666$, [Al{\sc \,ii}]$\,\lambda\,1670$). However, this actually makes it a stronger metallicity diagnostic, since the emission line features are metallicity dependent as well.

\subsection{Nebular Emission features in the UV}\label{sec:mod:em}

We distinguish which emission lines are the most promising diagnostics by identifying relatively bright lines that also show significant variation in strength with metallicity and ionization parameter. These are calculated based on variation from a fiducial 10\Myr CSFR stellar population with \logZeq{-0.5} and \logUeq{-2.5}. All emission lines tracked by \Cloudy are considered individually unless otherwise noted, even if they are considered to be part of a doublet or multiplet. For visual clarity, however, some figures may only include a label for the brightest line in the series.

We first calculate \femline, the line strength relative to \hb to identify emission lines that are relatively bright. We then characterize the sensitivity of the lines to the ionization parameter or metallicity by calculating \sigmaU and \sigmaZ, the variance in \femline over the full parameter range for \U and \logz, respectively. For the fiducial 10\Myr CSFR model, \sigmaZ is the variance in \femline at fixed ionization parameter, \logUeq{-2.5}, over $-2 \leq \logz \leq 0.5$. Similarly, \sigmaU is the variance in \femline at constant metallicity, \logZeq{-0.5}, over $-4 \leq \logU \leq -1$. We note that this definition means that some lines may be difficult to detect in some regions of parameter space.

\begin{figure*}
  \begin{center}
    \includegraphics[width=\linewidth]{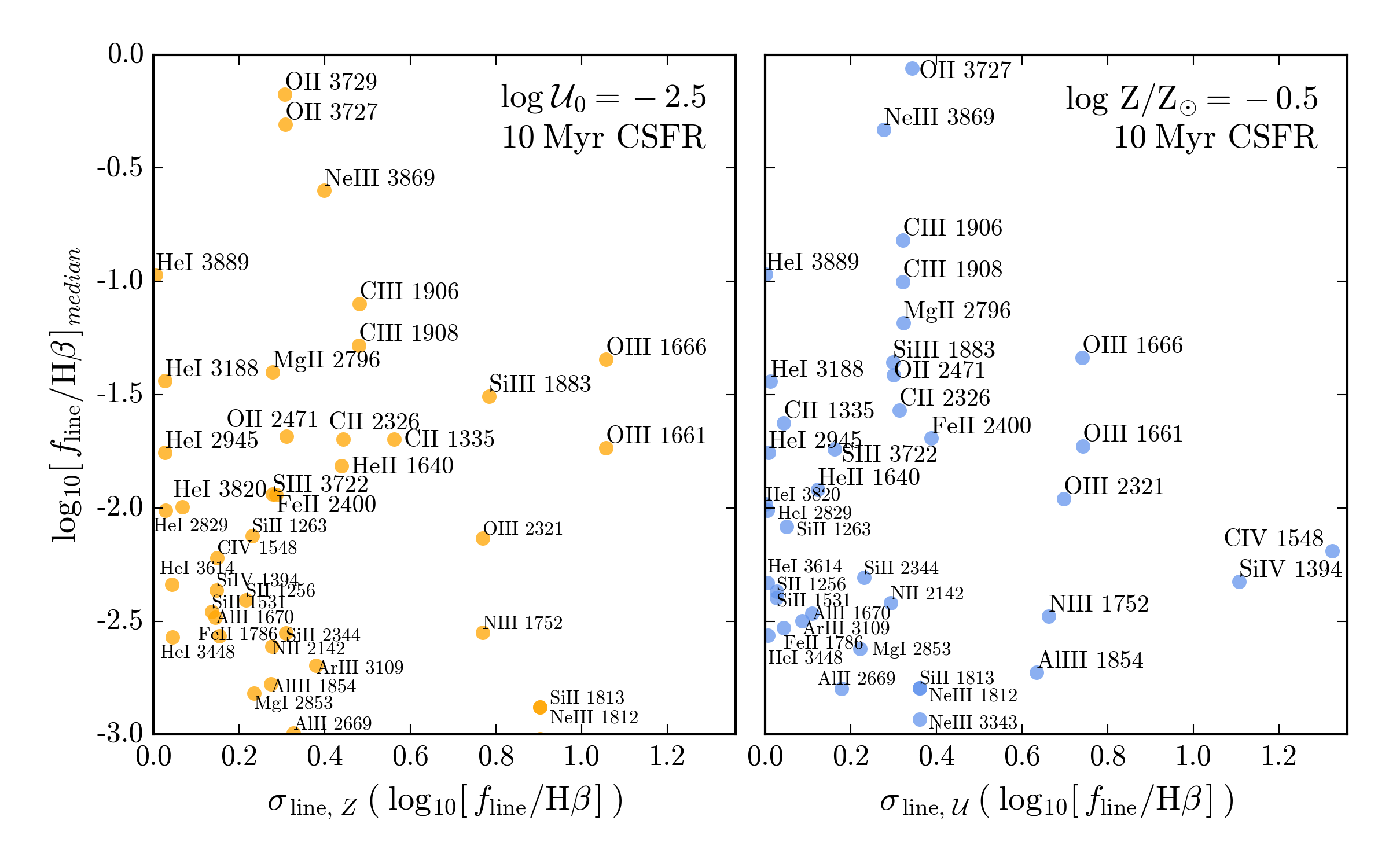}
    \caption{The variance in emission line strength, \sigmaL, from changes in metallicity (left) and ionization parameter (right) as a function of \femline, the median emission line strength relative to \hb. Emission lines with high \sigmaL and \femline are both relatively bright and show significant variation with model metallicity or ionization parameter. These emission lines are potentially useful diagnostics for the gas phase metallicity and ionization state of galaxies.}
    \label{fig:sigmaV}
  \end{center}
\end{figure*}


We show \sigmaU and \sigmaZ as a function of the median \femline in Fig.~\ref{fig:sigmaV} for the fiducial 10\Myr CSFR model. Emission lines that lie in the upper right hand of the plot are the most promising, since these lines are relatively bright and show significant variation with gas-phase metallicity (left panel) or ionization parameter (right panel). The well-known C{\sc \,iii}]$\,\lambda\,1909$\ang and O{\sc \,iii}]$\,\lambda\,1666$\ang lines fall into the promising regime of both panels of Fig.~\ref{fig:sigmaV}. Strong \ciii emission appears in spectra of low-metallicity galaxies at low and high redshift \citep[e.g.,][]{Shapley+2003, Erb+2010, Leitherer+2011, Bayliss+2014, Stark+2014, Stark+2015, Berg+2016, Senchyna+2017, Stroe+2017a}.

The Ne{\sc \,iii}]$\,\lambda3869$ and O{\sc \,ii}]$\,\lambda3729$ lines are also found in the promising region of Fig.~\ref{fig:sigmaV}; the Ne3O2 ratio was previously highlighted by \citet{Levesque+2014} as an excellent probe of gas cloud ionization state. Other promising emission lines in Fig.~\ref{fig:sigmaV} include [Al{\sc \,ii}]$\,\lambda2661$, [Mg{\sc \,ii}]$\,\lambda2796$, C{\sc \,iv}$\,\lambda1548$, C{\sc \,iv}$\,\lambda1551$, and [Si{\sc \,iii}]$\,\lambda1883$, which we will discuss at length in the following sections. [Mg{\sc \,ii}]$\,\lambda2796$ emission was detected in a stacked spectrum of a $z{\sim}2$ galaxies by \citet{Du+2016}. While C{\sc \,iv}$\,\lambda1548$ has been used to identify AGN sources \citep{Stroe+2017b, Feltre+2016}, sources with moderate \civ equivalent widths have been detected at high redshift that may be attributable to nebular emission and emission from massive star winds \citep{Stark+2014, Mainali+2017, Schmidt+2017}.

A full list of predicted UV emission line fluxes as a function of metallicity and ionization parameter is given in Table~\ref{tab:lineStrengths}\footnote{The table is available online in a machine readable format.}.

\begin{deluxetable*}{lccccccccccc}
\tabletypesize{\footnotesize}
\tablecolumns{12}
\tablecaption{UV line fluxes given by the \CloudyFSPS models, assuming a 10\Myr CSFR at solar metallicity. The lines are given as $\log_{10}[\, \mathrm{F}_{\mathrm{line}} / \mathrm{F}_{\mathrm{H}\beta} \,]$.}
\tablehead{
\colhead{SFH} &
\colhead{Age} &
\colhead{$\log_{10}\mathcal{U}_0$} &
\colhead{$\log_{10}(\mathrm{Z}/\mathrm{Z}_{\odot})$} &
\colhead{$12+\log_{10}(\mathrm{O}/\mathrm{H})$} &
\colhead{CIV} &
\colhead{CIV} &
\colhead{HeII} &
\colhead{OIII} &
\colhead{OIII} &
\colhead{AlII} &
\colhead{NIII}\\
\hline
\nocolhead{} &
\colhead{Myr} &
\nocolhead{} &
\nocolhead{} &
\nocolhead{} &
\colhead{1548\ang} &
\colhead{1551\ang} &
\colhead{1640\ang} &
\colhead{1661\ang} &
\colhead{1666\ang} &
\colhead{1671\ang} &
\colhead{1752\ang}
}
\startdata
csfr & 10.0 & -4.0 & 0.00 & 8.69 & -5.3259 & -5.6272 & -2.6206 & -4.5499 & -4.1581 & -2.6228 & -4.7133 \\
csfr & 10.0 & -3.5 & 0.00 & 8.69 & -3.7689 & -4.0643 & -2.4183 & -3.7273 & -3.3355 & -2.5325 & -3.9159 \\
csfr & 10.0 & -3.0 & 0.00 & 8.69 & -2.6854 & -2.8647 & -2.3427 & -3.2949 & -2.9033 & -2.4825 & -3.5258 \\
csfr & 10.0 & -2.5 & 0.00 & 8.69 & -2.4156 & -2.4729 & -2.3021 & -3.1025 & -2.7121 & -2.4571 & -3.3794 \\
csfr & 10.0 & -2.0 & 0.00 & 8.69 & -2.3026 & -2.3483 & -2.2641 & -2.9781 & -2.5905 & -2.4446 & -3.2914 \\
csfr & 10.0 & -1.5 & 0.00 & 8.69 & -2.1895 & -2.2539 & -2.2261 & -2.8917 & -2.5093 & -2.4448 & -3.2333 \\
csfr & 10.0 & -1.0 & 0.00 & 8.69 & -2.0432 & -2.1298 & -2.1901 & -2.8149 & -2.4401 & -2.4382 & -3.1827 \\
\enddata
\tablecomments{Only a portion of this table is shown here to demonstrate its form and content. A machine-readable version of the full table is available online.}
\label{tab:lineStrengths}
\end{deluxetable*}

\subsubsection{Emission line ratios}

We determine which combination of emission lines from \S\ref{sec:model:neb} will make useful emission line ratio diagnostics. We make a number of observationally motivated cuts to the list of emission lines considered for candidate line ratios. There are a total of 382 emission lines included in the \FSPS nebular model, 159 of which are in the UV. We remove 21 Hydrogen Lyman series lines, because they are notoriously difficult to interpret due to resonant scattering effects and do not consider emission lines blueward of Ly$\alpha$, which will be observationally difficult to access. We then restrict the line list to elemental species with abundance by number relative to hydrogen of $\log[ \mathrm{n} / \mathrm{H} ] > -6$, which leaves 60 remaining emission lines. Finally, we restrict ourselves to lines with $\log[ f_{\mathrm{line}} / \hb] > -4$, which is sufficient to catch the weak but temperature-sensitive auroral lines, leaving 42 candidate emission lines to consider for possible line ratio diagnostics.

We first determine all possible non-repeating combinations of the 42 emission lines. For each of these line ratios, we calculate the variance in the line ratio, \sigmaR, by taking the standard deviation of the line ratio across a range of model parameters. We consider variation in the line ratio due to ionization parameter and metallicity, using a 10\Myr CSFR model with \logZeq{-0.5} and \logUeq{-2.5} as the fiducial model. To find ratios that are sensitive to changes in metallicity, we calculate \sigmaRZ at fixed age and ionization parameter (10\Myr CSFR, \logUeq{-2.5}). Similarly, we calculate \sigmaRU at fixed age and metallicity to find ratios that are sensitive to changes in the ionization parameter (10\Myr CSFR, \logZeq{-0.5}). For each pair of emission lines, we use the brighter of the two emission lines as the denominator, as determined by the median brightness relative to \hb over the range of models considered.

\begin{figure*}
  \begin{center}
    \includegraphics[width=\linewidth]{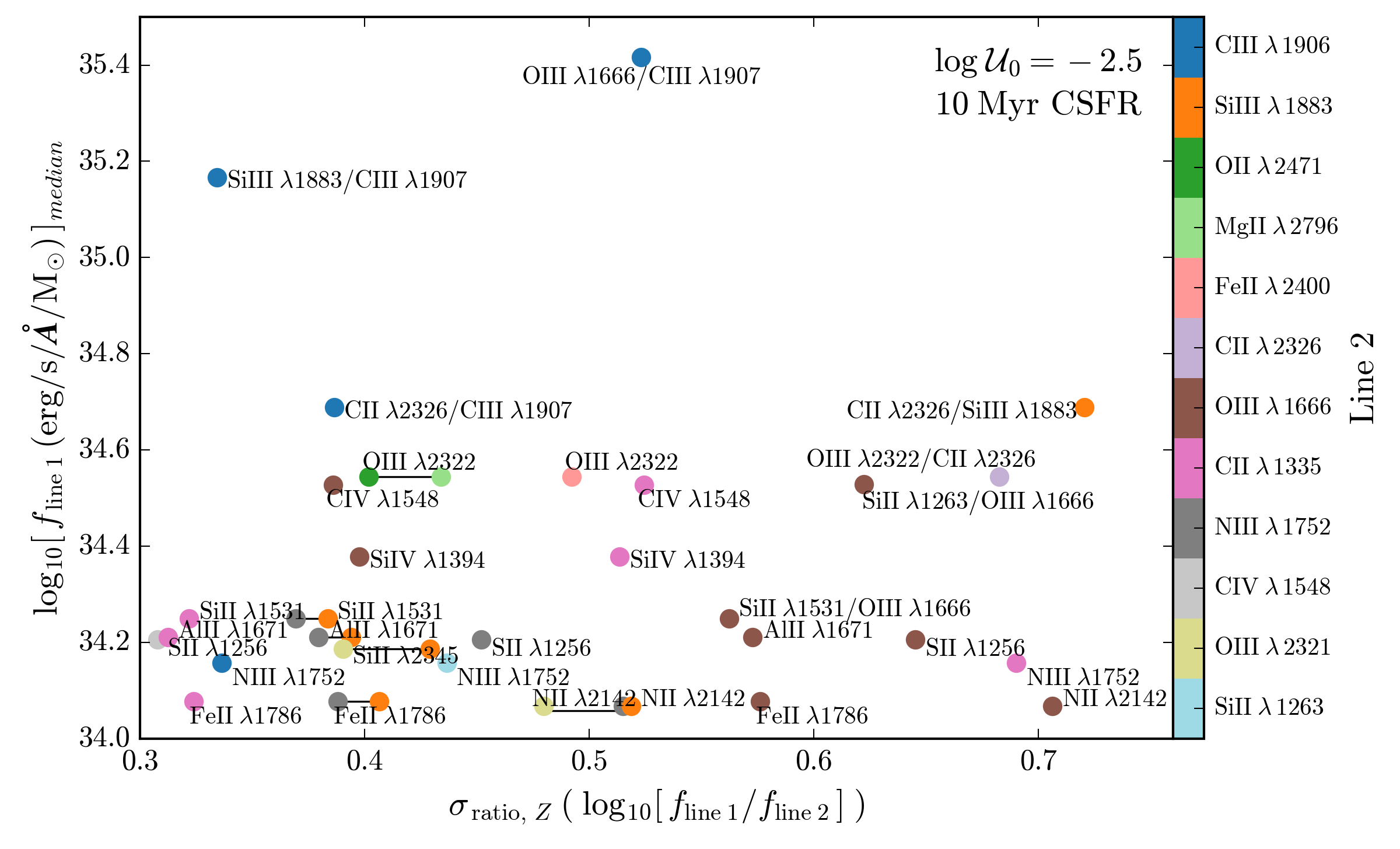}
    \caption{The $x$-axis shows \sigmaRZ, the variance in emission line ratio due to changes in metallicity, for a 10\Myr CSFR model at \logUeq{-2.5} and assuming 1\,\Msun of stars. The $y$-axis shows the median luminosity of the less luminous emission line used in the ratio, by definition the numerator, or ``line 1''. For visualization purposes, we have labeled each ratio by the emission line in the numerator, and color-coded the marker by the emission line in the denominator, ``line 2''.  Emission line ratios in the upper right corner of both plots will be the easiest to observe and should provide the most leverage in determining gas phase metallicities.}
    \label{fig:ratioSigmaZ}
  \end{center}
\end{figure*}

\begin{figure*}
  \begin{center}
    \includegraphics[width=\linewidth]{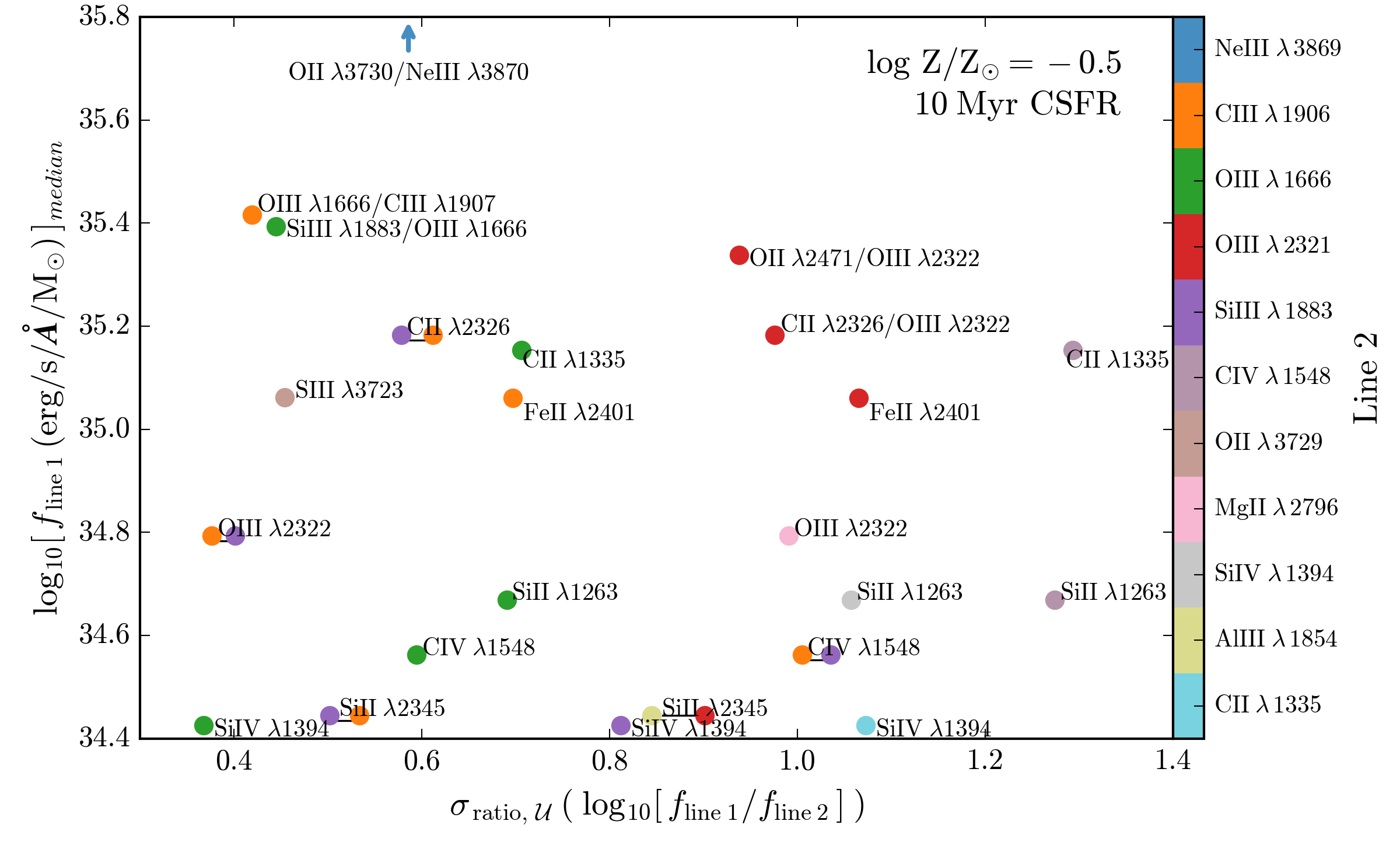}
    \caption{The $x$-axis shows \sigmaRU, the variance in emission line ratio due to changes in ionization parameter, for a 10\Myr CSFR model at \logZeq{-0.5} and assuming 1\,\Msun of stars. The $y$-axis shows the median luminosity of the less luminous emission line used in the ratio, by definition the numerator, or ``line 1''. For visualization purposes, we have labeled each ratio by the emission line in the numerator, and color-coded the marker by the emission line in the denominator, ``line 2''.  Emission line ratios in the upper right corner of both plots will be the easiest to observe and should provide the most leverage in determining ionization parameters.}
    \label{fig:ratioSigmaU}
  \end{center}
\end{figure*}

In Figs.~\ref{fig:ratioSigmaZ} \& \ref{fig:ratioSigmaU}, we plot the variance in the brightest of the candidate emission line ratios. Figs.~\ref{fig:ratioSigmaZ} \& \ref{fig:ratioSigmaU} show the line ratio variance, \sigmaR, versus the median brightness relative to \hb of the less luminous emission line in the pair (the ratio numerator, by definition), which would be the observationally limiting line. Fig.~\ref{fig:ratioSigmaZ} shows variance in emission line ratios driven by metallicity changes (\sigmaRZ)\ at constant age and ionization parameter, while Fig.~\ref{fig:ratioSigmaU} shows variance driven by ionization parameter changes (\sigmaRU) at constant age and metallicity, as noted in the upper-right corner of the figure. As in Fig.~\ref{fig:sigmaV}, emission line ratios in the upper right region of these figures will be the easiest to detect and should have the most sensitivity to changes in metallicity or ionization parameter. We limit our study to emission line ratios with \sigmaR $> 0.3$, which corresponds to line ratios that experience at least a factor of four difference over the parameters of interest.

As discussed in \S\ref{sec:mod:em}, the Ne3O2 emission line ratio was identified by \citet{Levesque+2014} as an excellent probe of gas ionization state. The Ne3O2 ratio appears in the upper-right part of Fig.~\ref{fig:ratioSigmaU}, confirming that this analysis can identify emission line ratios that are sensitive to metallicity and ionization parameter\footnote{The Ne3O2 ratio used in \citet{Levesque+2014} uses the sum of the 3726\ang and 3729\ang doublet. The ratio in Fig.~\ref{fig:ratioSigmaU} includes only the 3729\ang line, however, since all lines are tracked individually in our model. The 3729\ang line is the brighter line in the low-density regime considered here, and is thus visible in Fig.~\ref{fig:sigmaV}.}. Similarly, the metallicity-sensitive O3C3 ratio \citep[e.g.,][]{Stark+2015, Du+2016, Ding+2016, Gutkin+2016, Feltre+2016} appears in the upper-right portion of Fig.~\ref{fig:ratioSigmaZ}.

Other promising line ratios include [O{\sc \,iii}]$\,\lambda$2322 / [C{\sc \,ii}]$\,\lambda$2326 (O3C2), [Si{\sc \,ii}]$\,\lambda$1263 / [C{\sc \,ii}]$\,\lambda$1335 (Si2C2), [N{\sc \,iii}]$\,\lambda$1752 / [O{\sc \,iii}]$\,\lambda$1666 (N3O3), and [Si{\sc \,iii}]$\,\lambda$1883 / [C{\sc \,iii}]$\,\lambda$1907 (Si3C3); a full list is included in Table~\ref{tab:ratios}.

\begin{figure*}
  \begin{center}
    \includegraphics[width=\linewidth]{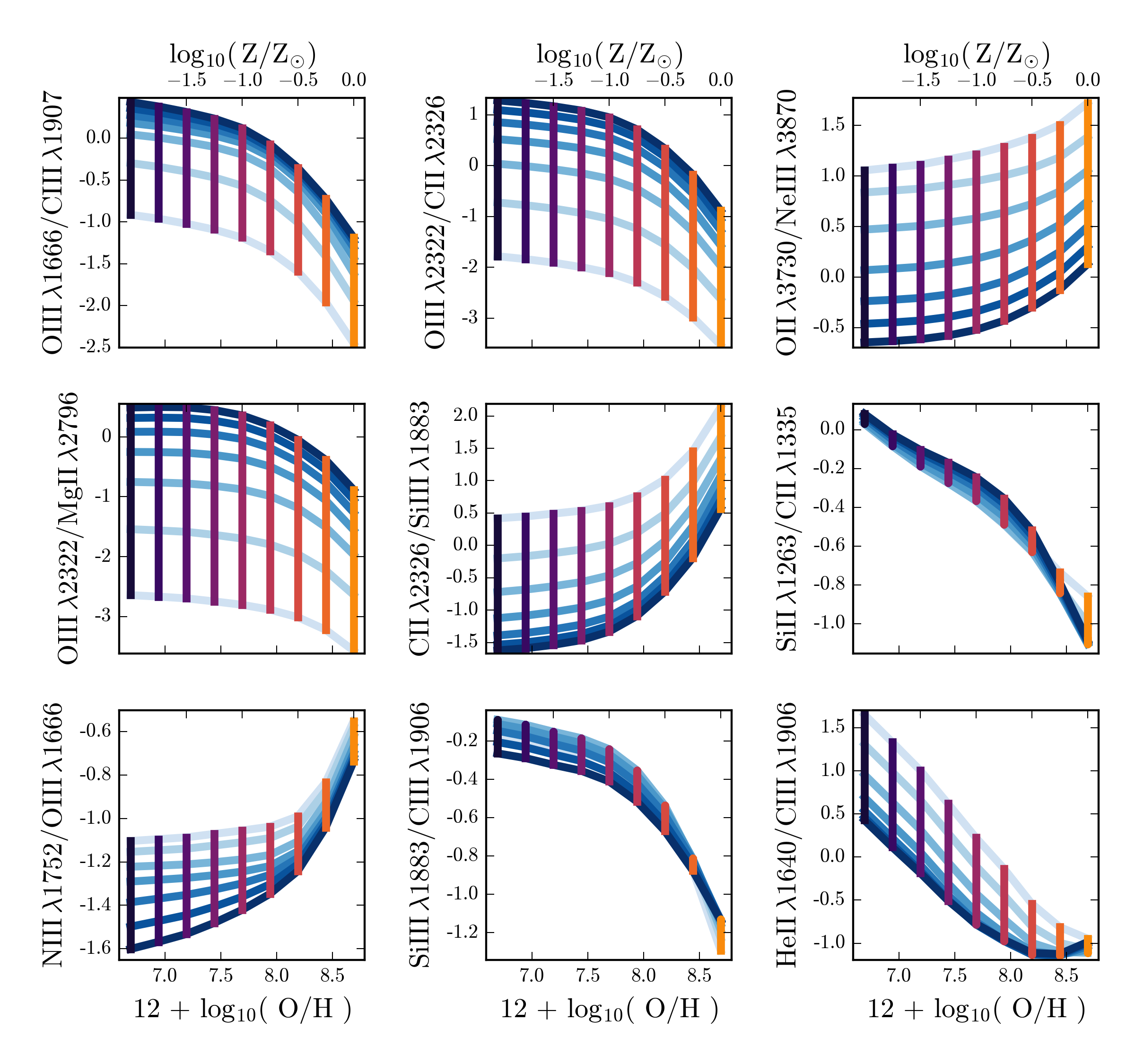}
    \caption{Several emission line ratios as a function of model metallicity (purple to yellow) and ionization parameter (blue) for 10\Myr constant SFR populations. The blue lines connect models of constant ionization parameter, with \logUeq{-1} in dark blue and \logUeq{-4} in light blue. Models of constant metallicity span from \logZeq{-2} in purple to \logZeq{0.0} in yellow.}
    \label{fig:RatioVsLogZ}
  \end{center}
\end{figure*}

In Fig.~\ref{fig:RatioVsLogZ} we show each of these line ratios as a function of metallicity to highlight their utility as metallicity indicators. These ratios show several orders of magnitude variation with metallicity and ionization parameter. These emission line ratios can be used by themselves to estimate the gas phase metallicity of a galaxy, but with uncertainty due to the unknown ionization parameter. Alternatively, they can be combined with other emission line ratios to create ``diagnostic diagrams'' in the UV, like the well-used Baldwin Phillips Terlevich (BPT) diagram in the optical \citep{BPT}. These combinations of emission line ratios may better distinguish the metallicity and ionization parameter of the observed population than the use of a single line.

\begin{figure*}
  \begin{center}
    \includegraphics[width=0.495\linewidth]{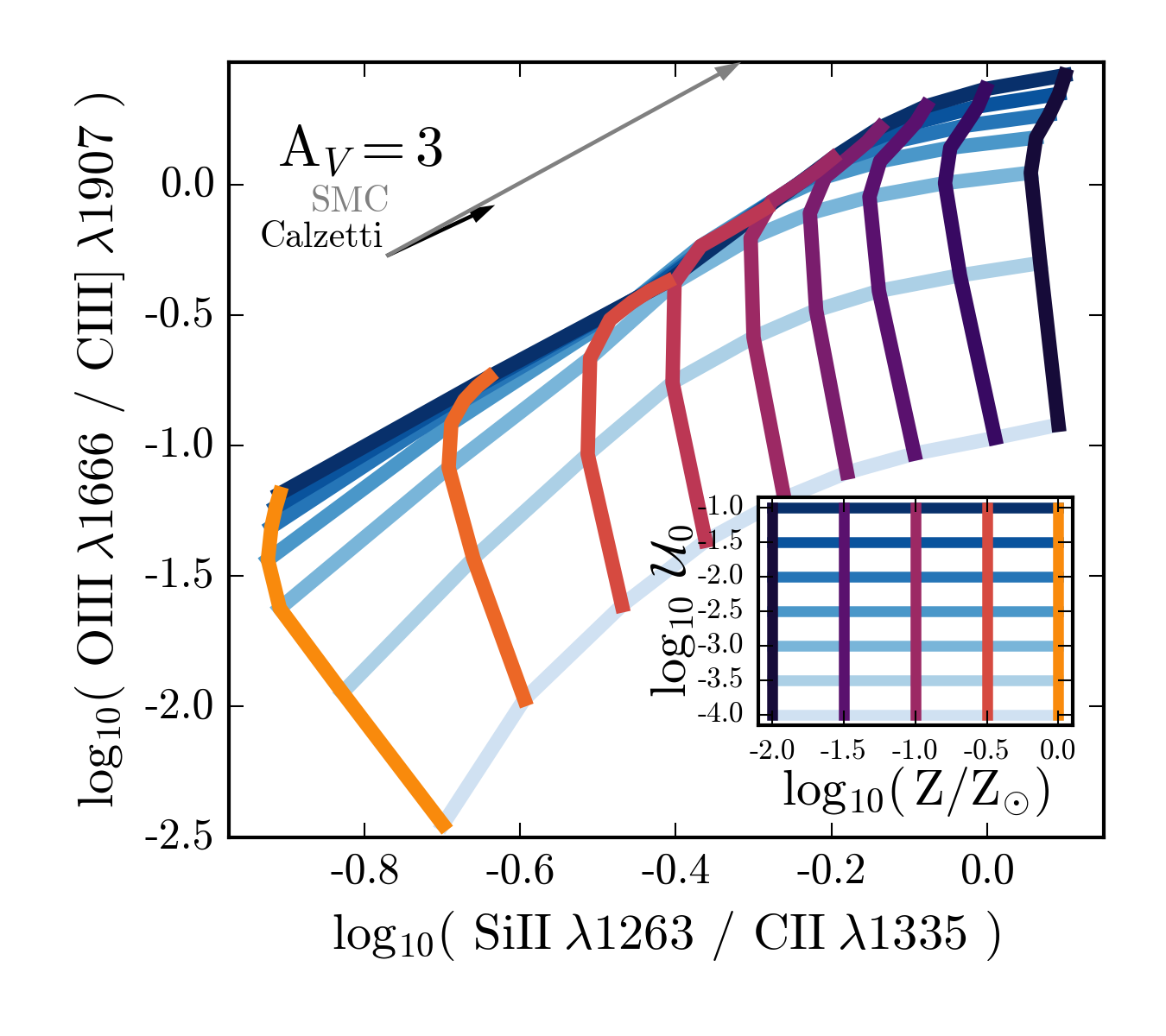}
    \includegraphics[width=0.495\linewidth]{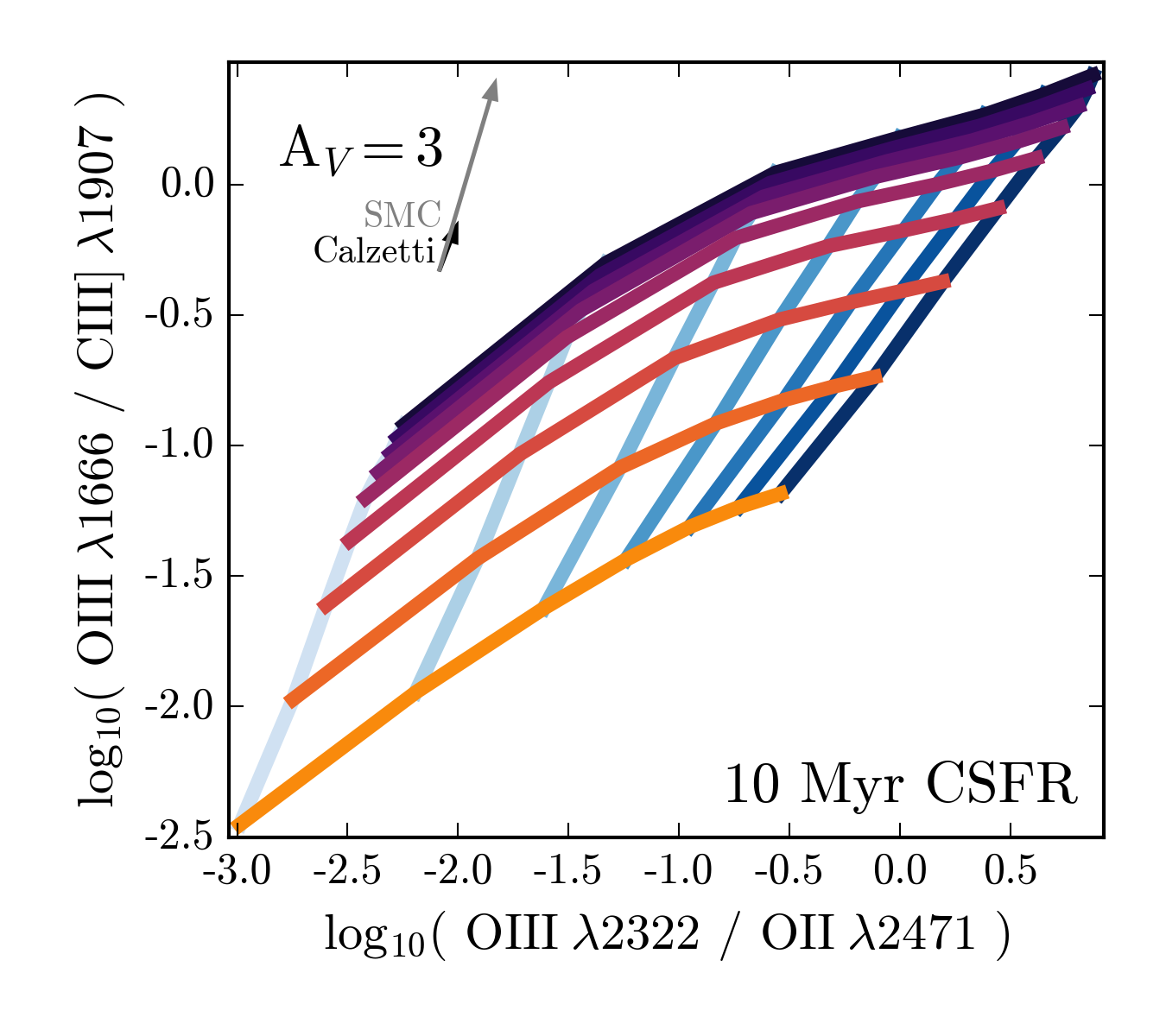}
    \caption{Two possible UV diagnostic diagrams using combinations of emission line ratios for 10\Myr constant SFR populations. The blue lines connect models of constant ionization parameter, from \logUeq{-1} in dark blue to \logUeq{-4} in light blue. Models of constant metallicity span from \logZeq{-2} in purple to \logZeq{0.0} in yellow. The grey and black arrows show one magnitude of extinction assuming SMC and Calzetti reddening laws, respectively.}
    \label{fig:RatioRatio}
  \end{center}
\end{figure*}

We show several promising emission line ratio combinations in Fig.~\ref{fig:RatioRatio}. The [Si{\sc \,ii}]$\,\lambda1263$ / [C{\sc \,ii}]$\,\lambda1335$ (Si2C2) \vs [O{\sc \,iii}]$\,\lambda1666$ / [C{\sc \,iii}]$\,\lambda1907$ (O3C3) diagnostic diagram (\emph{left}) is very sensitive to the gas phase metallicity while remaining relatively insensitive to ionization parameter, due to combining species from several ionization states. The O3C3 ratio also pairs well with the [O{\sc \,iii}]$\,\lambda2322$ / [O{\sc \,ii}]$\,\lambda2471$ ratio for a diagnostic diagram (\emph{right}) that has the potential to distinguish metallicity and ionization parameter, though the sensitivity to metallicity seems to saturate below \logZeq{-1.5}.

All of these diagnostic diagrams make use of emission lines that are relatively close in wavelength space to minimize the effects of reddening. In Fig.~\ref{fig:RatioRatio} we show the direction that 1 magnitude of reddening ($A_V=1$) would change the emission line ratios, assuming two different extinction laws, the canonical Calzetti extinction curve and the steeper SMC extinction curve, which may be more appropriate for low-metallicity galaxies.

\begin{deluxetable}{lcc}
\tabletypesize{\footnotesize}
\tablecolumns{2}
\tablecaption{UV Emission line ratios}
\tablehead{
\colhead{Ratio} &
\colhead{Definition}
}
\startdata
Si2C2 & $\log_{10}$( \SiuII$\lambda1263$ / \cii$\lambda1335$ ) \\
Si4C2 & $\log_{10}$( Si{\tt IV}$\lambda1394$ / \cii$\lambda1335$ ) \\
C4O3\tablenotemark{w,s} & $\log_{10}$( \civ$\lambda1551$ / \oiii$\lambda1666$ ) \\
He2Hb\tablenotemark{w,s} & $\log_{10}$( \heii$\lambda1640$ / H$\beta$ ) \\
O3C3 & $\log_{10}$( \oiii$\lambda1666$ / \ciii$\lambda1907$ ) \\
N3O3 & $\log_{10}$( \niii$\lambda1752$ / \oiii$\lambda1666$ ) \\
Fe2N3 & $\log_{10}$( Fe{\sc ii}$\lambda1786$ / \niii$\lambda1752$ ) \\
Si3C3 & $\log_{10}$( \SiuIII$\lambda1883$ / \ciii$\lambda1907$ ) \\
Si3N3 & $\log_{10}$( \SiuIII$\lambda1883$ / \niii$\lambda1752$ ) \\
O3C2 & $\log_{10}$( \oiii$\lambda2322$ / \cii$\lambda2326$ ) \\
O3C2 & $\log_{10}$( \oiii$\lambda2322$ / \oii$\lambda2471$ ) \\
O3Mg2 & $\log_{10}$( \oiii$\lambda2322$ / \mgii$\lambda2796$ ) \\
C2Si3 & $\log_{10}$( \cii$\lambda2326$ / \SiuIII$\lambda1883$ ) \\
Al2Mg2 & $\log_{10}$( \alII$\lambda2661$ / \mgii$\lambda2796$ ) \\
Ne3O2 & $\log_{10}$( \neiii$\lambda3870$ / \oii$\lambda3729$ ) \\
\enddata
\tablenotetext{w}{Ratio may include contribution from stellar wind emission}
\tablenotetext{s}{Ratio may be sensitive to shock contribution}
\label{tab:ratios}
\end{deluxetable}

\subsubsection{Comparisons with optical line ratios} \label{sec:mod:opt}

\begin{figure*}
  \begin{center}
    \includegraphics[width=0.495\linewidth]{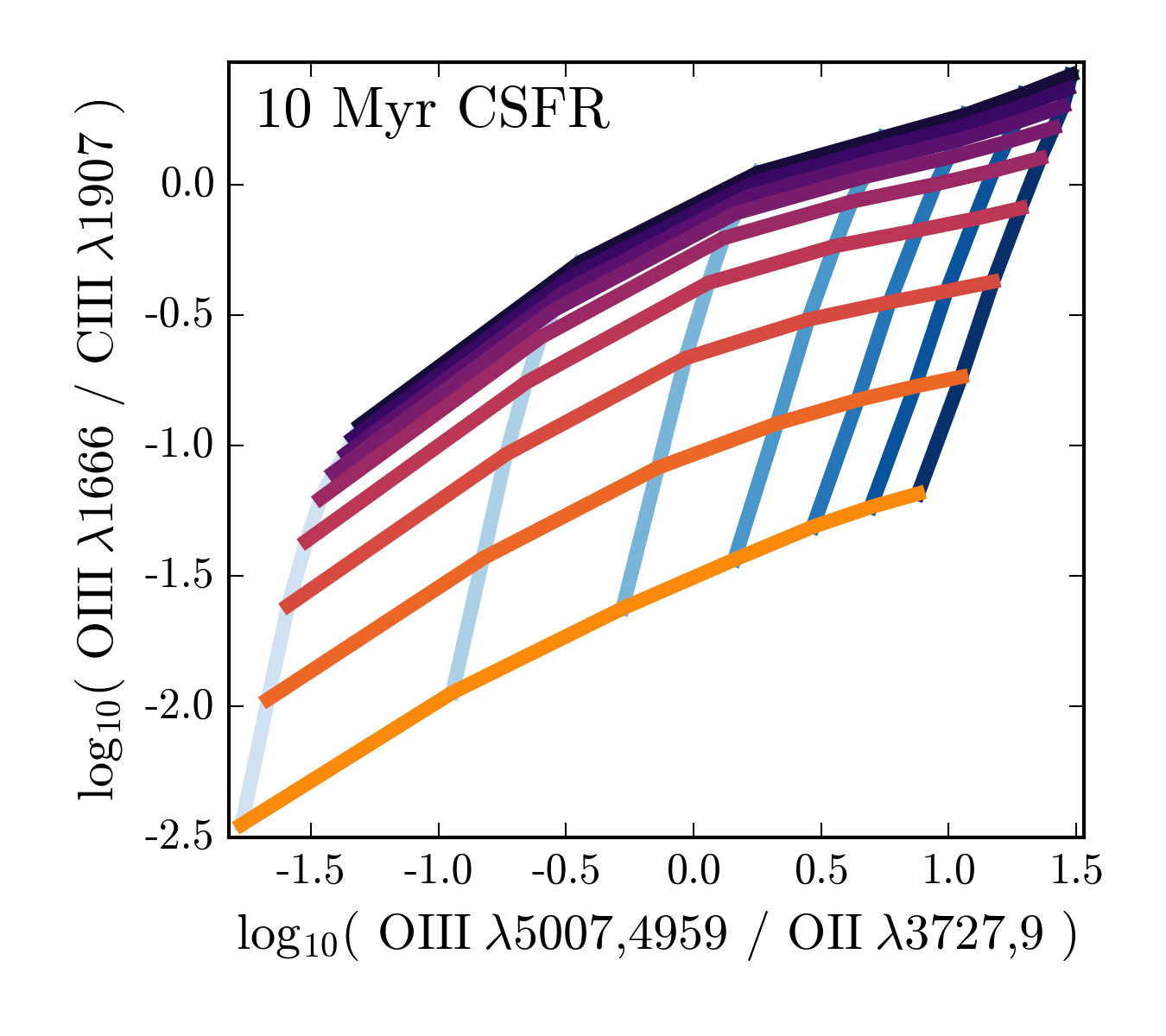}
    \includegraphics[width=0.495\linewidth]{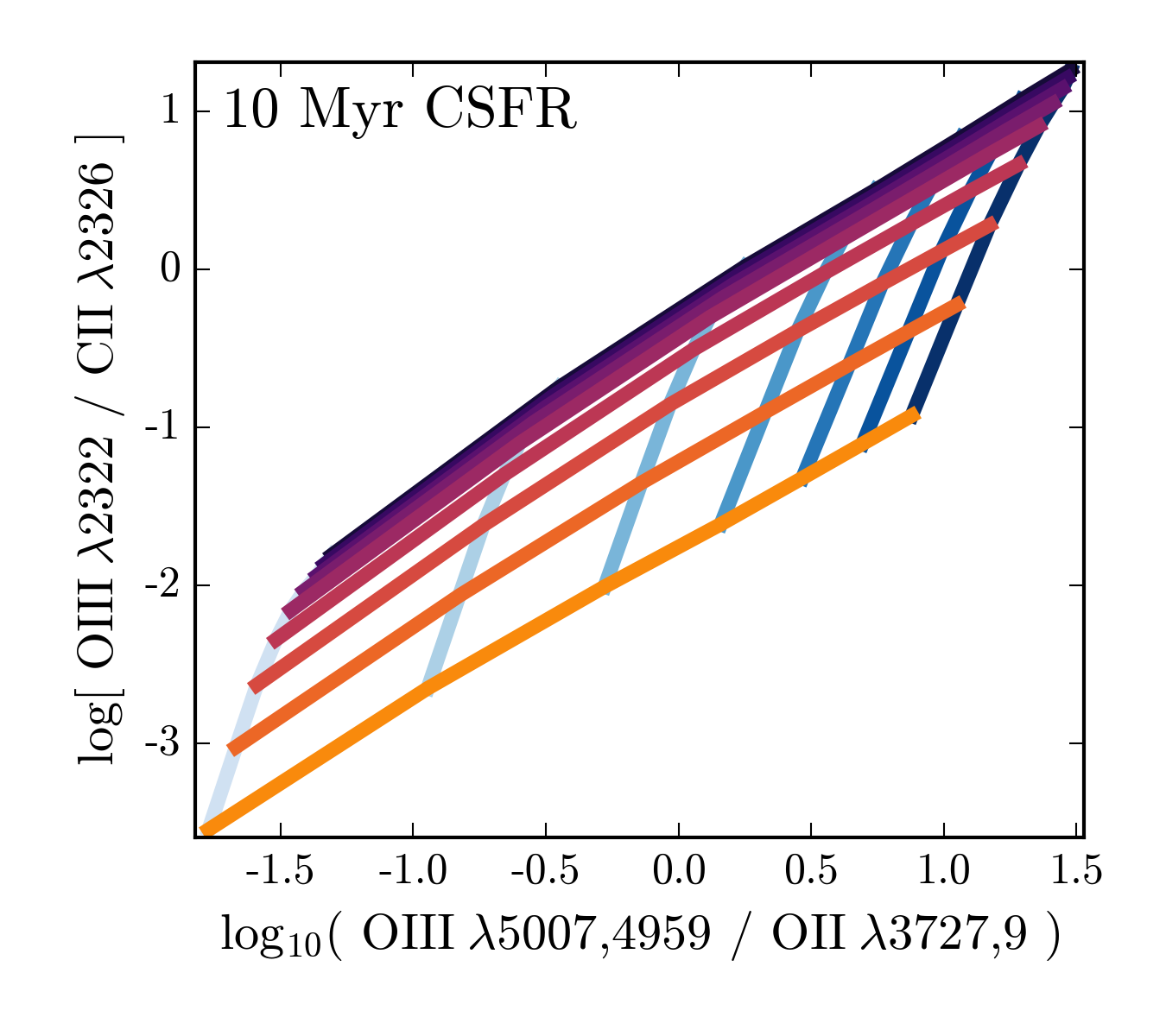}\\
    \includegraphics[width=0.495\linewidth]{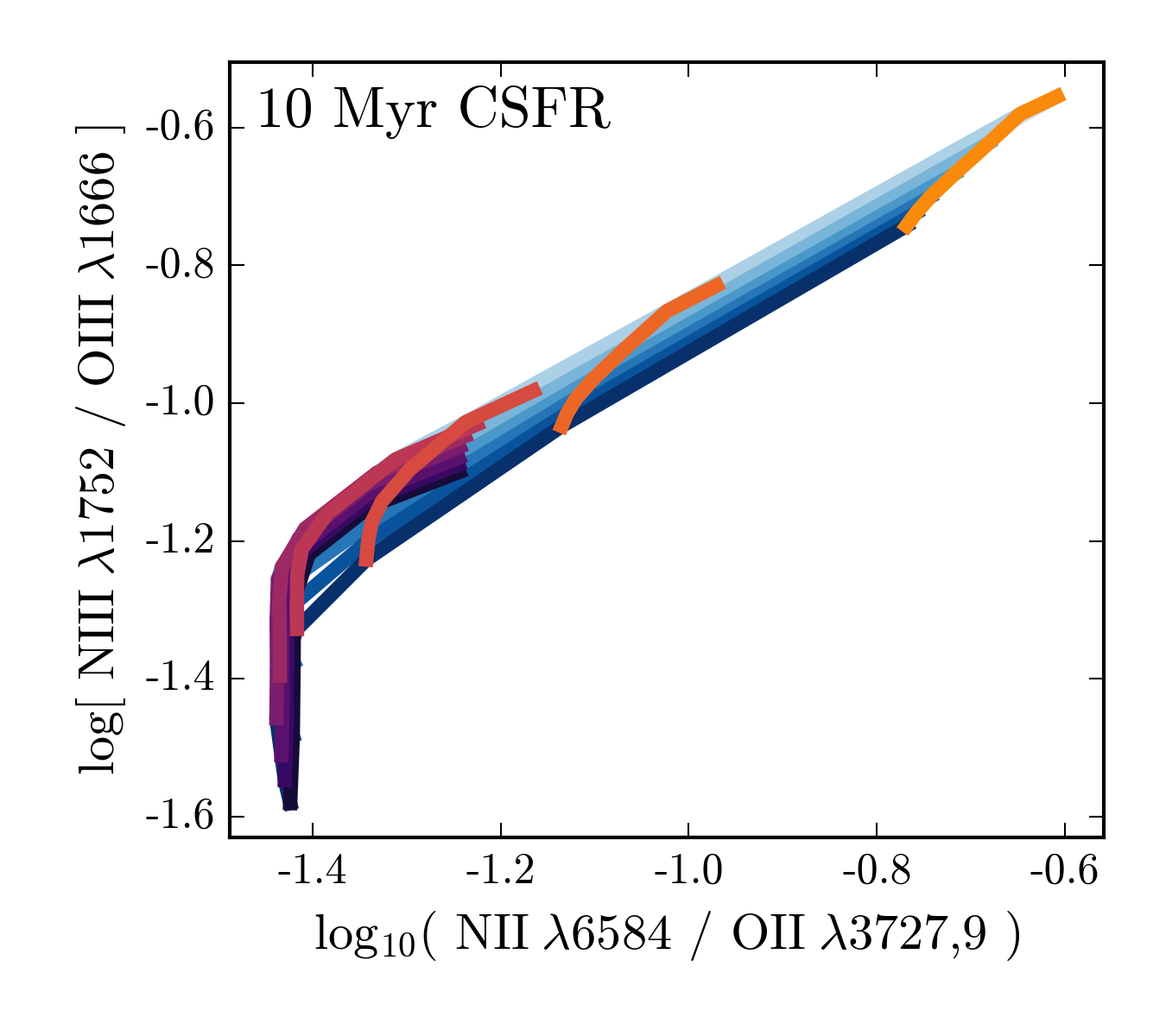}
    \includegraphics[width=0.495\linewidth]{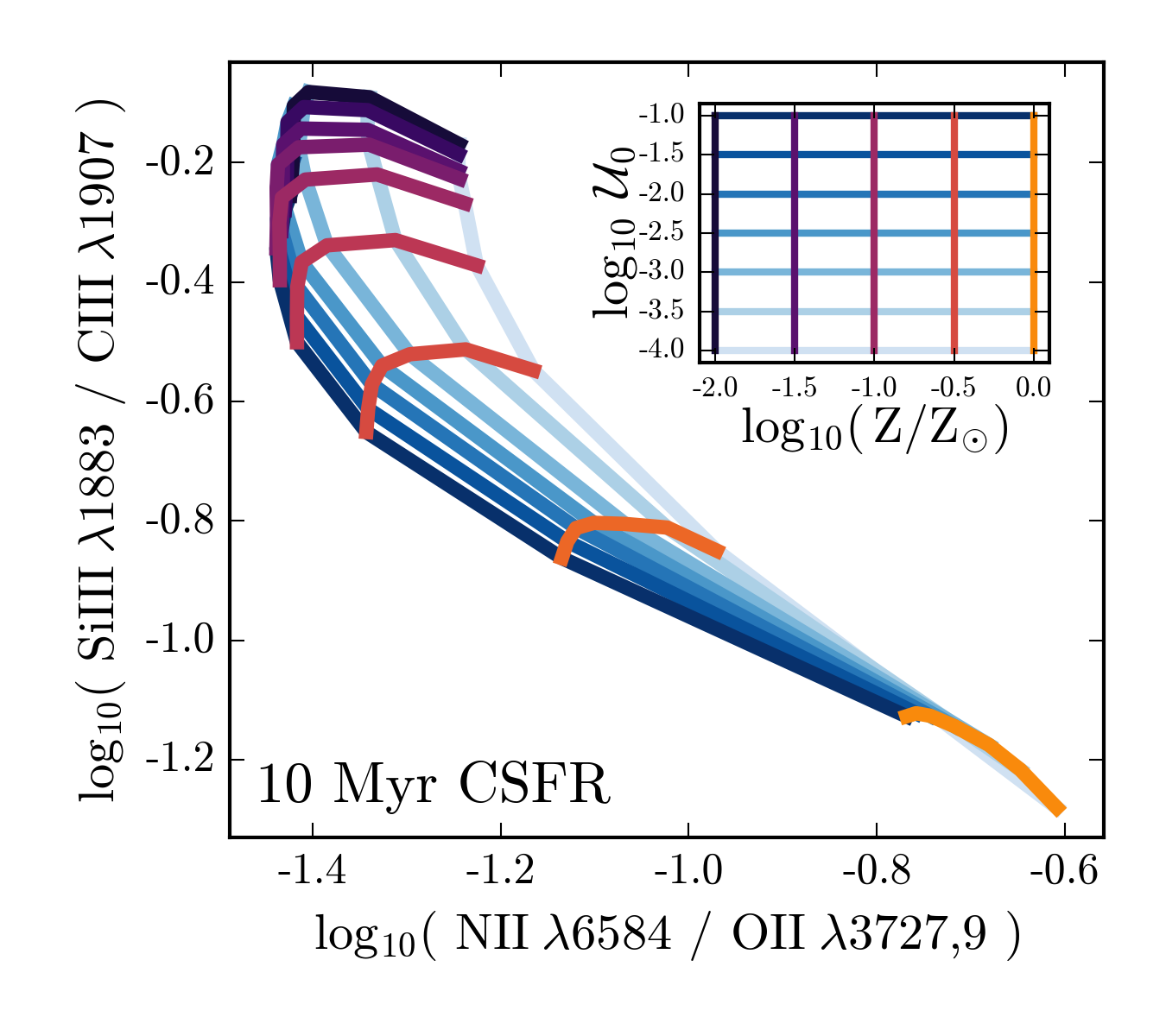}
    \caption{Comparisons of UV and optical emission line ratios. The model colors are the same as indicated in Fig.~\ref{fig:RatioRatio}. \emph{Top row:} Emission line ratios sensitive to ionization parameter: the UV O3C3 emission line ratio (\emph{top left}) and the UV O3C2 emission line ratio (\emph{top right}) compared to the O3O2 optical emission line ratio. \emph{Bottom row:} Emission line ratios sensitive to metallicity: the UV N3O3 emission line ratio (\emph{bottom left}) and the UV Si3C3 emission line ratio (\emph{bottom right}) compared to the optical N2O2 emission line ratio.}
    \label{fig:UVopt}
  \end{center}
\end{figure*}

In Fig.~\ref{fig:UVopt}, we compare several of the UV emission line ratios to well-known emission line ratios in the optical for the 10\Myr CSFR model grid. The top panel of Fig.~\ref{fig:UVopt} highlights emission line ratios that are sensitive to ionization parameters. On the $x$-axis, we show the the optical $\log_{10}$(\oiii$\lambda\,$5007 / \oii$\lambda\,$3726,9 ) ratio, also referred to as O3O2 or O$_{32}$, which is known to be sensitive to ionization parameter. We compare O$_{32}$ to two different UV line ratios: $\log_{10}$( \oiii$\lambda1666$ / \ciii$\lambda1907$ ) (O3C3; \emph{top left}) and $\log_{10}$( \oiii$\lambda2322$ / \ciii$\lambda2326$ ) (O3C2; \emph{top right}). Both UV ratios show similar sensitivity to ionization parameter and correlate positively with the optical O3O2 ratio. The UV ratio is more metal-dependent at fixed ionization parameter, but at low metallicity (below \logZeq{-1}) the sensitivities are comparable.

The bottom panel of Fig.~\ref{fig:UVopt} shows emission line ratios that are sensitive to metallicity. We compare UV metallicity-sensitive emission line ratios to the optical $\log_{10}$(\nii$\lambda\,$6584 / \oii$\lambda\,$3726,9 ) emission line ratio, which correlates strongly with metallicity and has little dependence on ionization parameter \citep{Kewley+2002}. The UV $\log_{10}$( \niii$\lambda1752$ / \oiii$\lambda1666$ ) ratio (N3O3; \emph{bottom left}) shows a strong correlation with the optical N2O2 ratio, where larger values of both indicate higher metallicities. Neither ratio does particularly well at low metallicity, though. The UV $\log_{10}$( \SiuIII$\lambda1883$ / \ciii$\lambda1907$ ) (Si3C3; \emph{bottom right}) is anti-correlated with the optical N2O2 ratio. The Si3C3 ratio is strongest at low-metallicity and will thus have slightly different optimal applications.

\subsection{UV diagnostics combining emission and absorption features} \label{sec:mod:comb}

For short-lived massive stars, the metallicity of the stars should be nearly identical to that of the surrounding natal gas cloud, but also see \citet{Steidel+2016} for a discussion of $\alpha$-enhanced systems. There are notably fewer nebular emission lines in the UV compared to the optical, but we can gain additional metallicity leverage by considering combinations of stellar absorption features and nebular emission features. In this section we identify useful combinations of emission and absorption lines that track well with the properties of the ionizing SSP and gas-phase metallicity. We use the strongest absorption features identified in \S\ref{sec:mod:abs} and combine them with the most promising diagnostic ratios determined in \S\ref{sec:mod:em}.

\begin{figure*}
  \begin{center}
    \includegraphics[width=0.495\linewidth]{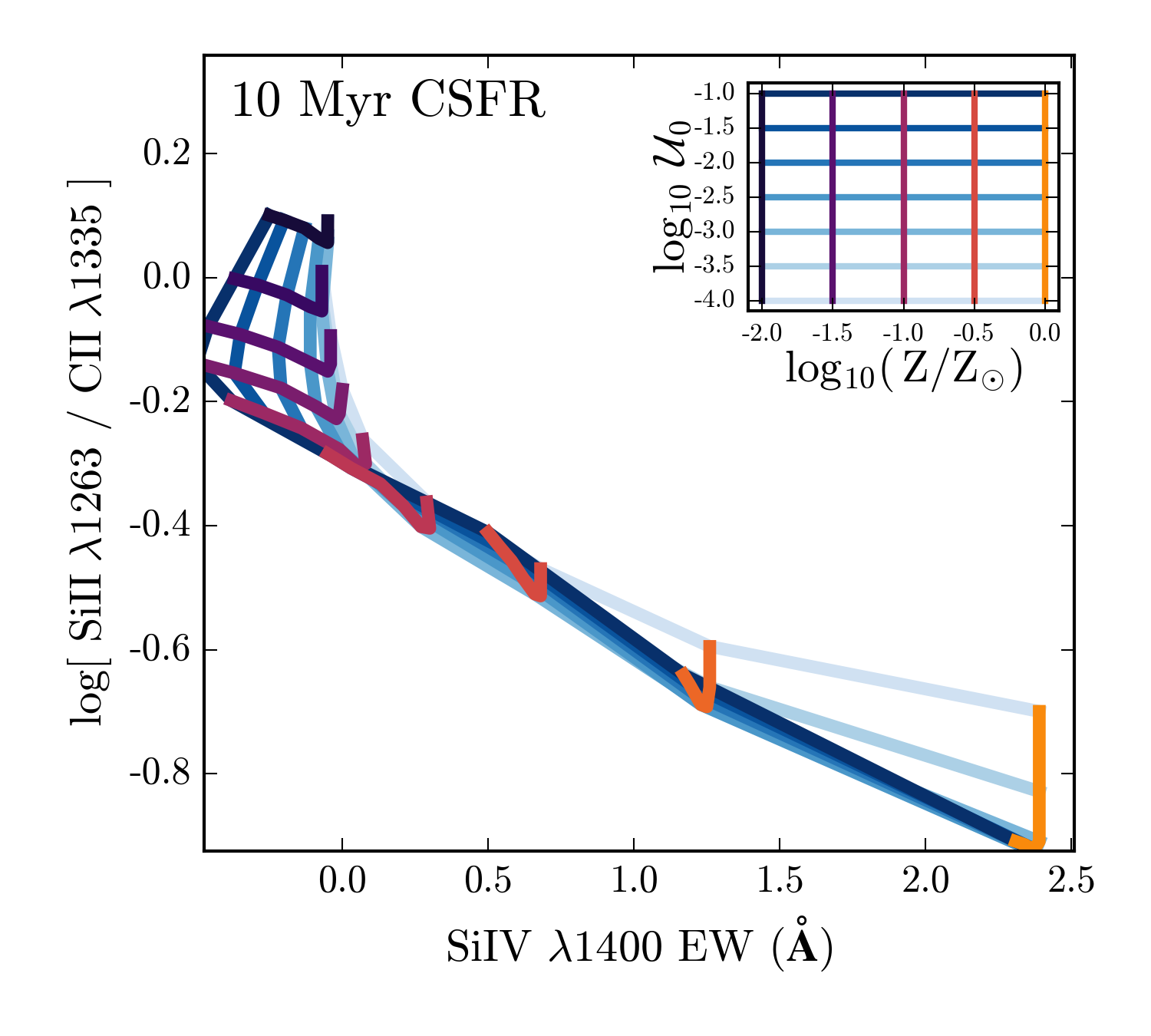}
    \includegraphics[width=0.495\linewidth]{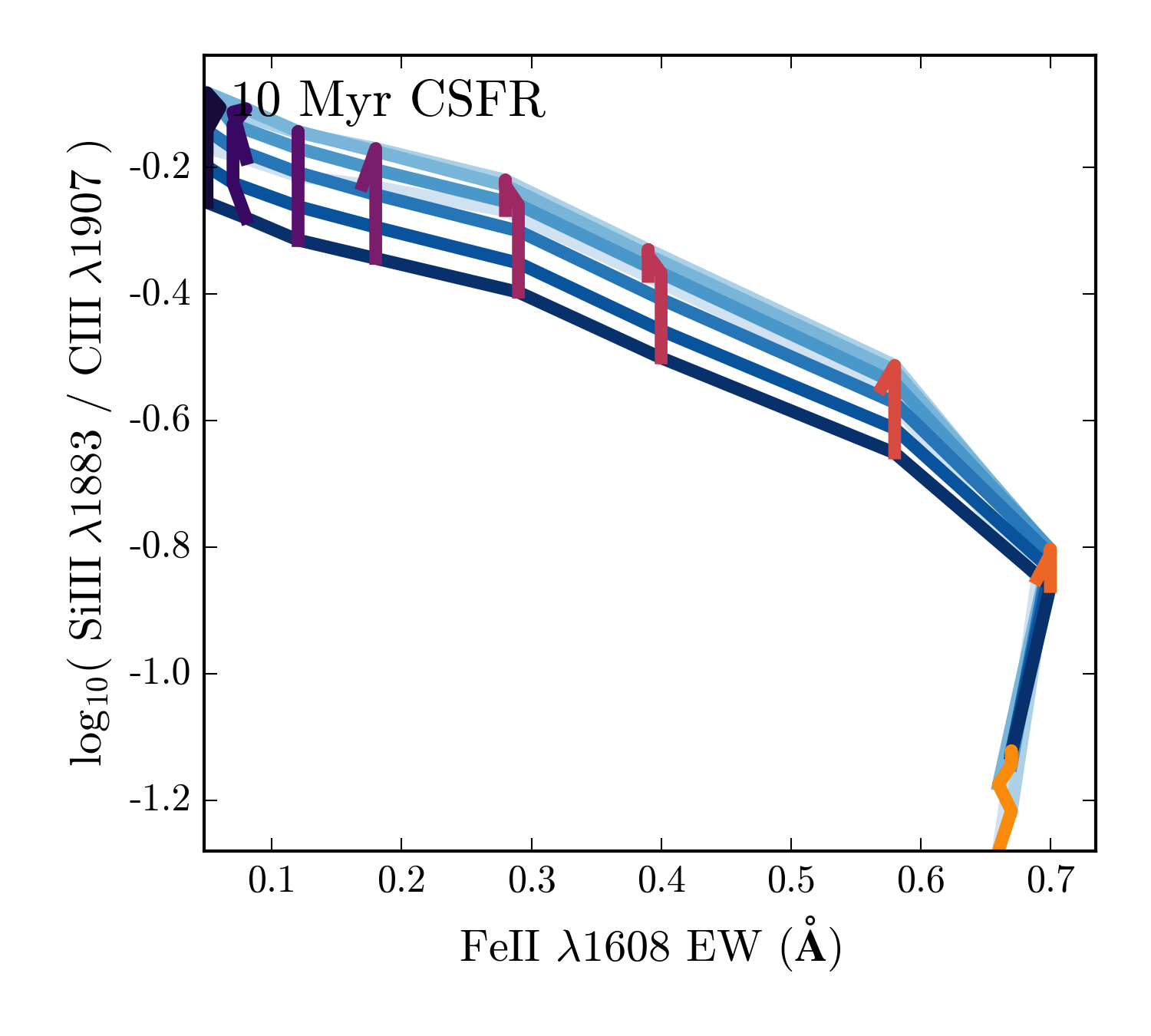}\\
    \includegraphics[width=0.495\linewidth]{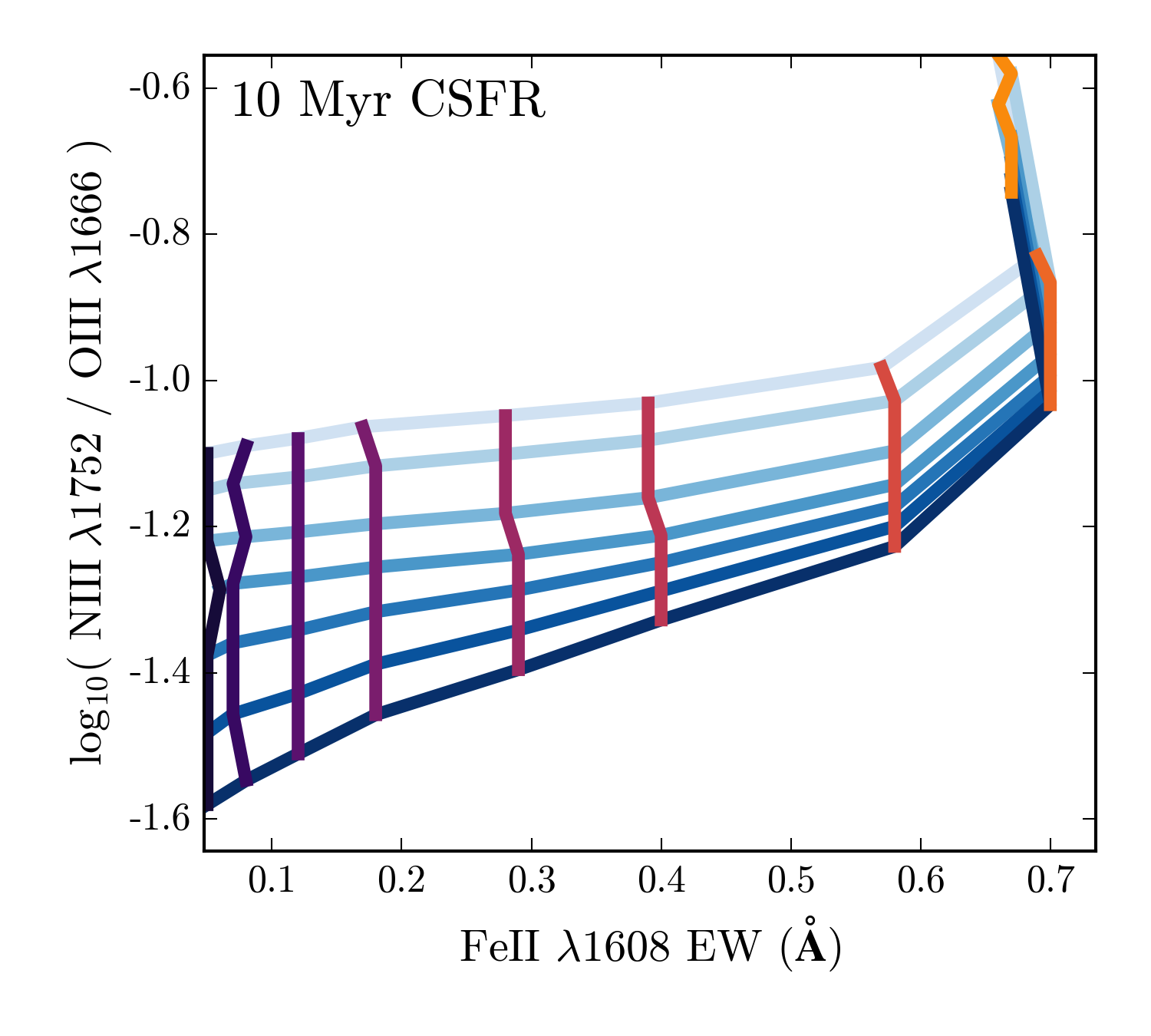}
    \includegraphics[width=0.495\linewidth]{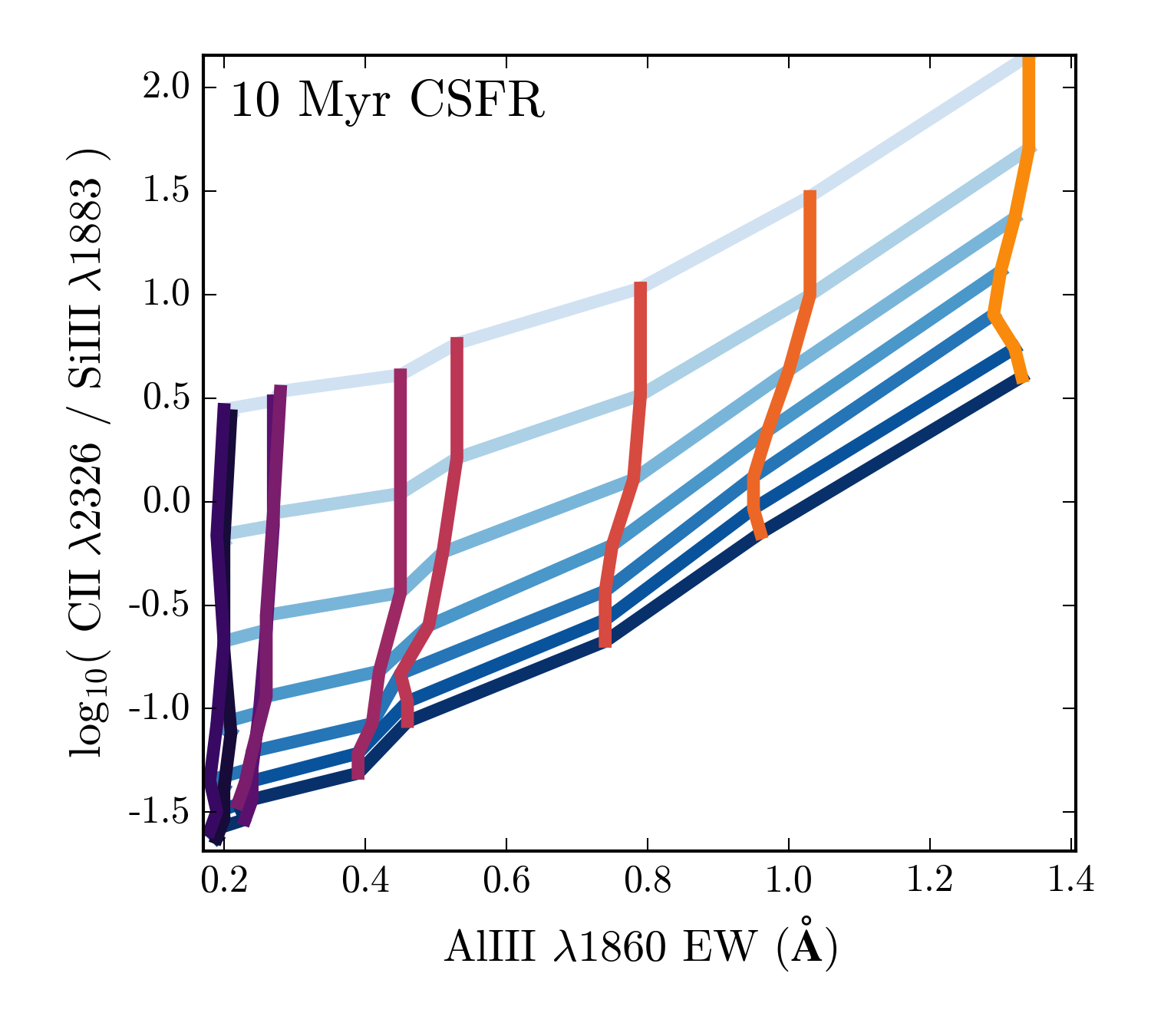}
    \caption{Combining the diagnostic capabilities of emission and absorption features for a 10\Myr constant SFR model. The $y$-axis shows an emission line ratio while the $x$-axis shows the equivalent width of an absorption feature. The blue lines connect models of constant ionization parameter. \logUeq{-1} is shown in dark blue and \logUeq{-4} is shown in light blue. Models of constant metallicity are connected by the colored lines, from \logZeq{-1} in purple and \logZeq{0.0} in yellow.}
    \label{fig:AbsEm}
  \end{center}
\end{figure*}


Fig.~\ref{fig:AbsEm} shows metallicity diagnostics using the combination of an emission line ratio and an absorption feature. Both of the absorption features strongly track the metallicity of the young stellar population and can provide complimentary metallicity information where the emission line ratios may be be less sensitive. For example, the Si4C2 emission line ratio excels at low metallicities where the \texttt{SIV\_1400} absorption feature is weakest (top left panel). In comparison, the \texttt{FeII\_1608} index has a better sensitivity to abundance changes at low-metallicity, pairing nicely with the Si3C3 and N3O3 emission line ratios (top right and bottom left panels of Fig.~\ref{fig:AbsEm}, respectively), which are most sensitive at moderate metallicities. The metallicity sensitivity of the \texttt{AlII\_1860} index correlates positively with the C2Si3 emission line ratio (bottom right), providing improved metallicity leverage.

We note that equivalent width measurements are much more sensitive to the continuum level than emission line ratios. The diagnostics in Fig.~\ref{fig:AbsEm} represent model predictions that are appropriate for young, massive, star clusters. The presence of a substantial old stellar population would increase model equivalent widths above those those presented here. To properly compare observations to the diagnostics that combine an equivalent width and an emission line ratio, the contribution from any underlying old stellar population should be removed.

\section{Observational Comparison} \label{sec:obs}

\subsection{UV spectra of local galaxies} \label{sec:obs:UV}

Blue Compact Dwarfs (BCDs\footnote{Blue Compact Dwarf is used here instead of Blue Compact Galaxy to avoid acronym confusion with Brightest Cluster Galaxies}), originally defined by \citet{Sargent+1970}, are characterized by blue optical colors, small sizes ($< 1$ kpc), and low luminosities (M$_{\mathrm{B}}>$-18). These galaxies have high specific-star formation rates and low metallicities, and have thus been proposed as nearby analogs of star formation in young galaxies observed at high redshift \citep{Thuan+2008}. The young ages associated with recent star formation make these objects very UV-bright, and their optical spectra are often indistinguishable from \hii regions.

\subsubsection{Emission lines} \label{sec:obs:emis}

We compare our models with data from \citet{Berg+2016}, who presented both UV and optical spectra for a sample of 7 nearby, low-metallicity, high-ionization BCDs. In this analysis to connect UV and optical emission line properties, \citet{Berg+2016} obtained UV spectra with the Cosmic Origins Spectrograph (COS) on the Hubble Space Telescope (HST) for a sample of galaxies with existing optical SDSS spectra and direct-method oxygen abundances. The UV spectra have ${\sim}0.6$\ang resolution at 1600\ang, which is sufficient for the emission line ratios considered in this paper. We include 13 additional galaxies from the upcoming Berg et al., 2018 (\emph{in prep}, priv. comm.) paper.

The galaxies in \citet{Berg+2016} are nearby ($0.003 < z < 0.040$), UV-bright ($m_{FUV} \leq 19.5$ AB), compact (D $< 5$''), low-metallicity ($7.2 \leq 12 + \log(O/H) \leq 8.0$) dwarf galaxies. These galaxies have relatively low masses (${\sim}10^7$\Msun) and high sSFRs (${\sim}10^{-8}$ yr$^{-1}$). All of the galaxies in the sample have auroral line detections in the optical for direct-method calculations of the nebular temperature, density, and metallicity.

Emission lines of \ciii 1907,1909\ang and \oiii 1666\ang are detected in all 20 galaxies, while the Si\textsc{III}]$\lambda$1883,1893\ang emission doublet is only detected in 13/20 galaxies. The non-detection is not surprising since these galaxies are reasonably metal-poor and the Si\textsc{III}]$\lambda$1893\ang emission feature is strongest around $12 + \log(O/H){\sim}8.3$ and gets weaker with decreasing metallicity. Only four galaxies had detected \oii$\lambda\lambda$3726,9  emission, for the remaining objects, we infer the total emission from the $\lambda$3726 and $\lambda$3729 line strengths using the $\lambda\lambda$7320,7330 lines.

\begin{figure*}
  \begin{center}
    \includegraphics[width=\linewidth]{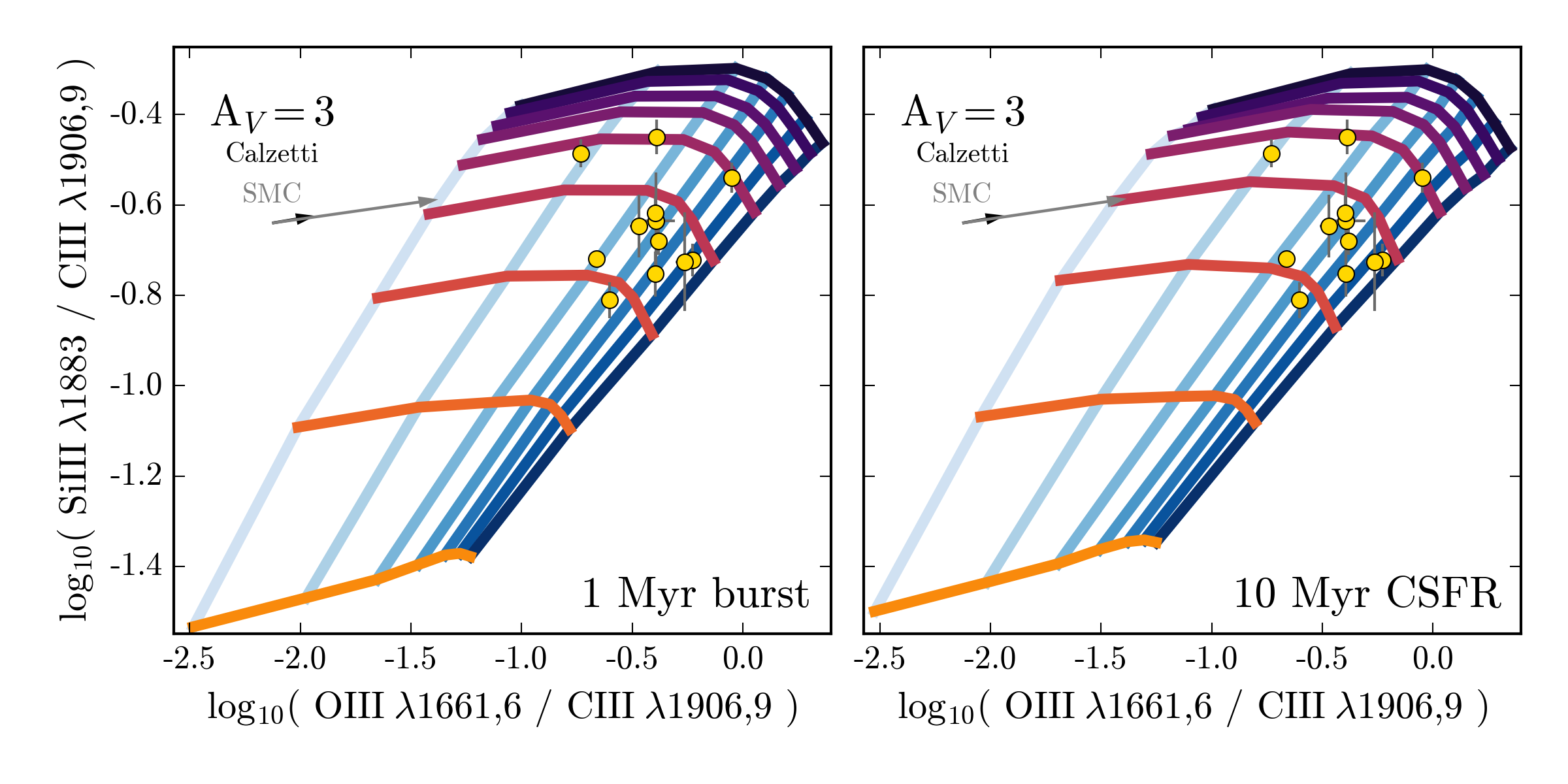}
    \caption{The \citet{Berg+2016} BCD sample compared to UV diagnostic diagrams for a 1\Myr burst (left) and a 10\Myr constant SFR (right). The blue lines connect models of constant ionization parameter. \logUeq{-1} is shown in dark blue and \logUeq{-4} is shown in light blue. Models of constant metallicity are connected by the colored lines, from \logZeq{-1} in purple and \logZeq{0.0} in yellow. The grey and black arrows show three magnitudes of extinction assuming SMC and Calzetti reddening laws, respectively.}
    \label{fig:dataDD1}
  \end{center}
\end{figure*}

\begin{figure*}
  \begin{center}
    \includegraphics[width=\linewidth]{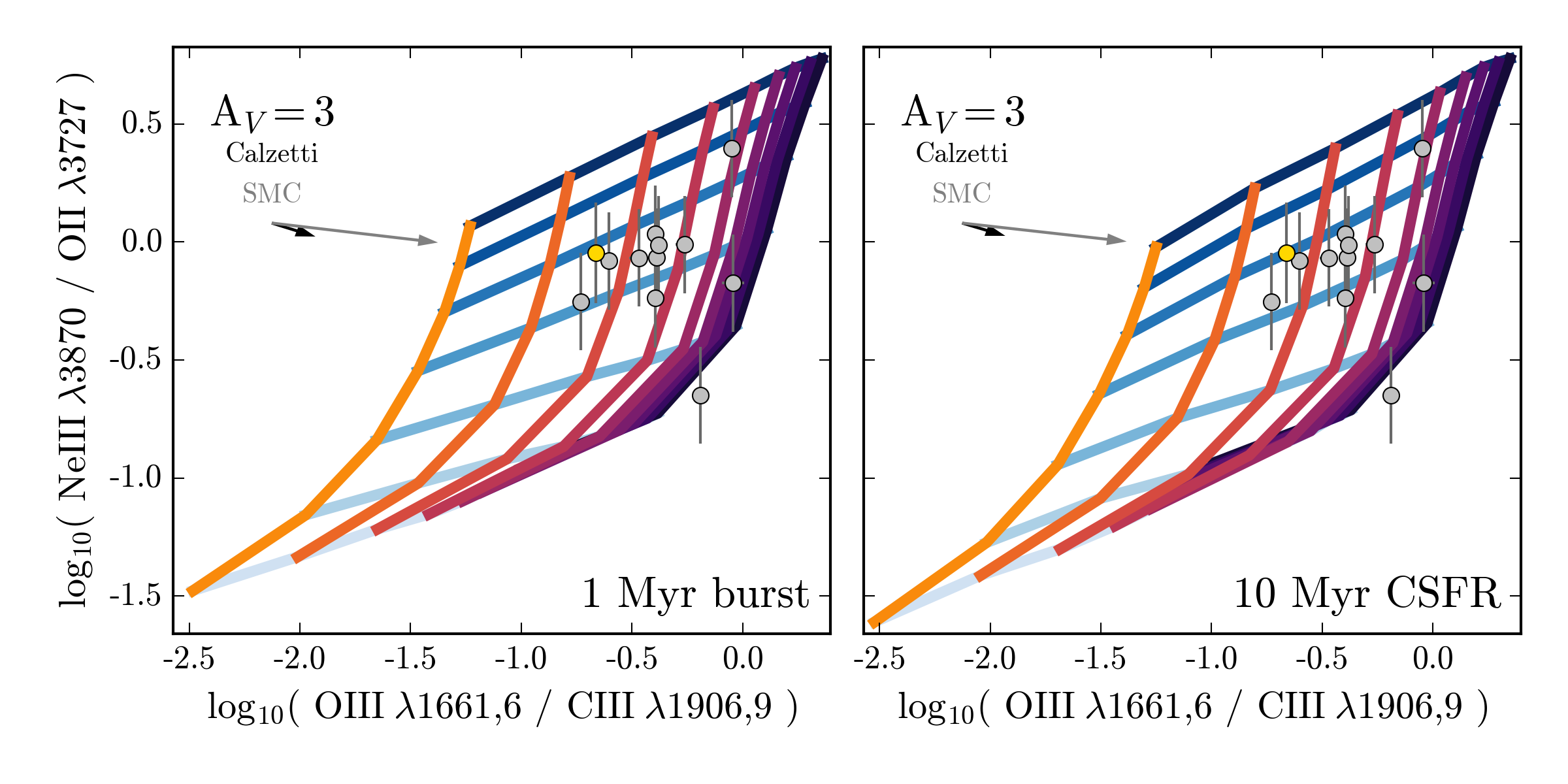}
    \caption{The \citet{Berg+2016} BCD sample compared to UV diagnostic diagrams for a 1\Myr burst (left) and a 10\Myr constant SFR (right). The blue lines connect models of constant ionization parameter. \logUeq{-1} is shown in dark blue and \logUeq{-4} is shown in light blue. Models of constant metallicity are connected by the colored lines, from \logZeq{-1} in purple and \logZeq{0.0} in yellow. The grey points represent BCDs where the \oii$\lambda$3727 strength was inferred from the \oii$\lambda\lambda$7320,7330 lines. The grey and black arrows show three magnitudes of extinction assuming SMC and Calzetti reddening laws, respectively.}
    \label{fig:dataDD2}
  \end{center}
\end{figure*}

\begin{figure*}
  \begin{center}
    \includegraphics[width=0.98\linewidth]{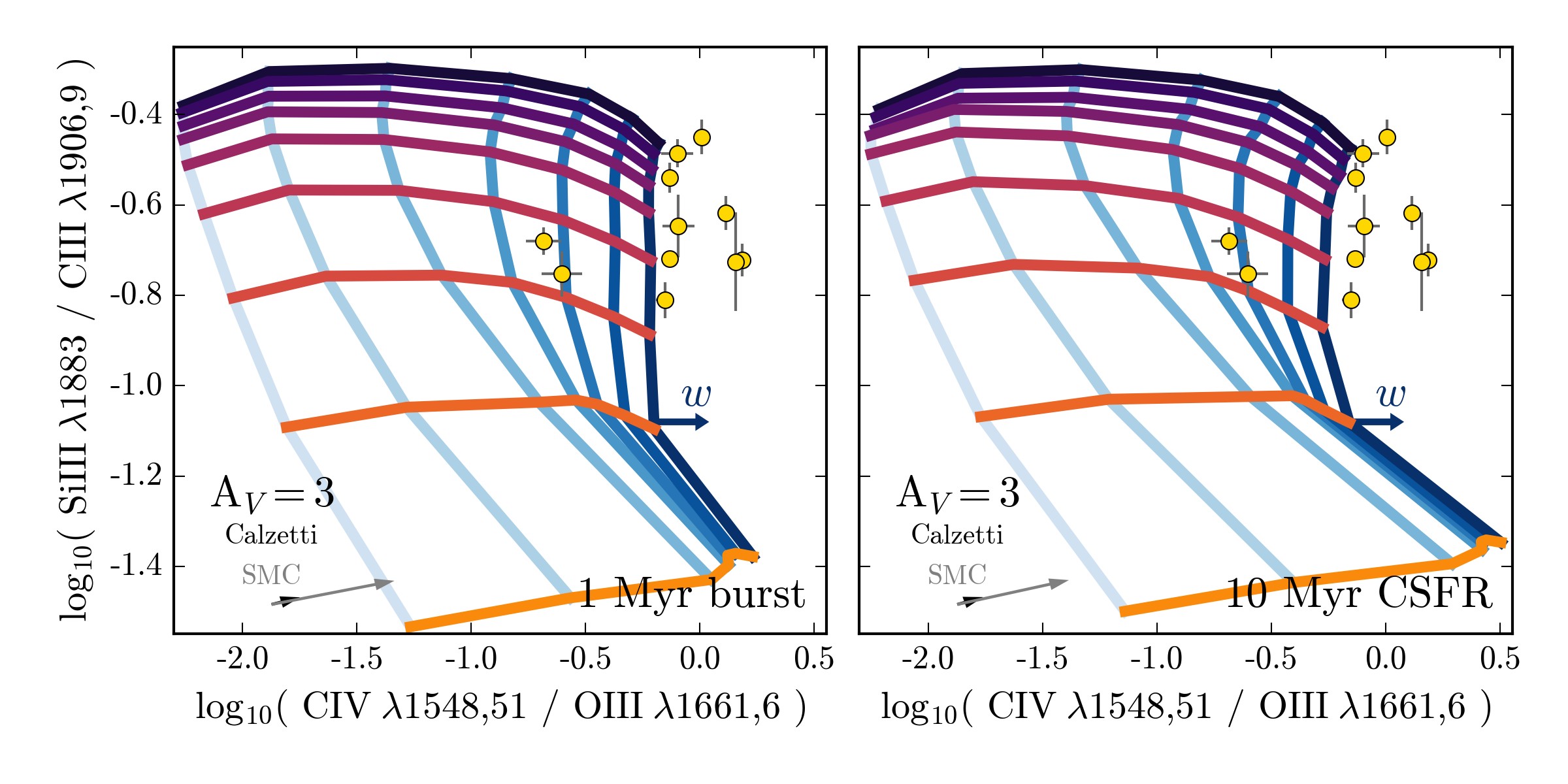}\\
    \includegraphics[width=0.98\linewidth]{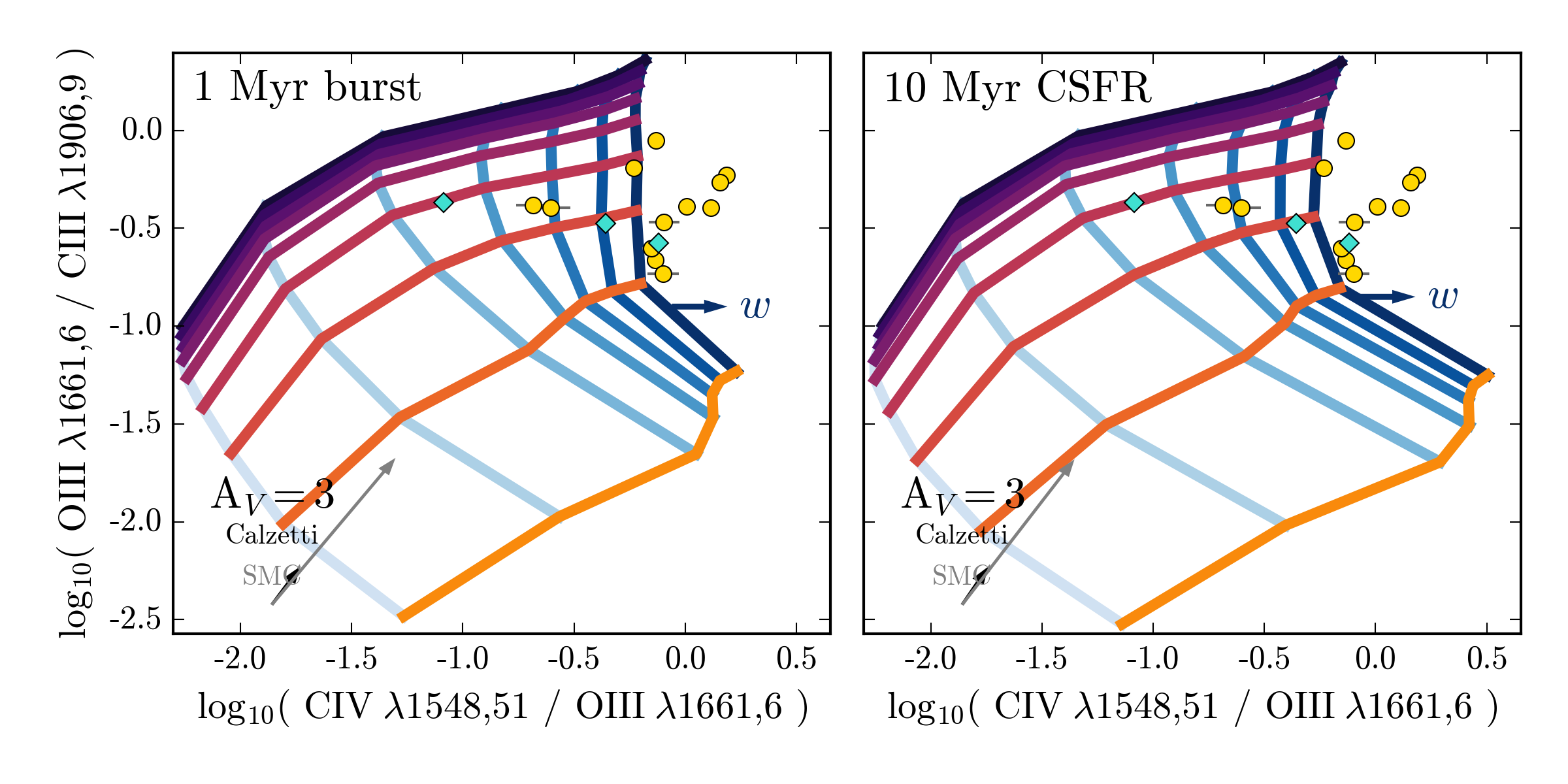}
    \caption{The \citet{Berg+2016} BCD sample (gold circles) compared to UV diagnostic diagrams for a 1\Myr burst (left) and a 10\Myr constant SFR (right). Observations from \citet{Senchyna+2017} are shown with blue diamonds on the bottom two plots. The blue lines connect models of constant ionization parameter. \logUeq{-1} is shown in dark blue and \logUeq{-4} is shown in light blue. Models of constant metallicity are connected by the colored lines, from \logZeq{-1} in purple and \logZeq{0.0} in yellow. Stellar wind emission can inflate the measured C\textsc{IV}1548, 1551 flux, we indicate the direction of this with the blue arrow (see Fig.~\ref{fig:CIVgrid}). The grey and black arrows show three magnitudes of extinction assuming SMC and Calzetti reddening laws, respectively.}
    \label{fig:dataDD3}
  \end{center}
\end{figure*}

In Figs.~\ref{fig:dataDD1}, \ref{fig:dataDD2}, \& \ref{fig:dataDD3}, we show measured emission line ratios from the \citet{Berg+2016} sample on various UV diagnostic diagrams. The diagnostic diagrams shown here are only a subset of those from \S\ref{sec:mod:comb}, since we are limited to those diagrams where all four emission lines were measurable in the observed galaxies. To better compare our diagnostics with observations where emission line doublets are not resolved, we use the sum of \civ$\lambda$1548,51, \oiii$\lambda$1661,6, \ciii$\lambda$1907,9, and \oii$\lambda$3726,9 doublets in the remaining figures, as noted on the axis labels.

Fig.~\ref{fig:dataDD1} shows \oiii$\lambda1661,6$/\ciii$\lambda1906,9$ (O3C3) \vs \SiuIII$\lambda1883$/\ciii$\lambda1906,9$ (Si3C3) for both a 1\Myr burst and a 10\Myr constant SFR model. The model grid is able to reproduce the full range of observed line ratios. The data points cover a region of the model grid with low metallicities (\logz${\sim}-1.25$ to $-0.5$, or $\log( \mathrm{O} / \mathrm{H} ) + 12{\sim}-7.5$ to $-8.0$) and high ionization parameters (between \logU${\sim}-2$ and $-1$), both of which are consistent with estimates from the optical spectroscopy.

Fig.~\ref{fig:dataDD2} shows O3C3 \vs \neiii$\lambda$3869/\oii$\lambda$3726,9 (Ne3O2) for stellar populations assuming a 1\Myr burst and a 10\Myr constant SFR. The young burst and the constant SFR models are very similar, and both grids are able to reproduce the observed line ratios. The O3C3 and Ne3O2 line ratios are sensitive to ionization parameter, and the combination of them provides the most leverage for discriminating among models of constant ionization parameter. The BCDs in the O3C3 \vs Ne3O2 diagram have line ratios consistent with ionization parameters around \logUeq{-2.0}. For the galaxies that appear in both Fig.~\ref{fig:dataDD1} and Fig.~\ref{fig:dataDD2}, the UV model grids predict ionization parameters that are consistent with one another and with the optically-derived ionization parameters.

Fig.~\ref{fig:dataDD3} shows O3C3 \vs \civ$\lambda$1548,51/\oiii$\lambda$1661,6 (C4O3; bottom row) and C4O3 \vs Si3C3 (top row) for a 1\Myr burst and 10\Myr constant SFR. In the bottom row of Fig.~\ref{fig:dataDD3} we include the sample of nearby dwarf galaxies presented in \citet{Senchyna+2017}, which agree well with the \citet{Berg+2016} sample. This comparison is only possible for Fig.~\ref{fig:dataDD3}, since the \citet{Senchyna+2017} sample only includes emission line measurements for \ciii$\lambda1906,9$ (as a blend), \oiii$\lambda$1661,6, and \civ$\lambda$1548,51.

In contrast to Figs.~\ref{fig:dataDD1} \& \ref{fig:dataDD2}, these diagnostics are not able to fully reproduce the range of observed line ratios. In particular, the majority of the C4O3 ratios lie beyond the highest ionization parameters in the model grid, which could indicate that the models do not reach high enough ionization states or do not have hard enough ionizing spectra. However, the consistency of the other grids with \logUeq{-2} suggest that instead this behavior is primarily driven by larger-than-predicted contributions from stellar wind emission to the \civ$\lambda1548,51$ flux. The grids in Fig.~\ref{fig:dataDD3} only include the nebular contribution, and the C4O3 ratio should thus be considered a lower limit when compared to the observed C4O3 ratios. We note that the \civ$\lambda1548,51$ lines in the \citet{Berg+2016} spectra are in fact broader than the other nebular emission lines, which is consistent with wind contamination.

Contribution from both nebular emission and stellar wind emission is a known issue for other UV emission lines as well (e.g., \heii$\lambda1640$). In theory, with sufficient resolution and signal-to-noise, one could distinguish the two components by their different line profiles, as the nebular emission component should be narrow while the stellar wind emission component should be fairly broad. However, this measurement is difficult in the more common low-to-mid-resolution spectra. For this reason,  we recommend limiting diagnostic diagrams to line ratios that include no more than one emission line with a well-known stellar emission contribution. For example, the commonly used \heii$\lambda1640$ / \civ$\lambda1548,51$ line ratio \citep[e.g.,][]{Feltre+2016, Nakajima+2017} may be difficult to interpret due to the unknown contribution from stellar and nebular emission in both the numerator and denominator.

\subsubsection{Stellar and Nebular CIV emission} \label{sec:obs:civ}

\begin{figure}
  \begin{center}
    \includegraphics[width=\linewidth]{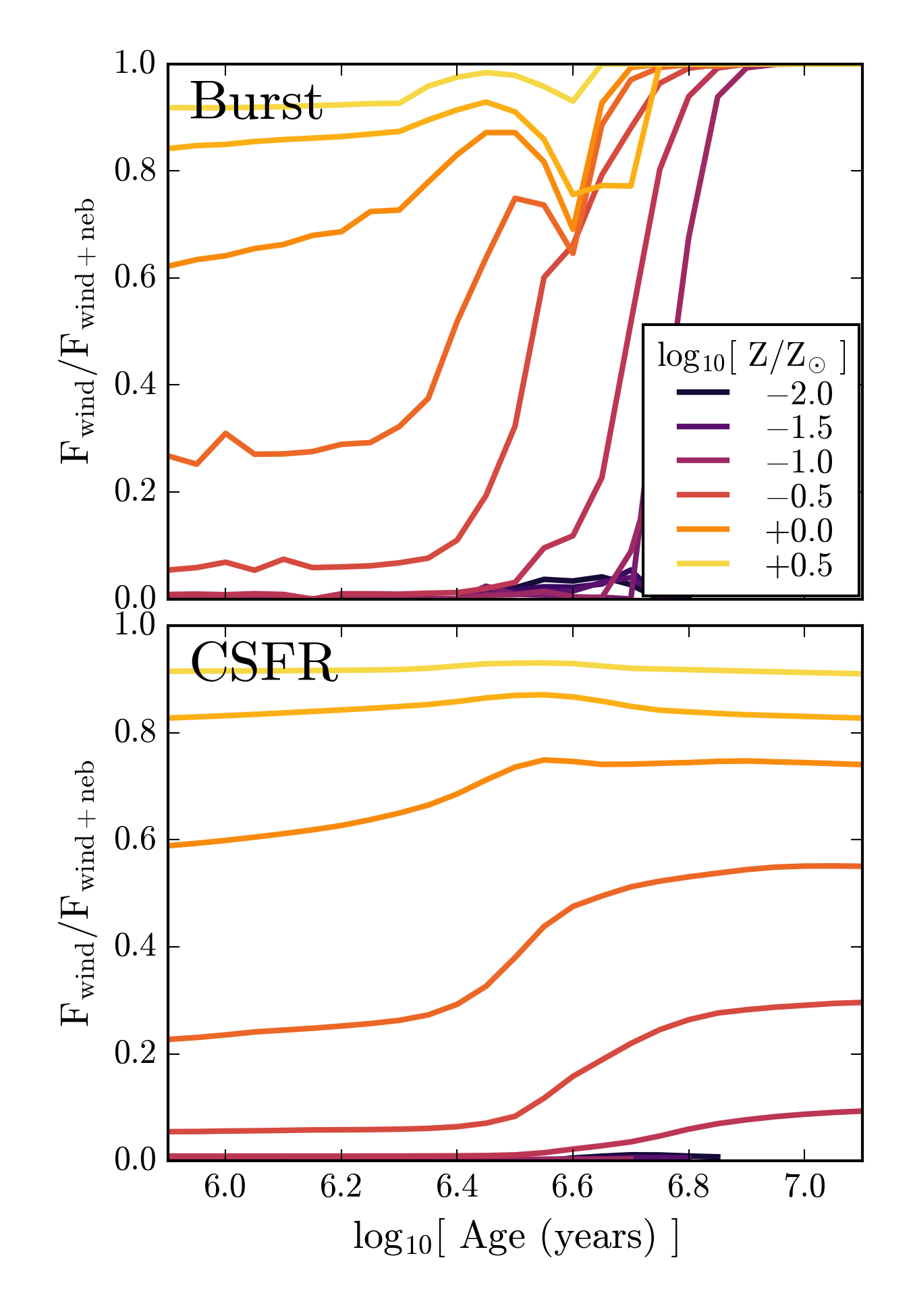}
    \caption{The fractional flux in the stellar \civ emission compared to the total \civ emission (nebular \civ$\,\lambda$1548 + nebular \civ$\,\lambda$1551 + stellar \civ$\,\lambda$1551) for populations assuming instantaneous bursts (top) and constant SFR (bottom). The lines are color coded by metallicity, from \logZeq{-2} in dark blue to \logZeq{0.5} in yellow. Stellar \civ wind emission is most important at high metallicities, \logZeq{0} and higher. The nebular \civ emission is most important at low metallicities, \logZeq{-0.5} and lower.}
    \label{fig:CIVflux}
  \end{center}
\end{figure}

The model grids in Fig.~\ref{fig:dataDD3} show the behavior expected for pure nebular emission. However, actual spectra will include both stellar and nebular components, along with strong features from stellar winds. These effects are particularly important for \civ, where the stellar \civ emission is shaped by the line-driven winds in massive stars, and will thus change strength with metallicity and time. Its line profile is also complex and has a P Cygni profile with a redshifted emission component and a blueshifted absorption component.

To estimate the total flux in the wind emission that would likely ``contaminate'' the nebular emission measurement at low spectral resolution, we fit a gaussian profile to the emission component of the P Cygni profile in our model spectra. We then add this flux to the total flux of the \civ$\lambda$1548 and \civ$\lambda$1551 nebular emission lines to quantify the change in the  \civ$\lambda$1548/\oiii$\lambda$1661,6 ratio. The fluxes are calculated assuming a total stellar mass of $10^7\,$\Msun and a distance of 10\,Mpc. The choice is ultimately unimportant, however, since we are primarily interested in comparing ratios of emission lines.

In Fig.~\ref{fig:CIVflux} we show the flux in the stellar \civ$\lambda$1551 wind emission as a fraction of the total nebular and stellar \civ flux as a function of time at different metallicities. The top panel shows instantaneous burst models while the bottom panel shows constant SFR models. At all metallicities, the nebular \civ emission decreases with time as the massive stars providing the ionizing radiation die. Nebular \civ emission is strongest at lower metallicities (\logz $\lesssim$ -0.5) and is relatively weak in solar-metallicity populations.

It is clear from Fig.~\ref{fig:CIVflux} that stellar wind contamination will be strongest at solar-like metallicity, where line-driven winds are strong and nebular \civ emission is relatively weak. In models at solar metallicity and higher, stellar wind emission contributes 60\% or more of the total \civ emission. While stellar wind emission also decreases with time, it decreases over slightly longer timescales than the nebular emission. The peak stellar wind emission changes with metallicity, with stellar emission peaking at later times in lower metallicity populations. Even at \logZeq{-1} stellar \civ emission can contribute as much as 30\% of the total \civ emission.

\begin{figure*}
  \begin{center}
    \includegraphics[width=0.98\linewidth]{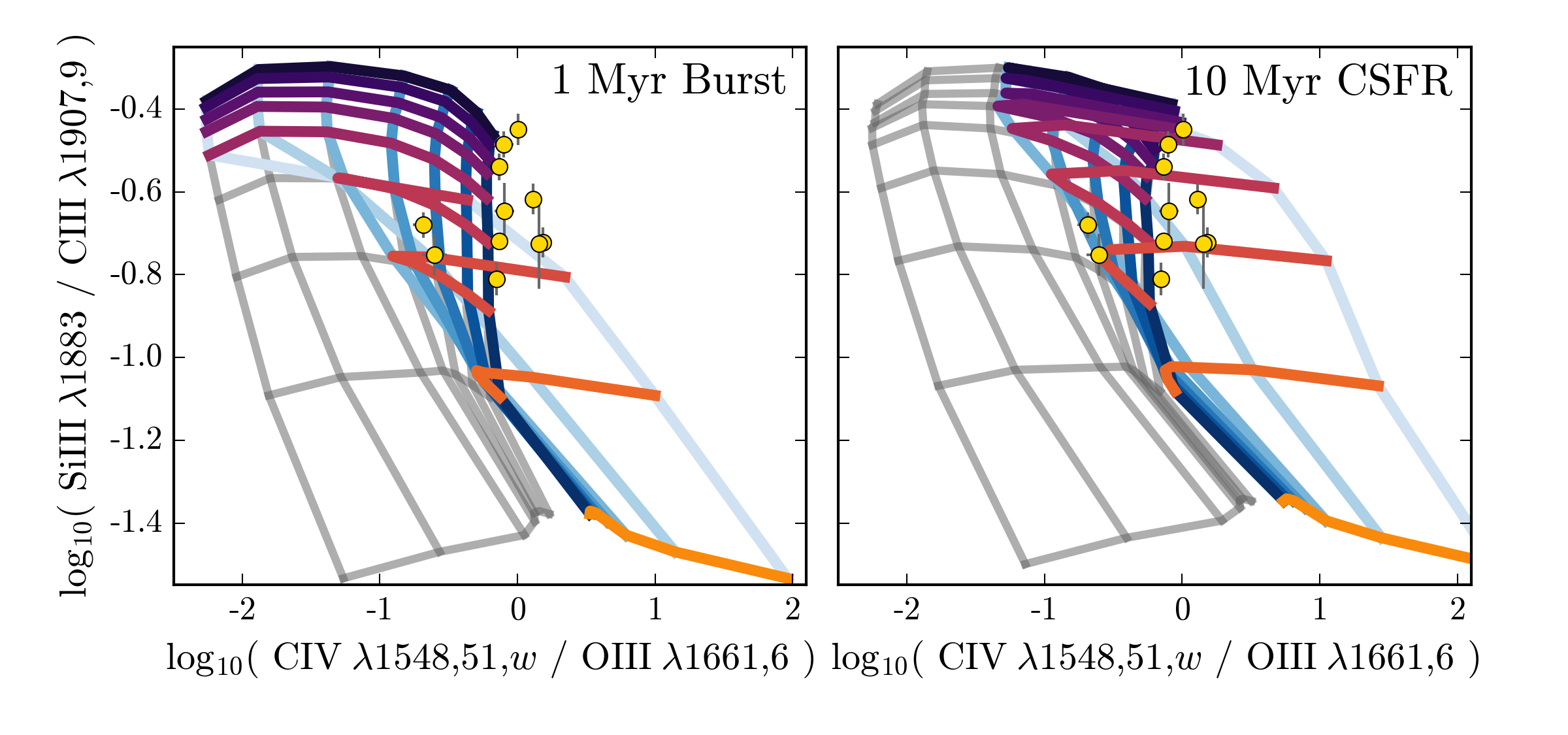}\\
    \includegraphics[width=0.98\linewidth]{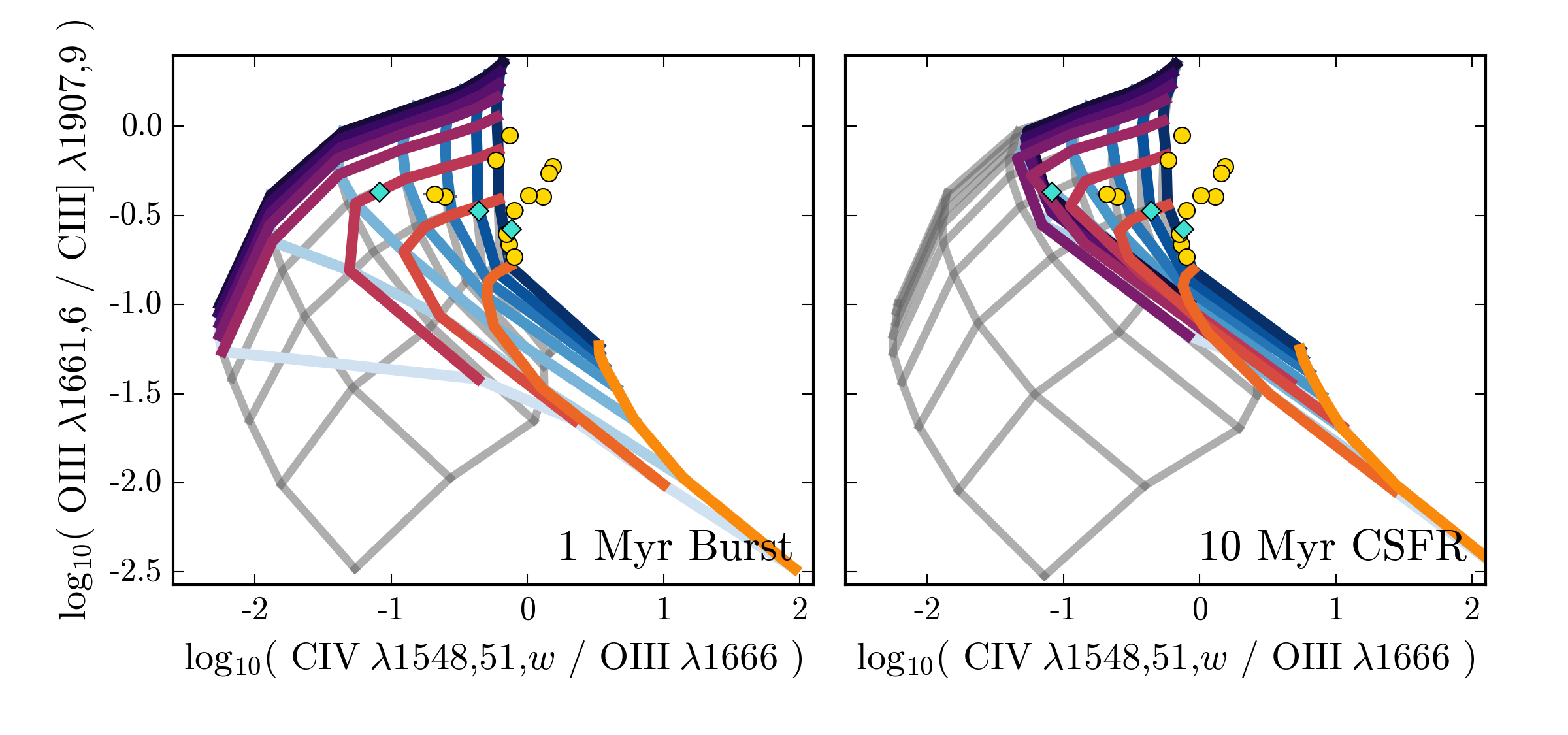}
    \caption{The effect of including \civ stellar wind emission in the \civ$\lambda$1548/\oiii$\lambda$1666 line ratio for a 1\Myr burst (left) and a 10\Myr constant SFR (right) for two different pairs of emission line diagnostics (top and bottom). The grey lines show the nebular-only contribution, identical to that shown in Fig.~\ref{fig:dataDD3}. The multi-colored lines show the model grid where nebular \civ$\lambda$1551 and stellar \civ$\lambda$1551 emission is included in the \civ$\lambda$1548/\oiii$\lambda$1666 ratio. The blue lines connect models of constant ionization parameter, from \logUeq{-1} (dark blue) to \logUeq{-4} (light blue). Models of constant metallicity are connected by the colored lines, from \logZeq{-1} in purple to \logZeq{0.0} in orange. The \citet{Berg+2016} BCD sample is shown with gold circles and the \citet{Senchyna+2017} sample is shown with blue diamonds.}
    \label{fig:CIVgrid}
  \end{center}
\end{figure*}

We show the resultant model grids in Fig.~\ref{fig:CIVgrid}, where we show the same O3C3 \vs \civ$\lambda$1548,51/\oiii$\lambda$1661,6 diagnostic diagram from Fig.~\ref{fig:dataDD3} for 1\Myr instantaneous burst and 10\Myr CSFR populations. We show the original ``nebular-only'' \civ$\lambda$1548,51/\oiii$\lambda$1661,6 grid in grey and the same observed \civ$\lambda$1548,51/\oiii$\lambda$1661,6 ratios from Fig.~\ref{fig:dataDD3} for visual reference. The new model grid includes \civ$\lambda$1548,1551 and the stellar \civ$\lambda$1551 wind emission, which we label as \civ$\lambda$1548,51,$w$ for clarity.

The model grids with winds show improved agreement with the data, due to the $\sim0.2-0.5$ dex shift to higher C4O3 ratios. As expected, the C4O3 ratio changes dramatically at solar metallicity, where the wind emission contributes significantly and changes the C4O3 ratio by 0.3 to 2.0 dex. For populations with metallicities -1 $\lesssim$ \logz $\lesssim$ -0.5, the line ratio changes are more modest, typically 0.2 to 0.5\,dex. At the lowest metallicities where wind emission is unimportant, the line ratio increases by at most 0.2\,dex, driven by the \civ$\lambda$1551 nebular emission.

There are slight differences between the instantaneous burst and constant SFR grids, but both do a good job at reproducing most of the observed data points. The differences between the model grids and data are not concerning, because these wind-corrected grids are essentially measuring uncertainties, due to difficulty of wind models. It is likely that the few off-grid points could be explained with modest perturbations to the wind model.

\subsubsection{Absorption Indices} \label{sec:obs:abs}

To evaluate the absorption line indices identified in \S\ref{sec:mod:abs}, we compare our models to measurements of 14 nearby ($0.003 < z < 0.03$) BCDs presented in \citet{Zetterlund+2015}, observed using the HST COS spectrograph. This sample has metallicities derived from optical emission line ratios from KK04 based on the R$_{23}$ diagnostic, giving an estimated accuracy of ${\sim}0.15$ dex \citep{Zetterlund+2015}. The accuracy of the R$_{23}$ metallicity diagnostic is worse near $12+\log(O/H)\sim 8.4$, where the value of R$_{23}$ rolls over, and may be biased high compared to stellar metallicities \citep[e.g.,][]{Kudritzki+2012}.

There were no nebular emission features detected in this sample, due to sample selection effects and low signal-to-noise (S/N${\sim}$1-5). However, \citet{Zetterlund+2015} did provide measurements of \texttt{AlII\_1670} and \texttt{CIV\_1550}, two of the L11 absorption indices.

The \citet{Zetterlund+2015} BCD spectra have not been corrected for the presence of an underlying old stellar population. For a typical dwarf galaxy with a burst of star formation in the last 300\Myr, most of the total stellar mass lies in old populations (3 Gyr or older), which contribute very little to the total UV flux, less than a percent. The assumption of continuous SFR over 4 Myr ultimately dominates the uncertainty in the stellar continuum, while SFR fluctuations over 10-100 Myr can produce continuum fluctuations at the 10\% level.

\begin{figure}
  \begin{center}
    \includegraphics[width=\linewidth]{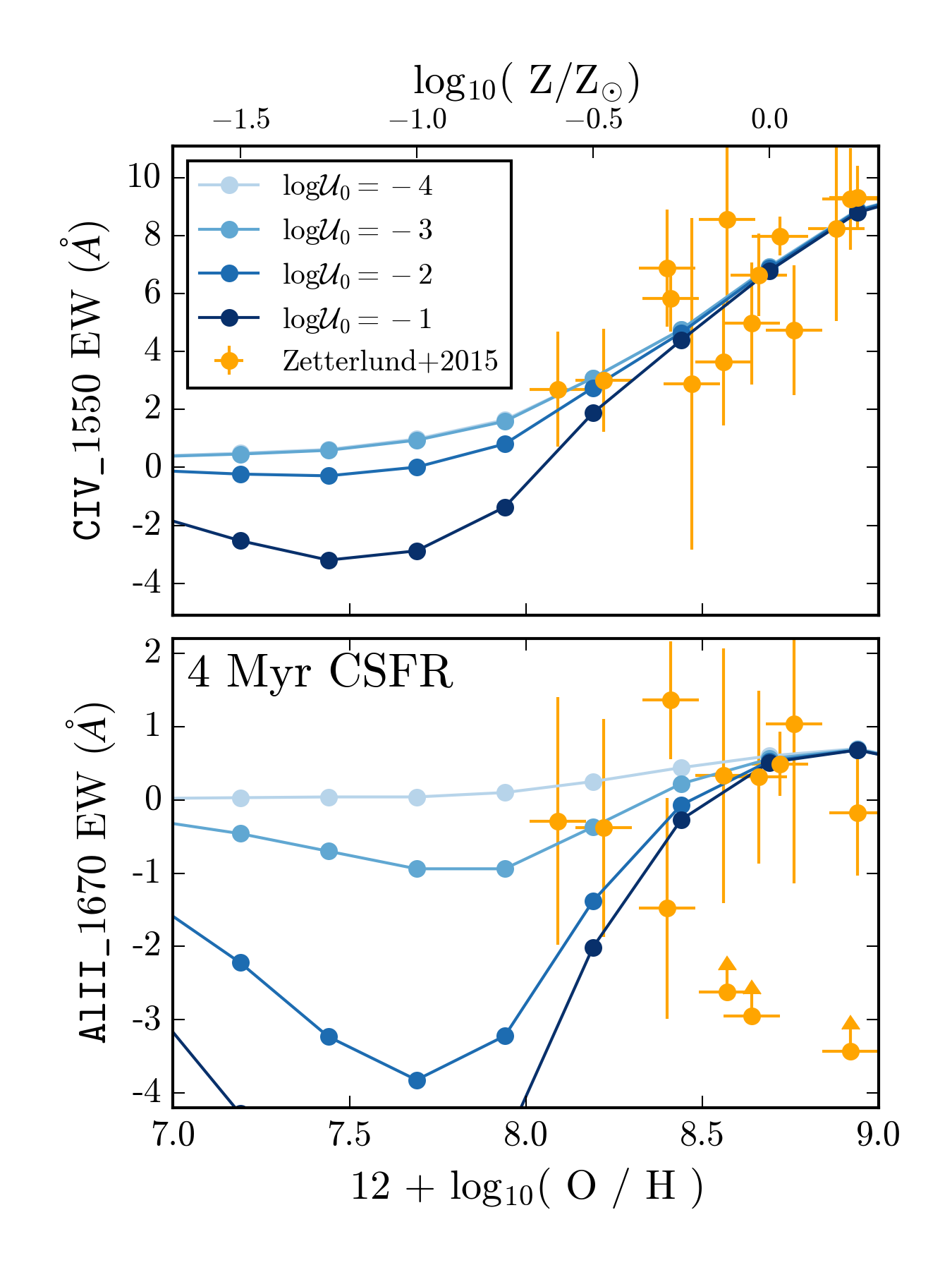}
    \caption{The equivalent widths of the L11 \texttt{AlII\_1670} and \texttt{CIV\_1550} absorption indices as a function of metallicity. The orange markers are measured from local BCD spectra, with metallicity estimates derived from optical emission line ratios. The blue lines show our emission model at several different ionization parameters for a 10\Myr CSFR.}
    \label{fig:BCDabs}
  \end{center}
\end{figure}

In Fig.~\ref{fig:BCDabs} we show equivalent widths of \texttt{AlII\_1670} and \texttt{CIV\_1550} as a function of galaxy metallicity. In general, the \texttt{CIV\_1550} index EWs are large and positive, as expected at high metallicities where stellar absorption dominates the index. The \texttt{AlII\_1670} index covers much weaker features and the index measurements are quite noisy due to the low S/N observations. The \texttt{AlII\_1670} takes both positive and negative values, potentially reflecting the influence of filling from emission lines (negative EWs), which is particularly problematic for this index, as discussed in \ref{sec:mod:abs}.

Although one \texttt{AlII\_1670} measurement lies above our model predictions, we note that our models do not include the effects of interstellar absorption. This was considered in detail in \citet{Vidal-Garcia+2017}, who found that interstellar line absorption near Al{\sc ii}$\lambda$1670 can be significant, especially at high metallicity.

\subsection{UV spectra of high redshift galaxies} \label{sec:obs:UV:LBGs}

Rest-frame UV spectra of distant galaxies are a powerful tool to study the star formation environment of young galaxies, and will become higher quality and increasingly commonplace with future optical and IR instruments on the next generation of telescopes. In this section we compare the UV diagnostic diagrams presented in \S\ref{sec:mod:comb} to the rest-frame UV spectra of galaxies at intermediate- and high-redshift obtained with ground-based optical and IR instruments.

Before proceeding, we note that that estimates of metallicity for these distant galaxies, when used, are determined from rest-frame optical emission line diagnostics observed in the IR. These diagnostics are calibrated against local galaxies, where the star forming environment could be drastically different from the star forming environment typical of distant galaxies, and for which the application of local metallicity diagnostics may not be appropriate. It is still interesting, however, to understand where and why the emission line models succeed or fail.

\begin{figure*}
  \begin{center}
    \includegraphics[width=0.495\linewidth]{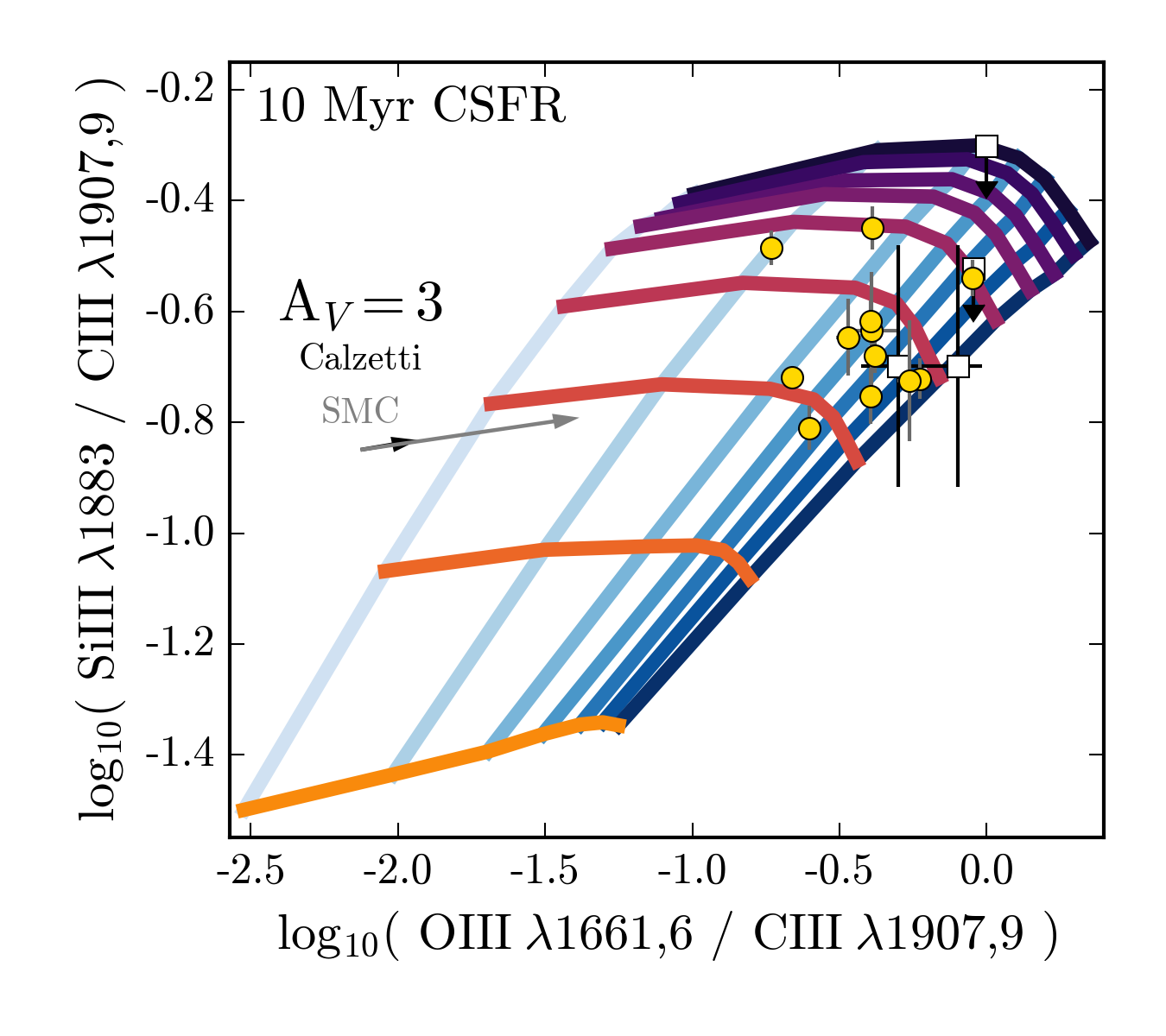}
    \includegraphics[width=0.495\linewidth]{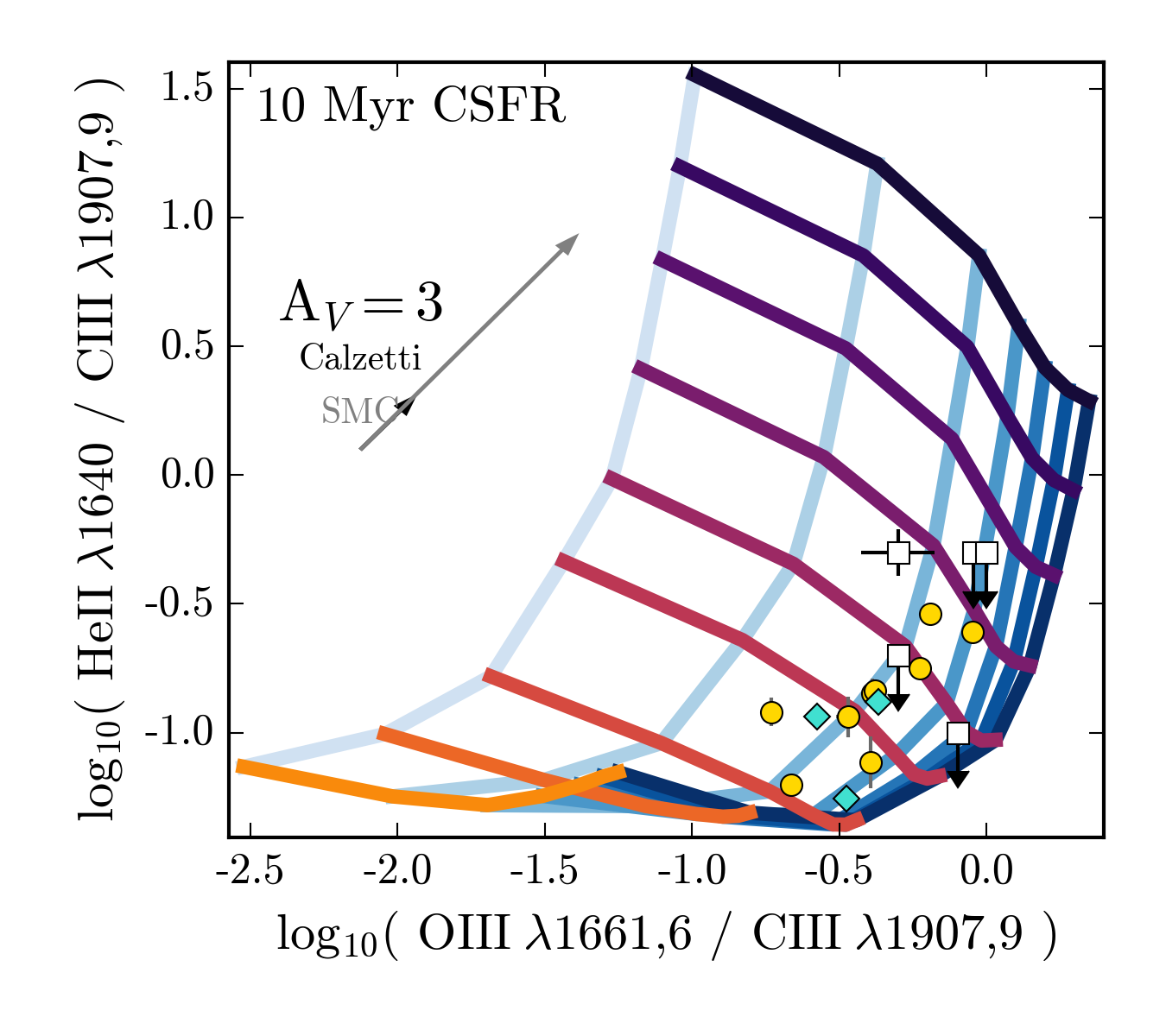}\\
    \includegraphics[width=0.495\linewidth]{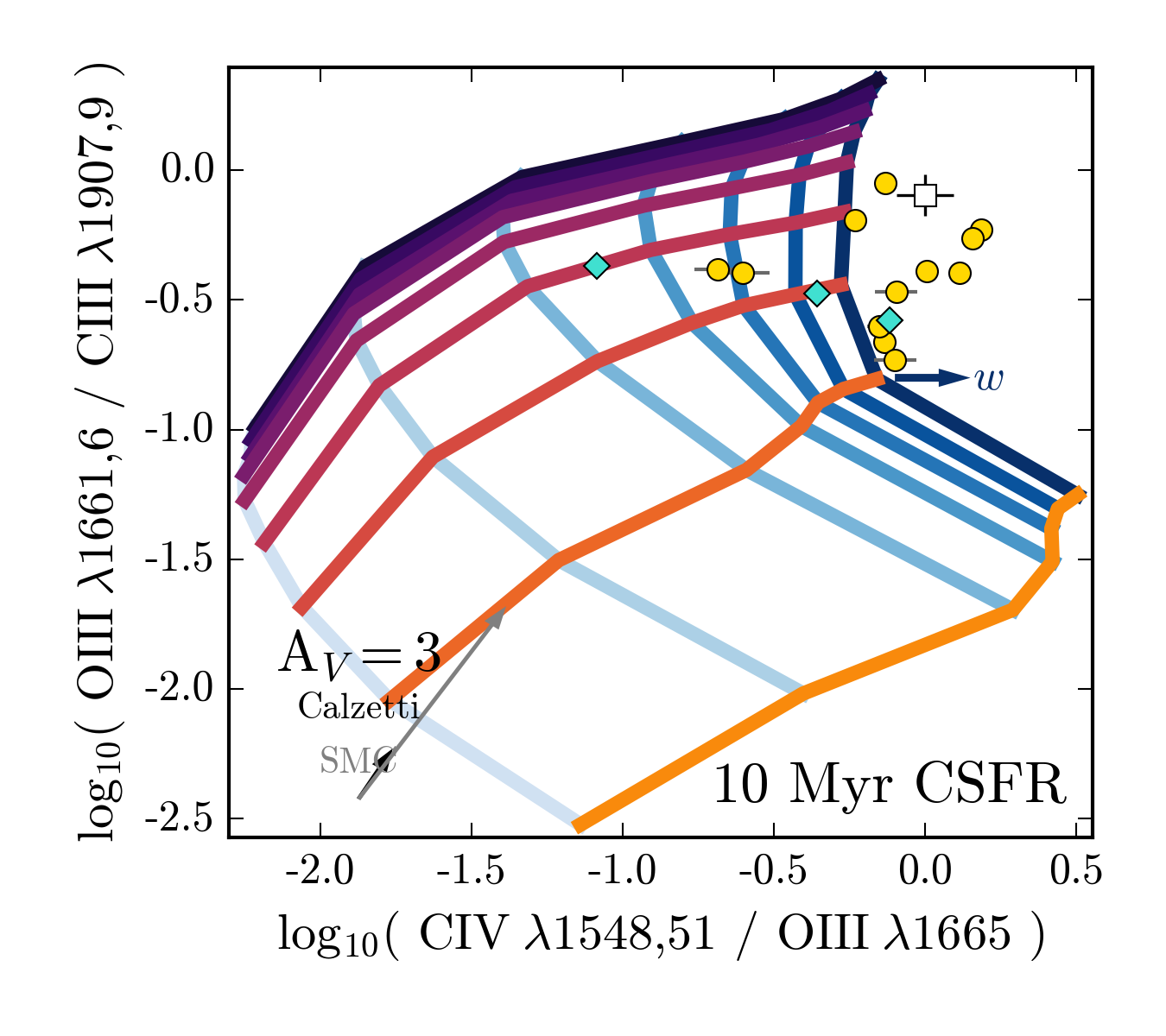}
    \includegraphics[width=0.495\linewidth]{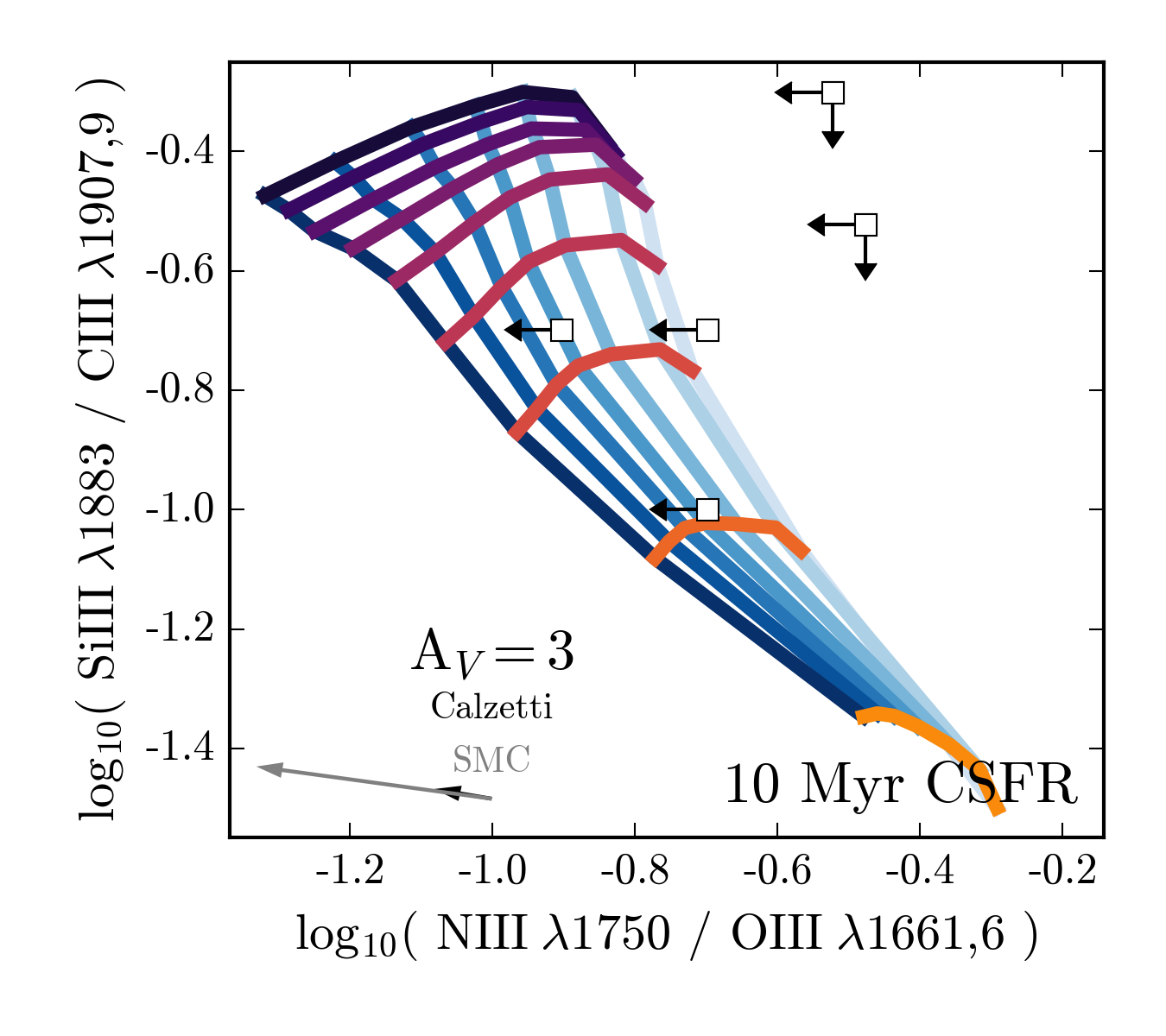}
    \caption{The \citet{Stark+2014} galaxies at $z{\sim}2$ (grey squares) compared to various UV diagnostic diagrams, assuming a 10\Myr stellar population with constant SFR. The blue lines connect models of constant ionization parameter. \logUeq{-1} is shown in dark blue and \logUeq{-4} is shown in light blue. Models of constant metallicity are connected by the colored lines, from \logZeq{-1} in purple and \logZeq{0.0} in yellow. The \citet{Berg+2016} local BCD sample (gold circles) is included in the upper left, upper right, and lower left diagrams; the \citet{Senchyna+2017} local BCD sample (blue diamonds) is included in the lower left diagram. The grey and black arrows show three magnitudes of extinction assuming SMC and Calzetti reddening laws, respectively. The model grids are mostly consistent with the \citet{Stark+2014} data within error. The disagreement in the C4O3 ratio (lower left) may be the result of stellar wind emission inflating the measured C\textsc{IV}1548 flux; we indicate the direction of this with the blue arrow.}
    \label{fig:dataHiZ}
  \end{center}
\end{figure*}

In Fig.~\ref{fig:dataHiZ}, we show several diagnostic diagrams with a sample of intermediate and high redshift galaxies compiled from \citet{Stark+2014} (black circles) and \citet{Christensen+2012} (white diamond). The \civ$\lambda1548,51$, \SiuIII$\lambda1883$, \oiii$\lambda1661,6$, \ciii$\lambda1906,9$ emission lines are measured in both the local galaxy sample from \citet{Berg+2016} and the lensed $z{\sim}3$ galaxy sample from \citet{Stark+2014}. We include the local galaxy observations in Fig.~\ref{fig:dataHiZ} as grey circles. \citet{Stark+2015} provides lower limits on the \SiuIII$\lambda1883$ emission line in cases where it was not detected based on observational limits.

In the Si3C3 \vs O3C3 diagram in Fig.~\ref{fig:dataHiZ} (\emph{upper left}), all of the high redshift galaxies have emission line ratios consistent with the model predictions. Overall, the line ratios for the high redshift galaxies tend toward higher ionization parameters and slightly higher metallicities than the local sample. This trend is likely driven entirely by the strength of the \ciii line, which is notably large in many of the high redshift galaxies \citep{Stark+2014}, but variations in \SiuIII$\lambda1883$ (an $\alpha$-element) could also play a role. Again, the shift towards higher ionization parameters and higher metallicities in the high redshift sample compared to local BCDs is not unexpected, given the bias toward more massive, high SFR galaxies in the former.

We show \heii$\lambda$1640/\ciii$\lambda$1906,9 ratio (He2C3) \vs the \oiii$\lambda$1661,6/\ciii$\lambda$1906,9 ratio (O3C3) in the upper right panel of Fig.~\ref{fig:dataHiZ}. All of the high redshift galaxies have emission line ratios consistent with the model predictions. The high redshift galaxies and the local galaxies occupy similar regions of the diagram, with no clear offset in metallicity or ionization parameter. However, all of the observations lie near models with higher metallicities than in previous diagnostic diagrams. The He2C3 ratio uses emission lines more widely spaced in wavelength than the other ratios in Fig.~\ref{fig:dataHiZ}, which could indicate that reddening corrections are responsible for the shift.

In the C4O3 \vs O3C3 diagram in Fig.~\ref{fig:dataHiZ} (\emph{lower left}), the high redshift galaxies occupy a region entirely beyond the model grid with large values of C4O3. The range of observed C4O3 and O3C3 values in the high redshift galaxies are consistent with values observed in the local galaxy sample from \citet{Berg+2016} and \citet{Senchyna+2017}, though the local sample extends to lower values of C4O3 and has a wider range in O3C3. These differences are not entirely unexpected, since the high redshift sample is smaller and necessarily probes the brightest observable targets. The high C4O3 ratios observed in the high redshift sample may support the idea that high redshift galaxies have on average ``harder'' ionizing spectra. Though we again note the important contribution stellar winds make to the C4O3 ratio, as noted by the blue arrow and discussed in \S\ref{sec:obs:civ} (Fig.~\ref{fig:CIVgrid}).

The N\textsc{ii}]$\lambda1750,1752$ emission doublet was not detected in any of the \citet{Stark+2014} sample, but lower limits were provided based on observational constraints. Despite the current non-detection, with more sensitive facilities the N\textsc{ii}]$\lambda1750$ / O\textsc{iii}]$\lambda1661,6$ line ratio (N2O3) could become a promising diagnostic when paired with Si3C3, which we show in the bottom right panel of Fig.~\ref{fig:dataHiZ}.

\section{Feasibility Estimates} \label{sec:feas}

In this section we discuss the feasibility of measuring important UV lines with existing and planned observatories. We quantify the utility of the diagnostics from \S\ref{sec:obs} with the launch of the James Webb Space Telescope (JWST), which will provide rest-frame UV spectra of galaxies at $z \geq 4$, and we identify potential diagnostics for the redshift ranges probed by the JWST spectrograph.

\begin{figure*}
  \begin{center}
    \includegraphics[width=\linewidth]{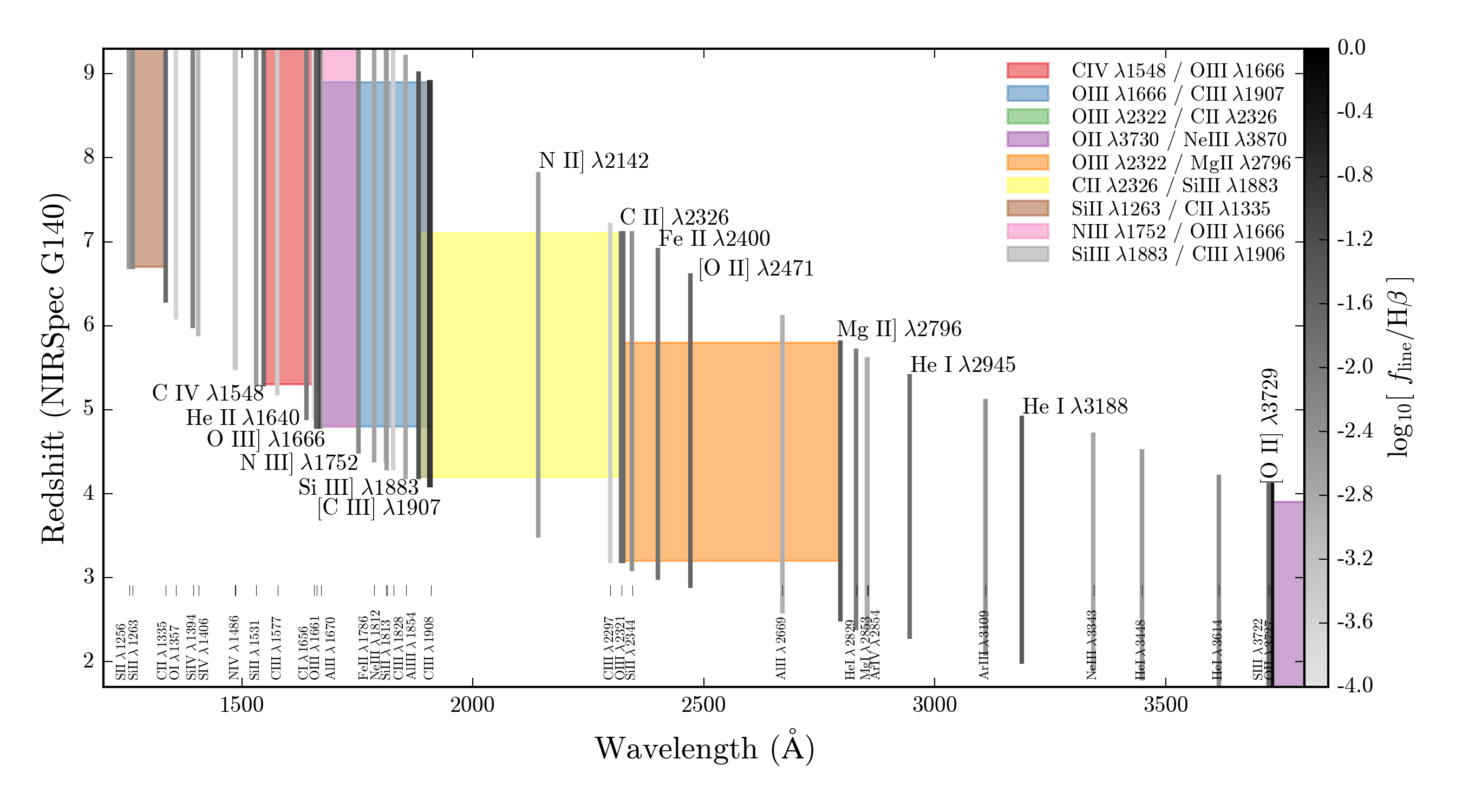}
    \caption{We show the relevant redshift ranges observable with JWST's NIRSpec instrument with the G140 grating. Each line is color-coded by its luminosity compared to the luminosity of \hb, assuming a 10\Myr constant SFR population with \logZeq{-0.5} and \logUeq{-2.5}.}
    \label{fig:jwst}
  \end{center}
\end{figure*}

In Fig.~\ref{fig:jwst}, we show observable UV emission lines as a function of the desired redshift range to be probed with JWST. Each emission line is color-coded by its strength relative to \hb for a fiducial model assuming a 10\Myr constant SFR, \logUeq{-2.5} and \logZeq{-0.5}, so users can determine the brightest emission lines in their desired redshift range. We also highlight the range of redshifts where each of the emission line ratios presented in Table~\ref{tab:ratios} are observable.

For the highest redshift galaxies expected to be discovered with JWST, $7 < z < 10$, low-resolution spectroscopy ($R\sim10-100$) can be obtained with NIRSpec \citep{Levesque+2015n}, allowing for the detection of the brightest emission lines like \ciii$\lambda\,1906,1909$ and \oiii$\lambda\,1661,1666$. In this regime, the O3C3 emission line ratio and the slope of the UV continuum are likely to be the most useful probes of the ISM properties. At these resolutions, the \ciii lines at 1906 and 1909\ang will be blended, and likely include the blended contribution from Si{\sc \,iii} line at 1883\ang.

For galaxies between $5 < z < 7$, NIRSpec will provide mid-resolution ($R\sim1000$) spectroscopy for reasonable exposure times. With sufficient S/N, the O3C3, Si3C3, and Ne3O2 diagnostics can provide constraints on the ionization state and metallicity of the gas to better understand the properties of galaxies during the epoch of their formation. Other emission line ratios presented in this work (e.g., N3O3, C4O3, He2C3, Si2C2, O3O2) can be used in this regime, however, the use of weak lines in these ratios is likely to prevent their application in all but the most extreme environments.

For a smaller subset of galaxies, high-resolution ($R\sim2700$) spectra can be obtained. These spectra will be able to resolve stellar absorption features, enabling the simultaneous study of the nebular emission and stellar population. With sufficient S/N to detect some of the weaker emission features, the diagnostics combining emission and absorption features will provide insight into the conversion of gas into stars, stellar feedback and chemical enrichment.

\section{Conclusions} \label{sec:conclusions}
We build upon the nebular model framework established in \citet{Byler+2017}, extending the self-consistent predictions for nebular line and continuum emission from stellar populations into the UV regime. With increasing sensitivity to high redshift galaxies, better tools to interpret rest-frame UV spectra will be increasingly important. Our conclusions are as follows:

\begin{itemize}
    \item We have identified UV emission lines that correlate with the gas phase metallicity and the hardness of the ionizing spectrum. We provide predicted UV emission line fluxes as a function of metallicity and ionization parameter in a machine-readable format. These will be fully integrated into \FSPS and included in future releases.
    \item We have identified combinations of emission lines that will be useful diagnostics for bulk properties like metallicity and ionization parameter. The combinations presented in this work are relatively bright and insensitive to reddening, and correlate with metallicity and ionization parameter.
    \item We present a joint analysis of stellar and nebular features in the UV. We determine which of the absorption line indices presented by \citet{Leitherer+2011} are the most useful metallicity diagnostics and quantify the contribution from nebular emission to each absorption index. We identify combinations of stellar absorption features and nebular emission lines that can be used as metallicity and ionization parameter diagnostics. We also quantify the contribution from stellar wind emission to nebular emission line ratios.
    \item We evaluate the emission and absorption diagnostics by comparing them with observed galaxies. We confirm that the diagnostics can reproduce observed emission and absorption from a sample of local BCDs. For line ratios where the contribution from stellar wind emission plays a role in one of the emission lines (e.g., \civ or \heii), the model predictions are unable to fully reproduce observations. We advise caution when interpreting line ratios where both emission lines could include a significant contribution from stellar winds, like \civ/\heii.
    \item We verify that the diagnostics hold across redshift by comparing them with observations of high redshift galaxies. In general, the model grids are consistent within errors of high redshift observations.
    \item We recommend diagnostics for the redshift ranges probed by the JWST spectrograph. The mid-resolution ($R\sim1000$) spectroscopy that NIRSpec will provide for galaxies between $5 < z < 7$ is ideal for the use of the emission line diagnostics presented here, including O3C3, Si3C3, S3C3, Si3N3, S3O3, and N3O3.
\end{itemize}

The analysis of both stellar and nebular features in the UV is a promising method to study the ISM in galaxies both locally and at high redshift. Future work fully considering the implications of variable star formation histories and gas phase abundances will anchor these diagnostics to physical conditions that may be more representative of those found in the early universe. We hope that these models can help guide future assessments of the UV spectra from star forming galaxies.

\acknowledgments

It is a pleasure to thank Grace Telford and Claus Leitherer for helpful discussions informing the ideas presented here. Special thanks to JJ Eldridge, Mason Ng, and Georgie Taylor for sharing with us WMBasic models \citep{Eldridge+2017}; to Erika Zetterlund and Charles Danforth for sharing with us the reduced spectra from \citet{Zetterlund+2015}; and to Danielle Berg, for sharing with us unpublished emission line measurements (Berg et al., in prep). C.C. acknowledges support from NASA grant NNX13AI46G, NSF grant AST- 1313280, and the Packard Foundation. N.B. acknowledges support from the University of Washington's Royalty Research Fund Grant 65-8055, and the Australian Research Council Centre of Excellence for All Sky Astrophysics in 3 Dimensions (ASTRO3D), through project number CE170100013. We would like to thank the anonymous referee for thorough and constructive feedback that greatly improved this work.

\software{\Cloudy\;v$13.03$ \citep{Ferland+2013},
          {\tt FSPS}\;v$3.0$ \citep{Conroy+2009, Conroy+2010},
          {\tt cloudyFSPS}\;v$1.0$ \citep{cloudyFSPSv1},
          {\tt python-fsps}\;v$0.1.1$ \citep{pythonFSPSdfm}}

\bibliographystyle{aasjournal}
\bibliography{main}

\begin{thebibliography}{}
\expandafter\ifx\csname natexlab\endcsname\relax\def\natexlab#1{#1}\fi
\providecommand{\url}[1]{\href{#1}{#1}}
\providecommand{\dodoi}[1]{doi:~\href{http://doi.org/#1}{\nolinkurl{#1}}}
\providecommand{\doeprint}[1]{\href{http://ascl.net/#1}{\nolinkurl{http://ascl.net/#1}}}
\providecommand{\doarXiv}[1]{\href{https://arxiv.org/abs/#1}{\nolinkurl{https://arxiv.org/abs/#1}}}

\bibitem[{{Abazajian} {et~al.}(2009){Abazajian}, {Adelman-McCarthy},
  {Ag{\"u}eros}, {Allam}, {Allende Prieto}, {An}, {Anderson}, {Anderson},
  {Annis}, {Bahcall}, \& et~al.}]{Abazajian+2009}
{Abazajian}, K.~N., {Adelman-McCarthy}, J.~K., {Ag{\"u}eros}, M.~A., {et~al.}
  2009, \apjs, 182, 543, \dodoi{10.1088/0067-0049/182/2/543}

\bibitem[{{Asplund} {et~al.}(2009){Asplund}, {Grevesse}, {Sauval}, \&
  {Scott}}]{Asplund+2009}
{Asplund}, M., {Grevesse}, N., {Sauval}, A.~J., \& {Scott}, P. 2009, \araa, 47,
  481, \dodoi{10.1146/annurev.astro.46.060407.145222}

\bibitem[{{Baldwin} {et~al.}(1981){Baldwin}, {Phillips}, \& {Terlevich}}]{BPT}
{Baldwin}, J.~A., {Phillips}, M.~M., \& {Terlevich}, R. 1981, \pasp, 93, 5,
  \dodoi{10.1086/130766}

\bibitem[{{Bayliss} {et~al.}(2014){Bayliss}, {Rigby}, {Sharon}, {Wuyts},
  {Florian}, {Gladders}, {Johnson}, \& {Oguri}}]{Bayliss+2014}
{Bayliss}, M.~B., {Rigby}, J.~R., {Sharon}, K., {et~al.} 2014, \apj, 790, 144,
  \dodoi{10.1088/0004-637X/790/2/144}

\bibitem[{{Berg} {et~al.}(2016){Berg}, {Skillman}, {Henry}, {Erb}, \&
  {Carigi}}]{Berg+2016}
{Berg}, D.~A., {Skillman}, E.~D., {Henry}, R.~B.~C., {Erb}, D.~K., \& {Carigi},
  L. 2016, \apj, 827, 126, \dodoi{10.3847/0004-637X/827/2/126}

\bibitem[{{Berg} {et~al.}(2012){Berg}, {Skillman}, {Marble}, {van Zee},
  {Engelbracht}, {Lee}, {Kennicutt}, {Calzetti}, {Dale}, \&
  {Johnson}}]{Berg+2012}
{Berg}, D.~A., {Skillman}, E.~D., {Marble}, A.~R., {et~al.} 2012, \apj, 754,
  98, \dodoi{10.1088/0004-637X/754/2/98}

\bibitem[{{Brinchmann} {et~al.}(2004){Brinchmann}, {Charlot}, {White},
  {Tremonti}, {Kauffmann}, {Heckman}, \& {Brinkmann}}]{Brinchmann+2004}
{Brinchmann}, J., {Charlot}, S., {White}, S.~D.~M., {et~al.} 2004, \mnras, 351,
  1151, \dodoi{10.1111/j.1365-2966.2004.07881.x}

\bibitem[{{Byler}(2018)}]{cloudyFSPSv1}
{Byler}, N. 2018, cloudyFSPS, 1.0.0,  Zenodo, \dodoi{10.5281/zenodo.1156412}.
\newblock \url{https://doi.org/10.5281/zenodo.1156412}

\bibitem[{{Byler} {et~al.}(2017){Byler}, {Dalcanton}, {Conroy}, \&
  {Johnson}}]{Byler+2017}
{Byler}, N., {Dalcanton}, J.~J., {Conroy}, C., \& {Johnson}, B.~D. 2017, \apj,
  840, 44, \dodoi{10.3847/1538-4357/aa6c66}

\bibitem[{{Canto Martins} {et~al.}(2011){Canto Martins}, {L{\`e}bre},
  {Palacios}, {de Laverny}, {Richard}, {Melo}, {Do Nascimento}, \& {de
  Medeiros}}]{Canto+2011}
{Canto Martins}, B.~L., {L{\`e}bre}, A., {Palacios}, A., {et~al.} 2011, \aap,
  527, A94, \dodoi{10.1051/0004-6361/201015015}

\bibitem[{{Cardelli} {et~al.}(1989){Cardelli}, {Clayton}, \&
  {Mathis}}]{Cardelli+1989}
{Cardelli}, J.~A., {Clayton}, G.~C., \& {Mathis}, J.~S. 1989, \apj, 345, 245,
  \dodoi{10.1086/167900}

\bibitem[{{Choi} {et~al.}(2017){Choi}, {Conroy}, \& {Byler}}]{Choi+2017}
{Choi}, J., {Conroy}, C., \& {Byler}, N. 2017, \apj, 838, 159,
  \dodoi{10.3847/1538-4357/aa679f}

\bibitem[{{Choi} {et~al.}(2016){Choi}, {Dotter}, {Conroy}, {Cantiello},
  {Paxton}, \& {Johnson}}]{Choi+2016}
{Choi}, J., {Dotter}, A., {Conroy}, C., {et~al.} 2016, \apj, 823, 102,
  \dodoi{10.3847/0004-637X/823/2/102}

\bibitem[{{Christensen} {et~al.}(2012){Christensen}, {Laursen}, {Richard},
  {Hjorth}, {Milvang-Jensen}, {Dessauges-Zavadsky}, {Limousin}, {Grillo}, \&
  {Ebeling}}]{Christensen+2012}
{Christensen}, L., {Laursen}, P., {Richard}, J., {et~al.} 2012, \mnras, 427,
  1973, \dodoi{10.1111/j.1365-2966.2012.22007.x}

\bibitem[{{Conroy} \& {Gunn}(2010)}]{Conroy+2010}
{Conroy}, C., \& {Gunn}, J.~E. 2010, \apj, 712, 833,
  \dodoi{10.1088/0004-637X/712/2/833}

\bibitem[{{Conroy} {et~al.}(2009){Conroy}, {Gunn}, \& {White}}]{Conroy+2009}
{Conroy}, C., {Gunn}, J.~E., \& {White}, M. 2009, \apj, 699, 486,
  \dodoi{10.1088/0004-637X/699/1/486}

\bibitem[{{Contini} {et~al.}(2002){Contini}, {Treyer}, {Sullivan}, \&
  {Ellis}}]{Contini+2002}
{Contini}, T., {Treyer}, M.~A., {Sullivan}, M., \& {Ellis}, R.~S. 2002, \mnras,
  330, 75, \dodoi{10.1046/j.1365-8711.2002.05042.x}

\bibitem[{{Cowley}(1995)}]{Cowley+1995}
{Cowley}, C.~R., ed. 1995, {An Introduction to Cosmochemistry}, ISBN0521415381

\bibitem[{{de Jager} {et~al.}(1988){de Jager}, {Nieuwenhuijzen}, \& {van der
  Hucht}}]{deJager+1988}
{de Jager}, C., {Nieuwenhuijzen}, H., \& {van der Hucht}, K.~A. 1988, \aaps,
  72, 259

\bibitem[{{de Mink} {et~al.}(2013){de Mink}, {Langer}, {Izzard}, {Sana}, \& {de
  Koter}}]{deMink+2013}
{de Mink}, S.~E., {Langer}, N., {Izzard}, R.~G., {Sana}, H., \& {de Koter}, A.
  2013, \apj, 764, 166, \dodoi{10.1088/0004-637X/764/2/166}

\bibitem[{{Ding} {et~al.}(2016){Ding}, {Cai}, {Fan}, {Stark}, {Bian}, {Jiang},
  {McGreer}, {Robertson}, \& {Siana}}]{Ding+2016}
{Ding}, J., {Cai}, Z., {Fan}, X., {et~al.} 2016, ArXiv e-prints.
\newblock \doarXiv{1612.00902}

\bibitem[{{Dopita} {et~al.}(2000){Dopita}, {Kewley}, {Heisler}, \&
  {Sutherland}}]{Dopita+2000}
{Dopita}, M.~A., {Kewley}, L.~J., {Heisler}, C.~A., \& {Sutherland}, R.~S.
  2000, \apj, 542, 224, \dodoi{10.1086/309538}

\bibitem[{{Dopita} {et~al.}(2013){Dopita}, {Sutherland}, {Nicholls}, {Kewley},
  \& {Vogt}}]{Dopita+2013}
{Dopita}, M.~A., {Sutherland}, R.~S., {Nicholls}, D.~C., {Kewley}, L.~J., \&
  {Vogt}, F.~P.~A. 2013, \apjs, 208, 10, \dodoi{10.1088/0067-0049/208/1/10}

\bibitem[{{Dopita} {et~al.}(2006){Dopita}, {Fischera}, {Sutherland}, {Kewley},
  {Leitherer}, {Tuffs}, {Popescu}, {van Breugel}, \& {Groves}}]{Dopita+2006}
{Dopita}, M.~A., {Fischera}, J., {Sutherland}, R.~S., {et~al.} 2006, \apjs,
  167, 177, \dodoi{10.1086/508261}

\bibitem[{{Dotter}(2016)}]{Dotter+2016}
{Dotter}, A. 2016, \apjs, 222, 8, \dodoi{10.3847/0067-0049/222/1/8}

\bibitem[{{Du} {et~al.}(2016){Du}, {Shapley}, {Martin}, \& {Coil}}]{Du+2016}
{Du}, X., {Shapley}, A.~E., {Martin}, C.~L., \& {Coil}, A.~L. 2016, ArXiv
  e-prints.
\newblock \doarXiv{1612.06866}

\bibitem[{{Ekstr{\"o}m} {et~al.}(2012){Ekstr{\"o}m}, {Georgy}, {Eggenberger},
  {Meynet}, {Mowlavi}, {Wyttenbach}, {Granada}, {Decressin}, {Hirschi},
  {Frischknecht}, {Charbonnel}, \& {Maeder}}]{Ekstrom+2012}
{Ekstr{\"o}m}, S., {Georgy}, C., {Eggenberger}, P., {et~al.} 2012, \aap, 537,
  A146, \dodoi{10.1051/0004-6361/201117751}

\bibitem[{{Eldridge} \& {Stanway}(2009)}]{Eldridge+2009}
{Eldridge}, J.~J., \& {Stanway}, E.~R. 2009, \mnras, 400, 1019,
  \dodoi{10.1111/j.1365-2966.2009.15514.x}

\bibitem[{{Eldridge} \& {Stanway}(2012)}]{Eldridge+2012}
---. 2012, \mnras, 419, 479, \dodoi{10.1111/j.1365-2966.2011.19713.x}

\bibitem[{{Eldridge} {et~al.}(2017){Eldridge}, {Stanway}, {Xiao}, {McClelland},
  {Taylor}, {Ng}, {Greis}, \& {Bray}}]{Eldridge+2017}
{Eldridge}, J.~J., {Stanway}, E.~R., {Xiao}, L., {et~al.} 2017, \pasa, 34,
  e058, \dodoi{10.1017/pasa.2017.51}

\bibitem[{{Erb} {et~al.}(2010){Erb}, {Pettini}, {Shapley}, {Steidel}, {Law}, \&
  {Reddy}}]{Erb+2010}
{Erb}, D.~K., {Pettini}, M., {Shapley}, A.~E., {et~al.} 2010, \apj, 719, 1168,
  \dodoi{10.1088/0004-637X/719/2/1168}

\bibitem[{{Esteban} {et~al.}(2014){Esteban}, {Garc{\'{\i}}a-Rojas}, {Carigi},
  {Peimbert}, {Bresolin}, {L{\'o}pez-S{\'a}nchez}, \&
  {Mesa-Delgado}}]{Esteban+2014}
{Esteban}, C., {Garc{\'{\i}}a-Rojas}, J., {Carigi}, L., {et~al.} 2014, \mnras,
  443, 624, \dodoi{10.1093/mnras/stu1177}

\bibitem[{{Feltre} {et~al.}(2016){Feltre}, {Charlot}, \&
  {Gutkin}}]{Feltre+2016}
{Feltre}, A., {Charlot}, S., \& {Gutkin}, J. 2016, \mnras, 456, 3354,
  \dodoi{10.1093/mnras/stv2794}

\bibitem[{{Ferland} {et~al.}(2013){Ferland}, {Porter}, {van Hoof}, {Williams},
  {Abel}, {Lykins}, {Shaw}, {Henney}, \& {Stancil}}]{Ferland+2013}
{Ferland}, G.~J., {Porter}, R.~L., {van Hoof}, P.~A.~M., {et~al.} 2013, \rmxaa,
  49, 137.
\newblock \doarXiv{1302.4485}

\bibitem[{{Foreman-Mackey} {et~al.}(2014){Foreman-Mackey}, {Sick}, \&
  {Johnson}}]{pythonFSPSdfm}
{Foreman-Mackey}, D., {Sick}, J., \& {Johnson}, B.~D. 2014, python-fsps: Python
  bindings to FSPS, 0.1.1,  Zenodo, \dodoi{10.5281/zenodo.12157}.
\newblock \url{https://doi.org/10.5281/zenodo.12157}

\bibitem[{{Garnett}(1990)}]{Garnett+1990}
{Garnett}, D.~R. 1990, \apj, 363, 142, \dodoi{10.1086/169324}

\bibitem[{{Garnett} {et~al.}(1999){Garnett}, {Shields}, {Peimbert},
  {Torres-Peimbert}, {Skillman}, {Dufour}, {Terlevich}, \&
  {Terlevich}}]{Garnett+1999}
{Garnett}, D.~R., {Shields}, G.~A., {Peimbert}, M., {et~al.} 1999, \apj, 513,
  168, \dodoi{10.1086/306860}

\bibitem[{{Georgy} {et~al.}(2013){Georgy}, {Ekstr{\"o}m}, {Granada}, {Meynet},
  {Mowlavi}, {Eggenberger}, \& {Maeder}}]{Georgy+2013}
{Georgy}, C., {Ekstr{\"o}m}, S., {Granada}, A., {et~al.} 2013, \aap, 553, A24,
  \dodoi{10.1051/0004-6361/201220558}

\bibitem[{{Grevesse} {et~al.}(2010){Grevesse}, {Asplund}, {Sauval}, \&
  {Scott}}]{Grevesse+2010}
{Grevesse}, N., {Asplund}, M., {Sauval}, A.~J., \& {Scott}, P. 2010, \apss,
  328, 179, \dodoi{10.1007/s10509-010-0288-z}

\bibitem[{{Gutkin} {et~al.}(2016){Gutkin}, {Charlot}, \&
  {Bruzual}}]{Gutkin+2016}
{Gutkin}, J., {Charlot}, S., \& {Bruzual}, G. 2016, \mnras, 462, 1757,
  \dodoi{10.1093/mnras/stw1716}

\bibitem[{{Henry} {et~al.}(2000){Henry}, {Edmunds}, \&
  {K{\"o}ppen}}]{Henry+2000}
{Henry}, R.~B.~C., {Edmunds}, M.~G., \& {K{\"o}ppen}, J. 2000, \apj, 541, 660,
  \dodoi{10.1086/309471}

\bibitem[{{Hillier} \& {Lanz}(2001)}]{Hillier+2001}
{Hillier}, D.~J., \& {Lanz}, T. 2001, in Astronomical Society of the Pacific
  Conference Series, Vol. 247, Spectroscopic Challenges of Photoionized
  Plasmas, ed. G.~{Ferland} \& D.~W. {Savin}, 343

\bibitem[{{Huang} {et~al.}(2010){Huang}, {Gies}, \& {McSwain}}]{Huang+2010}
{Huang}, W., {Gies}, D.~R., \& {McSwain}, M.~V. 2010, \apj, 722, 605,
  \dodoi{10.1088/0004-637X/722/1/605}

\bibitem[{{Jaskot} \& {Ravindranath}(2016)}]{Jaskot+2016}
{Jaskot}, A.~E., \& {Ravindranath}, S. 2016, \apj, 833, 136,
  \dodoi{10.3847/1538-4357/833/2/136}

\bibitem[{{Kauffmann} {et~al.}(2003{\natexlab{a}}){Kauffmann}, {Heckman},
  {White}, {Charlot}, {Tremonti}, {Brinchmann}, {Bruzual}, {Peng}, {Seibert},
  {Bernardi}, {Blanton}, {Brinkmann}, {Castander}, {Cs{\'a}bai}, {Fukugita},
  {Ivezic}, {Munn}, {Nichol}, {Padmanabhan}, {Thakar}, {Weinberg}, \&
  {York}}]{Kauffmann+2003a}
{Kauffmann}, G., {Heckman}, T.~M., {White}, S.~D.~M., {et~al.}
  2003{\natexlab{a}}, \mnras, 341, 33, \dodoi{10.1046/j.1365-8711.2003.06291.x}

\bibitem[{{Kauffmann} {et~al.}(2003{\natexlab{b}}){Kauffmann}, {Heckman},
  {Tremonti}, {Brinchmann}, {Charlot}, {White}, {Ridgway}, {Brinkmann},
  {Fukugita}, {Hall}, {Ivezi{\'c}}, {Richards}, \&
  {Schneider}}]{Kauffmann+2003b}
{Kauffmann}, G., {Heckman}, T.~M., {Tremonti}, C., {et~al.} 2003{\natexlab{b}},
  \mnras, 346, 1055, \dodoi{10.1111/j.1365-2966.2003.07154.x}

\bibitem[{{Kennicutt} {et~al.}(2003){Kennicutt}, {Bresolin}, \&
  {Garnett}}]{Kennicutt+2003}
{Kennicutt}, Jr., R.~C., {Bresolin}, F., \& {Garnett}, D.~R. 2003, \apj, 591,
  801, \dodoi{10.1086/375398}

\bibitem[{{Kewley} \& {Dopita}(2002)}]{Kewley+2002}
{Kewley}, L.~J., \& {Dopita}, M.~A. 2002, \apjs, 142, 35,
  \dodoi{10.1086/341326}

\bibitem[{{Kewley} {et~al.}(2001){Kewley}, {Dopita}, {Sutherland}, {Heisler},
  \& {Trevena}}]{Kewley+2001}
{Kewley}, L.~J., {Dopita}, M.~A., {Sutherland}, R.~S., {Heisler}, C.~A., \&
  {Trevena}, J. 2001, \apj, 556, 121, \dodoi{10.1086/321545}

\bibitem[{{Kobulnicky} {et~al.}(2014){Kobulnicky}, {Kiminki}, {Lundquist},
  {Burke}, {Chapman}, {Keller}, {Lester}, {Rolen}, {Topel}, {Bhattacharjee},
  {Smullen}, {Vargas {\'A}lvarez}, {Runnoe}, {Dale}, \&
  {Brotherton}}]{Kobulnicky+2014}
{Kobulnicky}, H.~A., {Kiminki}, D.~C., {Lundquist}, M.~J., {et~al.} 2014,
  \apjs, 213, 34, \dodoi{10.1088/0067-0049/213/2/34}

\bibitem[{{Kroupa}(2001)}]{Kroupa+2001}
{Kroupa}, P. 2001, \mnras, 322, 231, \dodoi{10.1046/j.1365-8711.2001.04022.x}

\bibitem[{{Kudritzki} \& {Puls}(2000)}]{Kudritzki+2000}
{Kudritzki}, R.-P., \& {Puls}, J. 2000, \araa, 38, 613,
  \dodoi{10.1146/annurev.astro.38.1.613}

\bibitem[{{Kudritzki} {et~al.}(2012){Kudritzki}, {Urbaneja}, {Gazak},
  {Bresolin}, {Przybilla}, {Gieren}, \& {Pietrzy{\'n}ski}}]{Kudritzki+2012}
{Kudritzki}, R.-P., {Urbaneja}, M.~A., {Gazak}, Z., {et~al.} 2012, \apj, 747,
  15, \dodoi{10.1088/0004-637X/747/1/15}

\bibitem[{{Kurucz}(2005)}]{Kurucz+2005}
{Kurucz}, R.~L. 2005, Memorie della Societa Astronomica Italiana Supplementi,
  8, 14

\bibitem[{{Langer}(1998)}]{Langer+1998}
{Langer}, N. 1998, \aap, 329, 551

\bibitem[{{Leitherer} {et~al.}(2011){Leitherer}, {Tremonti}, {Heckman}, \&
  {Calzetti}}]{Leitherer+2011}
{Leitherer}, C., {Tremonti}, C.~A., {Heckman}, T.~M., \& {Calzetti}, D. 2011,
  \aj, 141, 37, \dodoi{10.1088/0004-6256/141/2/37}

\bibitem[{{Levesque}(2015)}]{Levesque+2015n}
{Levesque}, E.~M. 2015, STSci Newsletter

\bibitem[{{Levesque} {et~al.}(2012){Levesque}, {Leitherer}, {Ekstrom},
  {Meynet}, \& {Schaerer}}]{Levesque+2012}
{Levesque}, E.~M., {Leitherer}, C., {Ekstrom}, S., {Meynet}, G., \& {Schaerer},
  D. 2012, \apj, 751, 67, \dodoi{10.1088/0004-637X/751/1/67}

\bibitem[{{Levesque} \& {Richardson}(2014)}]{Levesque+2014}
{Levesque}, E.~M., \& {Richardson}, M.~L.~A. 2014, \apj, 780, 100,
  \dodoi{10.1088/0004-637X/780/1/100}

\bibitem[{{Ma} {et~al.}(2016){Ma}, {Hopkins}, {Kasen}, {Quataert},
  {Faucher-Gigu{\`e}re}, {Kere{\v s}}, {Murray}, \& {Strom}}]{Ma+2016}
{Ma}, X., {Hopkins}, P.~F., {Kasen}, D., {et~al.} 2016, \mnras, 459, 3614,
  \dodoi{10.1093/mnras/stw941}

\bibitem[{{Mainali} {et~al.}(2017){Mainali}, {Kollmeier}, {Stark}, {Simcoe},
  {Walth}, {Newman}, \& {Miller}}]{Mainali+2017}
{Mainali}, R., {Kollmeier}, J.~A., {Stark}, D.~P., {et~al.} 2017, \apjl, 836,
  L14, \dodoi{10.3847/2041-8213/836/1/L14}

\bibitem[{{Maraston} {et~al.}(2009){Maraston}, {Nieves Colmen{\'a}rez},
  {Bender}, \& {Thomas}}]{Maraston+2009}
{Maraston}, C., {Nieves Colmen{\'a}rez}, L., {Bender}, R., \& {Thomas}, D.
  2009, \aap, 493, 425, \dodoi{10.1051/0004-6361:20066907}

\bibitem[{{McGaugh}(1991)}]{McGaugh+1991}
{McGaugh}, S.~S. 1991, \apj, 380, 140, \dodoi{10.1086/170569}

\bibitem[{{Nakajima} {et~al.}(2017){Nakajima}, {Schaerer}, {Le Fevre},
  {Amorin}, {Talia}, {Lemaux}, {Tasca}, {Vanzella}, {Zamorani}, {Bardelli},
  {Grazian}, {Guaita}, {Hathi}, {Pentericci}, \& {Zucca}}]{Nakajima+2017}
{Nakajima}, K., {Schaerer}, D., {Le Fevre}, O., {et~al.} 2017, ArXiv e-prints.
\newblock \doarXiv{1709.03990}

\bibitem[{{Nugis} \& {Lamers}(2000)}]{Nugis+2000}
{Nugis}, T., \& {Lamers}, H.~J.~G.~L.~M. 2000, \aap, 360, 227

\bibitem[{{Pagel} {et~al.}(1979){Pagel}, {Edmunds}, {Blackwell}, {Chun}, \&
  {Smith}}]{Pagel+1979}
{Pagel}, B.~E.~J., {Edmunds}, M.~G., {Blackwell}, D.~E., {Chun}, M.~S., \&
  {Smith}, G. 1979, \mnras, 189, 95, \dodoi{10.1093/mnras/189.1.95}

\bibitem[{{Pauldrach} {et~al.}(2001){Pauldrach}, {Hoffmann}, \&
  {Lennon}}]{Pauldrach+2001}
{Pauldrach}, A.~W.~A., {Hoffmann}, T.~L., \& {Lennon}, M. 2001, \aap, 375, 161,
  \dodoi{10.1051/0004-6361:20010805}

\bibitem[{{Paxton} {et~al.}(2011){Paxton}, {Bildsten}, {Dotter}, {Herwig},
  {Lesaffre}, \& {Timmes}}]{Paxton+2011}
{Paxton}, B., {Bildsten}, L., {Dotter}, A., {et~al.} 2011, \apjs, 192, 3,
  \dodoi{10.1088/0067-0049/192/1/3}

\bibitem[{{Paxton} {et~al.}(2013){Paxton}, {Cantiello}, {Arras}, {Bildsten},
  {Brown}, {Dotter}, {Mankovich}, {Montgomery}, {Stello}, {Timmes}, \&
  {Townsend}}]{Paxton+2013}
{Paxton}, B., {Cantiello}, M., {Arras}, P., {et~al.} 2013, \apjs, 208, 4,
  \dodoi{10.1088/0067-0049/208/1/4}

\bibitem[{{Paxton} {et~al.}(2015){Paxton}, {Marchant}, {Schwab}, {Bauer},
  {Bildsten}, {Cantiello}, {Dessart}, {Farmer}, {Hu}, {Langer}, {Townsend},
  {Townsley}, \& {Timmes}}]{Paxton+2015}
{Paxton}, B., {Marchant}, P., {Schwab}, J., {et~al.} 2015, \apjs, 220, 15,
  \dodoi{10.1088/0067-0049/220/1/15}

\bibitem[{{Pettini} \& {Pagel}(2004)}]{Pettini+2004}
{Pettini}, M., \& {Pagel}, B.~E.~J. 2004, \mnras, 348, L59,
  \dodoi{10.1111/j.1365-2966.2004.07591.x}

\bibitem[{{Rix} {et~al.}(2004){Rix}, {Pettini}, {Leitherer}, {Bresolin},
  {Kudritzki}, \& {Steidel}}]{Rix+2004}
{Rix}, S.~A., {Pettini}, M., {Leitherer}, C., {et~al.} 2004, \apj, 615, 98,
  \dodoi{10.1086/424031}

\bibitem[{{Rosen} {et~al.}(2012){Rosen}, {Krumholz}, \&
  {Ramirez-Ruiz}}]{Rosen+2012}
{Rosen}, A.~L., {Krumholz}, M.~R., \& {Ramirez-Ruiz}, E. 2012, \apj, 748, 97,
  \dodoi{10.1088/0004-637X/748/2/97}

\bibitem[{{Salim} {et~al.}(2007){Salim}, {Rich}, {Charlot}, {Brinchmann},
  {Johnson}, {Schiminovich}, {Seibert}, {Mallery}, {Heckman}, {Forster},
  {Friedman}, {Martin}, {Morrissey}, {Neff}, {Small}, {Wyder}, {Bianchi},
  {Donas}, {Lee}, {Madore}, {Milliard}, {Szalay}, {Welsh}, \&
  {Yi}}]{Salim+2007}
{Salim}, S., {Rich}, R.~M., {Charlot}, S., {et~al.} 2007, \apjs, 173, 267,
  \dodoi{10.1086/519218}

\bibitem[{{Sana} {et~al.}(2012){Sana}, {de Mink}, {de Koter}, {Langer},
  {Evans}, {Gieles}, {Gosset}, {Izzard}, {Le Bouquin}, \&
  {Schneider}}]{Sana+2012}
{Sana}, H., {de Mink}, S.~E., {de Koter}, A., {et~al.} 2012, Science, 337, 444,
  \dodoi{10.1126/science.1223344}

\bibitem[{{Sargent} \& {Searle}(1970)}]{Sargent+1970}
{Sargent}, W.~L.~W., \& {Searle}, L. 1970, \apjl, 162, L155,
  \dodoi{10.1086/180644}

\bibitem[{{Schmidt} {et~al.}(2017){Schmidt}, {Huang}, {Treu}, {Hoag}, {Brada{\v
  c}}, {Henry}, {Jones}, {Mason}, {Malkan}, {Morishita}, {Pentericci},
  {Trenti}, {Vulcani}, \& {Wang}}]{Schmidt+2017}
{Schmidt}, K.~B., {Huang}, K.-H., {Treu}, T., {et~al.} 2017, \apj, 839, 17,
  \dodoi{10.3847/1538-4357/aa68a3}

\bibitem[{{Senchyna} {et~al.}(2017){Senchyna}, {Stark}, {Vidal-Garc{\'{\i}}a},
  {Chevallard}, {Charlot}, {Mainali}, {Jones}, {Wofford}, {Feltre}, \&
  {Gutkin}}]{Senchyna+2017}
{Senchyna}, P., {Stark}, D.~P., {Vidal-Garc{\'{\i}}a}, A., {et~al.} 2017,
  \mnras, 472, 2608, \dodoi{10.1093/mnras/stx2059}

\bibitem[{{Shapley} {et~al.}(2003){Shapley}, {Steidel}, {Pettini}, \&
  {Adelberger}}]{Shapley+2003}
{Shapley}, A.~E., {Steidel}, C.~C., {Pettini}, M., \& {Adelberger}, K.~L. 2003,
  \apj, 588, 65, \dodoi{10.1086/373922}

\bibitem[{{Smith} {et~al.}(2002){Smith}, {Norris}, \& {Crowther}}]{Smith+2002}
{Smith}, L.~J., {Norris}, R.~P.~F., \& {Crowther}, P.~A. 2002, \mnras, 337,
  1309, \dodoi{10.1046/j.1365-8711.2002.06042.x}

\bibitem[{{Smith}(2014)}]{Smith+2014}
{Smith}, N. 2014, \araa, 52, 487, \dodoi{10.1146/annurev-astro-081913-040025}

\bibitem[{{Stanway} {et~al.}(2014){Stanway}, {Eldridge}, {Greis}, {Davies},
  {Wilkins}, \& {Bremer}}]{Stanway+2014}
{Stanway}, E.~R., {Eldridge}, J.~J., {Greis}, S.~M.~L., {et~al.} 2014, \mnras,
  444, 3466, \dodoi{10.1093/mnras/stu1682}

\bibitem[{{Stark} {et~al.}(2014){Stark}, {Richard}, {Siana}, {Charlot},
  {Freeman}, {Gutkin}, {Wofford}, {Robertson}, {Amanullah}, {Watson}, \&
  {Milvang-Jensen}}]{Stark+2014}
{Stark}, D.~P., {Richard}, J., {Siana}, B., {et~al.} 2014, \mnras, 445, 3200,
  \dodoi{10.1093/mnras/stu1618}

\bibitem[{{Stark} {et~al.}(2015){Stark}, {Richard}, {Charlot}, {Cl{\'e}ment},
  {Ellis}, {Siana}, {Robertson}, {Schenker}, {Gutkin}, \&
  {Wofford}}]{Stark+2015}
{Stark}, D.~P., {Richard}, J., {Charlot}, S., {et~al.} 2015, \mnras, 450, 1846,
  \dodoi{10.1093/mnras/stv688}

\bibitem[{{Steidel} {et~al.}(2016){Steidel}, {Strom}, {Pettini}, {Rudie},
  {Reddy}, \& {Trainor}}]{Steidel+2016}
{Steidel}, C.~C., {Strom}, A.~L., {Pettini}, M., {et~al.} 2016, \apj, 826, 159,
  \dodoi{10.3847/0004-637X/826/2/159}

\bibitem[{{Steidel} {et~al.}(2014){Steidel}, {Rudie}, {Strom}, {Pettini},
  {Reddy}, {Shapley}, {Trainor}, {Erb}, {Turner}, {Konidaris}, {Kulas}, {Mace},
  {Matthews}, \& {McLean}}]{Steidel+2014}
{Steidel}, C.~C., {Rudie}, G.~C., {Strom}, A.~L., {et~al.} 2014, \apj, 795,
  165, \dodoi{10.1088/0004-637X/795/2/165}

\bibitem[{{Stroe} {et~al.}(2017{\natexlab{a}}){Stroe}, {Sobral}, {Matthee},
  {Calhau}, \& {Oteo}}]{Stroe+2017a}
{Stroe}, A., {Sobral}, D., {Matthee}, J., {Calhau}, J., \& {Oteo}, I.
  2017{\natexlab{a}}, \mnras, 471, 2575, \dodoi{10.1093/mnras/stx1713}

\bibitem[{{Stroe} {et~al.}(2017{\natexlab{b}}){Stroe}, {Sobral}, {Matthee},
  {Calhau}, \& {Oteo}}]{Stroe+2017b}
---. 2017{\natexlab{b}}, \mnras, 471, 2558, \dodoi{10.1093/mnras/stx1712}

\bibitem[{{Telford} {et~al.}(2016){Telford}, {Dalcanton}, {Skillman}, \&
  {Conroy}}]{Telford+2016}
{Telford}, O.~G., {Dalcanton}, J.~J., {Skillman}, E.~D., \& {Conroy}, C. 2016,
  \apj, 827, 35, \dodoi{10.3847/0004-637X/827/1/35}

\bibitem[{{Thuan}(2008)}]{Thuan+2008}
{Thuan}, T.~X. 2008, in IAU Symposium, Vol. 255, Low-Metallicity Star
  Formation: From the First Stars to Dwarf Galaxies, ed. L.~K. {Hunt}, S.~C.
  {Madden}, \& R.~{Schneider}, 348--360

\bibitem[{{van Zee} {et~al.}(1998){van Zee}, {Salzer}, \&
  {Haynes}}]{vanZee+1998}
{van Zee}, L., {Salzer}, J.~J., \& {Haynes}, M.~P. 1998, \apjl, 497, L1,
  \dodoi{10.1086/311263}

\bibitem[{{Vidal-Garc{\'{\i}}a} {et~al.}(2017){Vidal-Garc{\'{\i}}a}, {Charlot},
  {Bruzual}, \& {Hubeny}}]{Vidal-Garcia+2017}
{Vidal-Garc{\'{\i}}a}, A., {Charlot}, S., {Bruzual}, G., \& {Hubeny}, I. 2017,
  \mnras, 470, 3532, \dodoi{10.1093/mnras/stx1324}

\bibitem[{{Vila Costas} \& {Edmunds}(1993)}]{VilaCostas+1993}
{Vila Costas}, M.~B., \& {Edmunds}, M.~G. 1993, \mnras, 265, 199,
  \dodoi{10.1093/mnras/265.1.199}

\bibitem[{{Vink} {et~al.}(2000){Vink}, {de Koter}, \& {Lamers}}]{Vink+2000}
{Vink}, J.~S., {de Koter}, A., \& {Lamers}, H.~J.~G.~L.~M. 2000, \aap, 362, 295

\bibitem[{{Vink} {et~al.}(2001){Vink}, {de Koter}, \& {Lamers}}]{Vink+2001}
---. 2001, \aap, 369, 574, \dodoi{10.1051/0004-6361:20010127}

\bibitem[{{Wolff} \& {Simon}(1997)}]{Wolff+1997}
{Wolff}, S., \& {Simon}, T. 1997, \pasp, 109, 759, \dodoi{10.1086/133942}

\bibitem[{{Xiao} {et~al.}(2018){Xiao}, {Stanway}, \& {Eldridge}}]{Xiao+2018}
{Xiao}, L., {Stanway}, E., \& {Eldridge}, J.~J. 2018, ArXiv e-prints.
\newblock \doarXiv{1801.07068}

\bibitem[{{York} {et~al.}(2000){York}, {Adelman}, {Anderson}, {Anderson},
  {Annis}, {Bahcall}, {Bakken}, {Barkhouser}, {Bastian}, {Berman}, {Boroski},
  {Bracker}, {Briegel}, {Briggs}, {Brinkmann}, {Brunner}, {Burles}, {Carey},
  {Carr}, {Castander}, {Chen}, {Colestock}, {Connolly}, {Crocker}, {Csabai},
  {Czarapata}, {Davis}, {Doi}, {Dombeck}, {Eisenstein}, {Ellman}, {Elms},
  {Evans}, {Fan}, {Federwitz}, {Fiscelli}, {Friedman}, {Frieman}, {Fukugita},
  {Gillespie}, {Gunn}, {Gurbani}, {de Haas}, {Haldeman}, {Harris}, {Hayes},
  {Heckman}, {Hennessy}, {Hindsley}, {Holm}, {Holmgren}, {Huang}, {Hull},
  {Husby}, {Ichikawa}, {Ichikawa}, {Ivezi{\'c}}, {Kent}, {Kim}, {Kinney},
  {Klaene}, {Kleinman}, {Kleinman}, {Knapp}, {Korienek}, {Kron}, {Kunszt},
  {Lamb}, {Lee}, {Leger}, {Limmongkol}, {Lindenmeyer}, {Long}, {Loomis},
  {Loveday}, {Lucinio}, {Lupton}, {MacKinnon}, {Mannery}, {Mantsch}, {Margon},
  {McGehee}, {McKay}, {Meiksin}, {Merelli}, {Monet}, {Munn}, {Narayanan},
  {Nash}, {Neilsen}, {Neswold}, {Newberg}, {Nichol}, {Nicinski}, {Nonino},
  {Okada}, {Okamura}, {Ostriker}, {Owen}, {Pauls}, {Peoples}, {Peterson},
  {Petravick}, {Pier}, {Pope}, {Pordes}, {Prosapio}, {Rechenmacher}, {Quinn},
  {Richards}, {Richmond}, {Rivetta}, {Rockosi}, {Ruthmansdorfer}, {Sandford},
  {Schlegel}, {Schneider}, {Sekiguchi}, {Sergey}, {Shimasaku}, {Siegmund},
  {Smee}, {Smith}, {Snedden}, {Stone}, {Stoughton}, {Strauss}, {Stubbs},
  {SubbaRao}, {Szalay}, {Szapudi}, {Szokoly}, {Thakar}, {Tremonti}, {Tucker},
  {Uomoto}, {Vanden Berk}, {Vogeley}, {Waddell}, {Wang}, {Watanabe},
  {Weinberg}, {Yanny}, {Yasuda}, \& {SDSS Collaboration}}]{York+2000}
{York}, D.~G., {Adelman}, J., {Anderson}, Jr., J.~E., {et~al.} 2000, \aj, 120,
  1579, \dodoi{10.1086/301513}

\bibitem[{{Zetterlund} {et~al.}(2015){Zetterlund}, {Levesque}, {Leitherer}, \&
  {Danforth}}]{Zetterlund+2015}
{Zetterlund}, E., {Levesque}, E.~M., {Leitherer}, C., \& {Danforth}, C.~W.
  2015, \apj, 805, 151, \dodoi{10.1088/0004-637X/805/2/151}

\end{thebibliography}
\appendix

\section{Emission Line List}\label{appdx:lines}

In Table~\ref{tab:emLines} we provide a list of the emission lines included in the \CloudyFSPS nebular model. The list of emission lines used in this work contains a total of 381 emission lines, expanded from the original \citet{Byler+2017} list of 128 emission lines. For each emission line, Table~\ref{tab:emLines} provides the element name and ionization state, the wavelength in vacuum, and the named used to track the emission line within \Cloudy. Only a portion of the list is given in Table~\ref{tab:emLines}. The table is available online in a machine readable format, and is available for download at the \CloudyFSPS project website (\url{http://nell-byler.github.io/cloudyfsps/}).

\startlongtable
\begin{deluxetable}{lll}
\tabletypesize{\footnotesize}
\tablecaption{Emission lines included in the \FSPS nebular model\label{tab:emLines}}
\tablehead{
\colhead{Vacuum Wavelength (\ang)} &
\colhead{Line ID} &
\colhead{\Cloudy ID}}
\startdata
917.473 & O {\sc\,i} 917.473A & \texttt{O  1 917.473A}\\
917.726 & O {\sc\,i} 917.726A & \texttt{O  1 917.726A}\\
917.97 & O {\sc\,i} 917.970A & \texttt{O  1 917.970A}\\
918.147 & P {\sc\,iii} 918.147A & \texttt{P  3 918.147A}\\
918.493 & Ar {\sc\,i} 918.493A & \texttt{Ar 1 918.493A}\\
918.704 & O {\sc\,i} 918.704A & \texttt{O  1 918.704A}\\
919.912 & O {\sc\,i} 919.912A & \texttt{O  1 919.912A}\\
920.243 & O {\sc\,i} 920.243A & \texttt{O  1 920.243A}\\
920.969 & H {\sc\,i} 920.969A & \texttt{H  1 920.969A}\\
922.509 & Ar {\sc\,i} 922.509A & \texttt{Ar 1 922.509A}\\
922.969 & O {\sc\,i} 922.969A & \texttt{O  1 922.969A}\\
923.148 & Ly-8 923 & \texttt{H  1 923.156A}\\
923.438 & O {\sc\,i} 923.438A & \texttt{O  1 923.438A}\\
924.009 & Cr {\sc\,iii} 924.009A & \texttt{Cr 3 924.009A}\\
926.249 & Ly-7 926 & \texttt{H  1 926.231A}\\
927.472 & O {\sc\,i} 927.472A & \texttt{O  1 927.472A}\\
928.186 & O {\sc\,i} 928.186A & \texttt{O  1 928.186A}\\
929.014 & Al {\sc\,ii} 929.014A & \texttt{Al 2 929.014A}\\
930.751 & Ly-6 930 & \texttt{H  1 930.754A}\\
934.501 & O {\sc\,i} 934.501A & \texttt{O  1 934.501A}\\
935.672 & O {\sc\,i} 935.672A & \texttt{O  1 935.672A}\\
937.814 & Ly-5 937 & \texttt{H  1 937.809A}\\
942.258 & N {\sc\,i} 942.258A & \texttt{N  1 942.258A}\\
943.939 & C {\sc\,i} 943.939A & \texttt{C  1 943.939A}\\
\enddata
\tablecomments{Only a portion of this table is shown here to demonstrate its form and content. A machine-readable version of the full table is available online. The line list is also available for download at the \CloudyFSPS project website: \url{http://nell-byler.github.io/cloudyfsps/}.}
\end{deluxetable}

\section{Sensitivity to C/O prescription} \label{appdx:CO}

Photoionization models must make various assumptions about the relative gas phase abundances of different elements. In general, nebular models assume that the abundances of all elements increase linearly with increasing metallicity. Specific choices are usually motivated by the desire to match the observed properties of samples of \hii regions. Most emission line strengths will depend linearly on their elemental abundances, to first order, but variations in important gas coolants like C, N, and O have additional effects on the \hii region structure and temperature, which in turn complicates the response of emission lines to changes in their abundances.

The treatment of relative elemental abundances is further complicated by the likelihood that both C and N scale non-linearly with metallicity. Nitrogen has known secondary nucleosynthetic production at high metallicity, wherein nitrogen is dredged up during the bottleneck step of the CNO cycle, which is directly dependent on metallicity. In the case of carbon, the triple-$\alpha$ process is the only known production process, but it does not depend on metallicity. Instead, additional carbon is produced through metallicity-dependent processes such as stellar winds, rather than a nucleosynthetic process. Thus, carbon is said to have a ``pseudo-secondary'' production process.

Nitrogen abundances are relatively easy to determine for local galaxies with optical spectra, and the relationship between N/O with metallicity has been well-studied \citep[e.g.,][]{Garnett+1990, vanZee+1998, Berg+2012}. Based on existing data, most photoionization models adopt a functional form for the relationship between the oxygen abundance and the N/O ratio that matches observations of local dwarf galaxies, massive extragalactic \hii regions, and starburst nucleus galaxies.

The C/O relationship is particularly relevant to this work, since \ciii is one of the strongest emission lines in the UV and is included in many of the emission line diagnostics presented in this work. Historically, carbon has been a difficult element to derive absolute abundances for, since few strong transitions exist in the optical or IR, and the optical recombination lines (RLs) become too faint to detect below $12+\log(O/H)=8$ \citep{Esteban+2014}.

\begin{figure*}
  \begin{center}
    \plotone{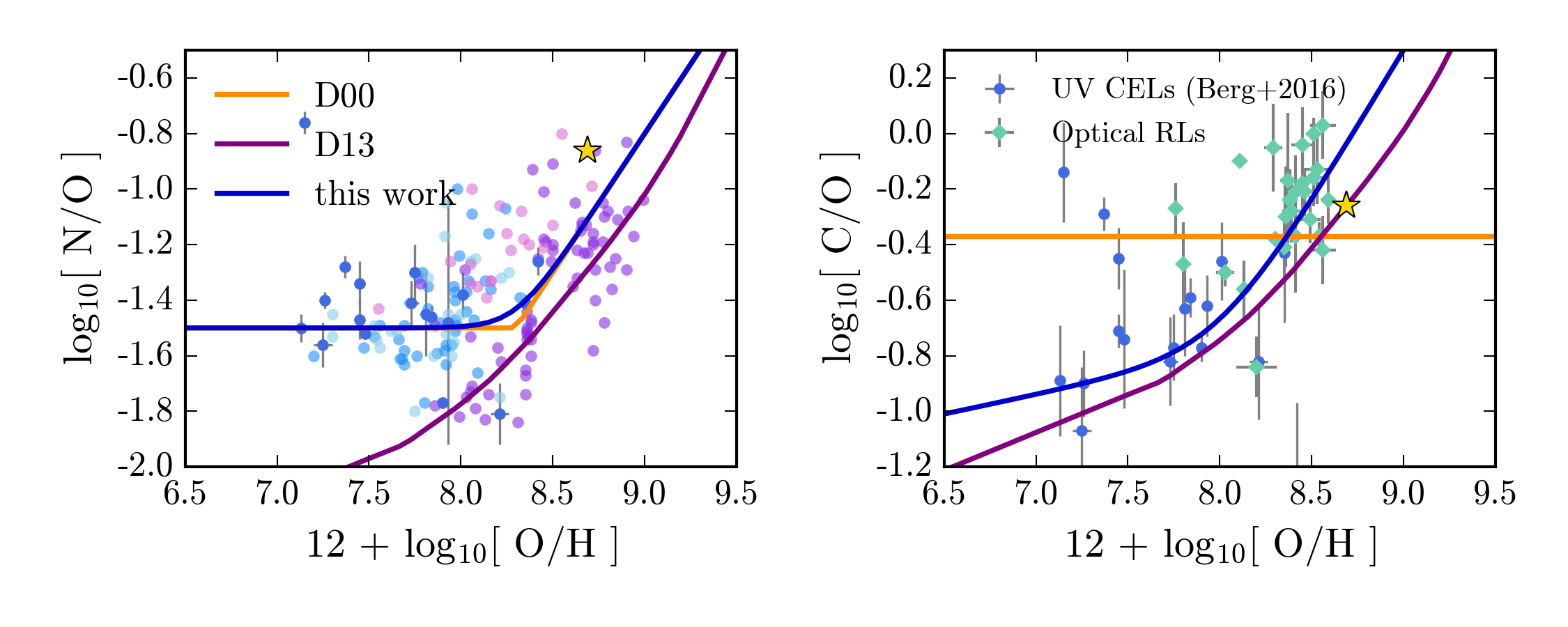}
    \caption{\emph{Left:} N/O relationships used by different nebular models, including the abundance prescription used in this work and those used in \citet{Dopita+2000} (``D00'') and \citet{Dopita+2013} (``D13''). The dark and light blue points are from \citet{Berg+2016} and \citet{Berg+2012}, respectively. The purple points are starburst galaxies from \citet{Contini+2002}, while the pink points show \hii regions in M101 from \citet{Kennicutt+2003}. \emph{Right:} C/O relationships used by different nebular models. The blue points are from \citet{Berg+2016}, with C/O abundances derived from UV collisionally excited lines. The teal points are extragalactic recombination line estimates from \citet{Esteban+2014}. The gold star in both plots represents where the Sun would be located according to the \citet{Asplund+2009} solar abundance set.}
    \label{fig:CNO}
  \end{center}
\end{figure*}

Recently, \citet{Berg+2016} analyzed the relationship of C/O with metallicity in low-metallicity galaxies using collisionally-excited lines (CELs) for C and O in the UV. \citet{Berg+2016} found that the C/O ratio was roughly constant across the metallicity range of their sample ($7<12+\log(O/H)<8$). However, when combined with C/O measurements based on optical recombination lines from galaxies at higher metallicities, C/O appears to increase with metallicity above $12+\log(O/H)=8$. These data are shown in the right panel of Fig.~\ref{fig:CNO}.

In Fig.~\ref{fig:CNO}, we show the existing N/O (\emph{left}) and C/O (\emph{right}) measurements along with the relationships between N/O and C/O used here and in various other nebular models. For both N/O and C/O, we have modified the \citet{Dopita+2013} prescription such that it plateaus for $12+\log[ O/H ] \lesssim 8$ to better match observations. The exact functional form is given in Eqs.~\ref{eq:nitrogen} \& \ref{eq:carbon} in \S\ref{sec:model:neb}, and shown as the blue line in Fig.~\ref{fig:CNO}. As noted in \citet{Steidel+2014}, the precise behavior of the N/O ratio with metallicity depends on the calibration sample and the details of the methods used to measure the abundances, and unsurprisingly there is a substantial range in the N/O versus O/H relationships applied in recent literature.

\subsection{Affect on UV and optical emission lines}

\begin{figure*}
  \begin{center}
    \plottwo{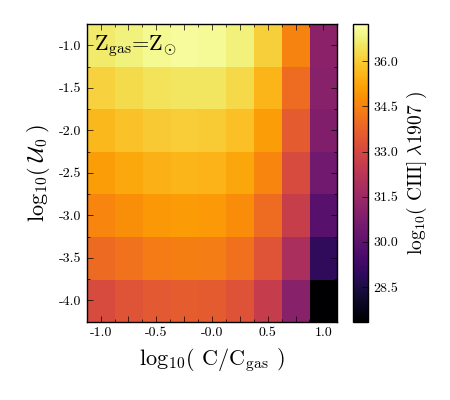}{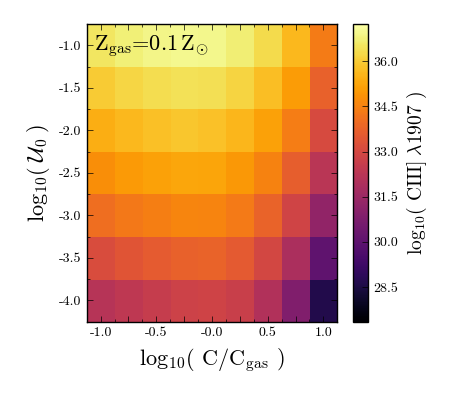}
    \caption{The variation in \ciii$\lambda1907$ line strength as a function of ionization parameter as the abundance of C is changed at constant $12+\log_{10}(\mathrm{O}/\mathrm{H})$ abundance for a 10\Myr CSFR model at solar (\emph{left}) and 10\% solar (\emph{right}) metallicity. \ciii is sensitive to ionization parameter and moderately sensitive to gas density. \ciii is sensitive to the overall gas phase metallicity; at fixed oxygen abundance, a two order of magnitude change in carbon abundance produces relatively little variation in line strength.}
    \label{fig:COvarA}
  \end{center}
\end{figure*}

The behavior of both the N/O and C/O ratio are important for the UV emission line diagnostics presented in this work. In Fig.~\ref{fig:COvarA} we show how the \ciii$\lambda$1907 emission line changes while varying C/O at constant oxygen abundance. The left panel of Fig.~\ref{fig:COvarA} demonstrates that at constant $12 + \log (\mathrm{O}/\mathrm{H})$, while \ciii$\lambda$1907 emission depends strongly on ionization parameter (varying from $10^{32}$ to $10^{36}$ergs$\cdot$s$^{-1}$), the strength of the line is essentially independent of the actual carbon abundance.

\begin{figure*}
  \begin{center}
    \plottwo{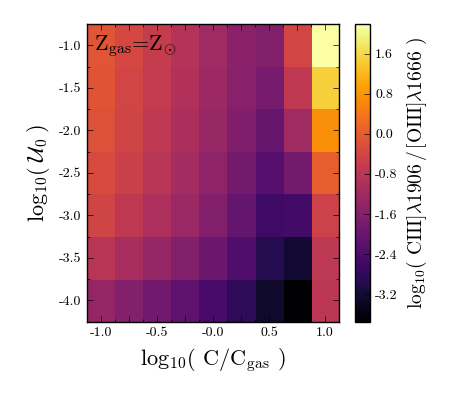}{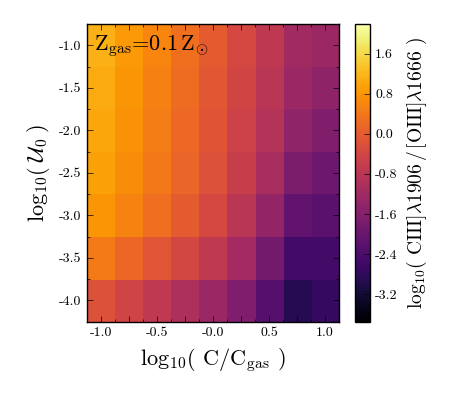}
    \caption{The variation in the \oiii$\lambda1666$/\ciii$\lambda1907$ (O3C3) emission line ratio as a function of ionization parameter as the abundance of C is decreased at constant $12+\log[O/H]$ abundance for a 10\Myr CSFR model at solar (\emph{left}) and 10\% solar (\emph{right}) metallicity. The decrease in C abundance inhibits cloud cooling, raising the temperature of the nebula slightly, producing variations in oxygen line strength at constant $12+\log[O/H]$. The O3C3 line ratio is very sensitive to ionization parameter, metallicity, and the C/O ratio.}
    \label{fig:COvarB}
  \end{center}
\end{figure*}

While \ciii$\lambda$1907 is not a direct measure of the carbon abundance on its own, when \ciii$\lambda$1907 is paired with \oiii$\lambda$1666 it becomes much more sensitive to the relative abundance of carbon. Fig.\ref{fig:COvarB} shows the \oiii/\ciii emission line ratio as a function of ionization parameter and carbon abundance at constant $12 + \log_{10} (\mathrm{O}/\mathrm{H})$. The decreasing abundance of carbon ultimately inhibits cloud cooling, which raises the temperature of the nebula slightly and produces variations in oxygen line strength at constant $12 + \log_{10} (\mathrm{O}/\mathrm{H})$. The O3C3 line ratio is very sensitive to ionization parameter and the C/O ratio while being relatively insensitive to gas density.

\begin{figure*}
  \begin{center}
    \plotone{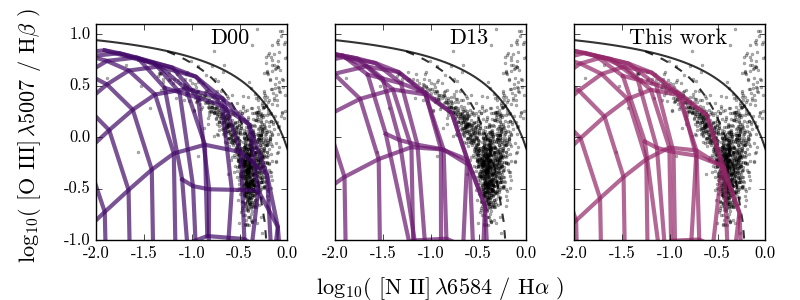}
    \plotone{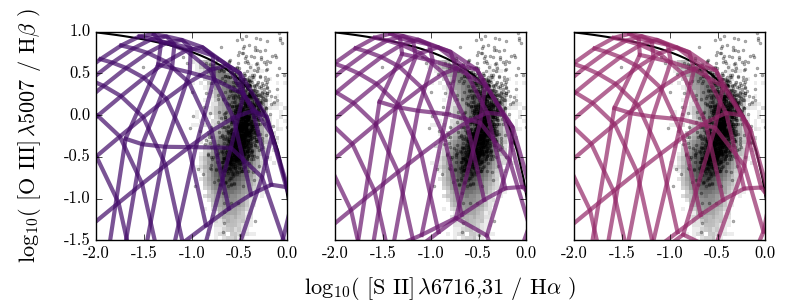}
    \plotone{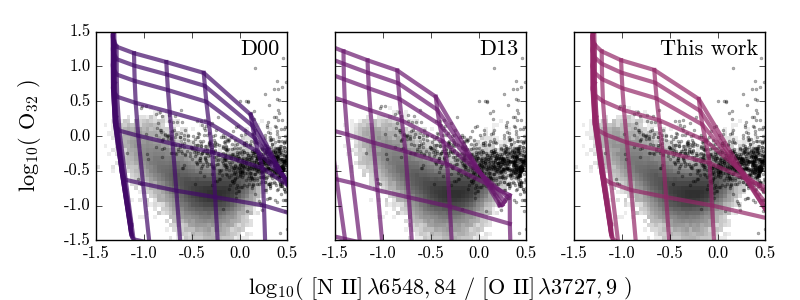}
    \caption{Model grids on the standard BPT diagram (top row) and optical diagnostic diagrams (middle, bottom rows). The greyscale 2D histogram shows the number density of SDSS star-forming galaxies from the \citet{Telford+2016} sample, the black points show a random sample of SDSS galaxies. The solid black line is the extreme starburst classification line from \citet{Kewley+2001} and the dashed line is the pure star formation classification line from \citet{Kauffmann+2003a}. All models assume a 1 Myr instantaneous burst and otherwise identical input parameters. Each panel represents a model run with a different abundance prescription, as noted in the upper right corner of each plot, and shown in Fig.~\ref{fig:CNO}.}
    \label{fig:BPT}
  \end{center}
\end{figure*}

Despite our modifications to previous abundance prescriptions, our nebular model still produces BPT diagram line ratios that are consistent with observations. In Fig.~\ref{fig:BPT} we show model grids on several diagnostic diagrams assuming the three different abundance prescriptions considered in Fig.~\ref{fig:CNO}: the abundance prescription used in this work, the \citet{Dopita+2013} prescription (``D13''), and the \citet{Dopita+2000} prescription (``D00''). The models are run for identical input parameters and the only modification is the mixture of gas-phase abundances specified in \Cloudy.

To show where observed galaxies lie in each of these diagnostic diagrams, we include a 2D histogram showing the number density of star forming galaxies, as done in \citet{Byler+2017}. The star-forming galaxy sample is derived from galaxy spectra from the Sloan Digital Sky Survey Data Release 7 \citep[SDSS DR7;][]{York+2000, Abazajian+2009} and emission line fluxes measured from the publicly available SDSS DR7 MPA/JHU catalog \citep{Kauffmann+2003a, Brinchmann+2004, Salim+2007}. We use the emission line sample presented in \citet{Telford+2016}, briefly summarized here. The sample includes $\sim 135,000$ galaxies with redshifts between 0.07 and 0.30. Galaxies are required to have S/N of 25 in the \ha{} line, 5 in the \hb{} line, and 3 in the \sii{} lines. Emission line fluxes are corrected for dust extinction using the Balmer decrement and the \citet{Cardelli+1989} extinction law, assuming $R_{\mathrm{V}} = 3.1$ and an intrinsic Balmer decrement of 2.86. AGNs are removed from the sample according to the empirical BPT diagram classification of \citet{Kauffmann+2003b}. We also show a random sample of SDSS galaxies (i.e., including AGN) with black circles.

The top panel in Fig.~\ref{fig:BPT} shows the standard BPT diagram. The model grids shown are run using identical input parameters, differing only in the mixture of gas-phase abundances specified in \Cloudy. However, the resultant model grids vary by 0.4 dex in \nii/\ha. The model presented here is able to reproduce the star-forming sequence quite well, while simultaneously producing UV emission line ratios consistent with observed galaxies. The other abundance prescriptions struggle with this: the \citet{Dopita+2013} nitrogen abundances are too low to correctly reproduce the optical star-forming sequence, though they do well at predicting the UV \ciii behavior. The \citet{Dopita+2000} model is able to reproduce the optical star-forming sequence, but does not include the additional pseudo-secondary contribution for carbon needed to explain the UV observations.

We note that Fig.~\ref{fig:BPT} is only intended to highlight how changes in the adopted relative elemental abundances can change emission line ratio predictions. The behavior displayed in Fig.~\ref{fig:BPT} is not representative of the other models' ability to reproduce observed line ratios, as each of these models can and does adequately reproduce the locus of star-forming galaxies through adopting different physical parameters than that used in \CloudyFSPS. The grids shown in Fig.~\ref{fig:BPT} use the elemental abundance prescriptions from the above models, but are all run using the specific combination of input parameters chosen for the \CloudyFSPS nebular model.

\end{document}